\documentclass[12pt]{article}
\pdfoutput=1
\usepackage{amssymb,amsmath}
\usepackage{mathtools}
\usepackage{graphicx}
\usepackage{epsfig}
\usepackage{epstopdf}
\usepackage{setspace}
\usepackage{latexsym}
\usepackage{tikz}
\usepackage{psfrag}
\usepackage{booktabs}
\usepackage{bbm}
\usepackage[toc]{appendix}
\usepackage{color}
\usepackage{datetime}
\usepackage{tensor}
\usepackage[
      colorlinks=true,
      linkcolor=darkblue,  
      urlcolor=blue,    
      filecolor=blue,     
      citecolor=red,
      linktocpage=true,
      pdfstartview=FitV,
      bookmarksopen=true    
      ]{hyperref}

\definecolor{darkblue}{rgb}{0.2, 0, 0.8}

\numberwithin{equation}{section}

\newcommand{\sq}[1]{\left[#1\right]}
\newcommand{\ang}[1]{\langle #1\rangle}

\usepackage{cite,dsfont}
\usepackage{amsfonts}
\usepackage{amssymb}
\usepackage{amsmath}
\usepackage{graphicx}
\usepackage{slashed}
\usepackage{setspace}

\usepackage{color}

\newcommand{\reef}[1]{(\ref{#1})}

\newcommand{\be}{\begin{equation}}
\newcommand{\ee}{\end{equation}}

\def\be{\begin{equation}}
\def\ee{\end{equation}}
\def\bea{\begin{eqnarray}}
\def\eea{\end{eqnarray}}
\def\ba{\begin{array}}
\def\ea{\end{array}}
\def\bd{\begin{displaymath}}
\def\ed{\end{displaymath}}

\def\tr{{\rm tr}}
\def\ra{\rangle}
\def\la{\langle}

\def\pa{\partial}                              
\def\>{\rangle} 
\def\<{\langle} 
\def\Dsl{D \hskip-.6em \raise1pt\hbox{$ / $ } }
\def\to{\rightarrow}
\def\pa{\partial}

\newcommand{\eps}{\epsilon}
\newcommand{\lra}{\leftrightarrow}

\begin{document}  


\begin{titlepage}

 \begin{flushright}
{\tt LCTP-18-16} \\
\end{flushright}

\vspace*{2cm}

\begin{center}
  \setstretch{1.5}
{\Large \bf Soft Bootstrap and Supersymmetry} \\
  \setstretch{1.0}
\vspace*{1.2cm}

\hspace{-0.14in}{\bf Henriette Elvang, Marios Hadjiantonis, \\ Callum R.~T.~Jones, 
and Shruti Paranjape}
\medskip

Leinweber Center for Theoretical Physics,\\ 
Randall Laboratory of Physics, Department of Physics,\\
University of Michigan, Ann Arbor, MI 48109, USA

\bigskip
{\small elvang, mhadjian, jonescal, shrpar@umich.edu}  \\
\end{center}

\vspace*{-.6cm}

\begin{abstract}  
The soft bootstrap is an on-shell method to constrain the landscape of effective field theories (EFTs) of massless particles via the consistency of the low-energy S-matrix. 
Given assumptions on the on-shell data (particle spectra, linear
symmetries, and low-energy theorems), the soft bootstrap is an
efficient algorithm for determining the possible consistency of an EFT
with those properties. 
The implementation of the soft bootstrap uses the recently discovered method of soft subtracted recursion. We derive a precise criterion for the validity of these recursion relations and show that they fail exactly when the assumed symmetries can be trivially realized by independent operators in the effective action. We use this to show that the possible pure (real and complex) scalar, fermion, and vector exceptional EFTs are highly constrained. Next, we prove how the soft behavior of states in a supermultiplet must be related and illustrate the results in extended supergravity. We demonstrate the power of the soft bootstrap in two applications. First, for the $\mathcal{N} = 1$ and $\mathcal{N}=2$ $\mathbb{CP}^1$ nonlinear sigma models, we show that on-shell constructibility  establishes the emergence of accidental IR symmetries. 
This includes a new on-shell perspective on the interplay between $\mathcal{N}=2$ supersymmetry, low-energy theorems, and electromagnetic duality.  
We also show that $\mathcal{N}=2$ supersymmetry requires 3-point interactions with the photon that make the soft behavior of the scalar $O(1)$ instead of vanishing, despite the underlying symmetric coset. 
Second, we study Galileon theories, including aspects of supersymmetrization, the possibility of a vector-scalar Galileon EFT, and the existence of higher-derivative corrections preserving the enhanced special Galileon symmetry. The latter is addressed both by soft bootstrap and by application of double-copy/KLT relations applied to higher-derivative corrections of chiral perturbation theory. 

\end{abstract}

\end{titlepage}

\setcounter{tocdepth}{2}
{\small
\setlength\parskip{-0.5mm}
\tableofcontents
}

\newpage

\section{Introduction}
\label{sec:Introduction}

Effective field theories (EFTs) encode the low-energy dynamics of the light degrees of freedom in a physical system. The general principle of EFTs is to include all possible local interaction terms permissible by symmetries up to a certain order in the derivative expansion. Irrelevant operators are suppressed by powers of the UV cutoff and have dimensionless Wilson coefficients that parameterize the (possibly unknown) UV physics. Of particular interest, both for formal and phenomenological applications, are the EFTs describing the low-energy interactions of Goldstone modes of spontaneously broken symmetries. Traditionally, such effective actions are constructed explicitly from the underlying symmetry breaking pattern using the method of \textit{nonlinear realization} \cite{Coleman:1969sm,Callan:1969sn,Volkov:1973vd}. 

However, constructing effective actions one by one is not an efficient approach to the  problem 
of classifying such models and studying the properties of the associated scattering amplitudes. Similar to gauge and gravity theories, the Lagrangian description of EFTs  
has an enormous redundancy in the form of nonlinear field redefinitions which are completely invisible in the S-matrix\cite{Georgi:1991ch,Arzt:1993gz}. The modern \textit{on-shell} approach completely avoids both the redundant description and the associated process of calculating observables from explicitly given Lagrangians. 
Instead one uses the required physical and mathematical properties of the on-shell scattering amplitudes to constrain the underlying models and directly calculate the physical scattering amplitudes. 

The effective actions for Goldstone modes 
typically have the unusual property that while there may be an \textit{infinite} number of gauge invariant local operators at a fixed order in the derivative expansion, the associated infinite set of Wilson coefficients is determined in terms of a \textit{finite} number of independent parameters. How can this be understood in purely on-shell terms? The traditional explanation is that the spontaneously broken symmetries are nonlinearly realized on the fundamental fields and therefore mix operators in the effective action of different valence. From a more physical perspective, the  spontaneously broken  symmetries manifest themselves on the physical observables via low-energy or soft theorems. The non-independence of the Wilson coefficients is required to produce a cancellation between Feynman diagrams  that ensures  the low-energy theorem to hold. This is a redundant statement: while the number of independent parameters required to specify the effective action at a given order is reparametrization invariant, the actual Wilson coefficients are not. 
As we will see, from a purely on-shell perspective the collapse from an infinite number of free parameters to a finite number is a symptom of the underlying \textit{recursive constructiblility} of the S-matrix, which itself can be understood as a consequence of the low-energy theorems.

It is instructive to consider an explicit example that illustrates these ideas. Consider a flat $3$-brane in 5d Minkowski space. There is a Goldstone mode $\phi$ associated with the spontaneous breaking of translational symmetry in the direction transverse to the brane, and it is well-known that the leading low-energy dynamics is governed by the Dirac-Born-Infeld (DBI) action. In static gauge, it takes the form 
\be
  S_\text{DBI} = \Lambda^4 \int d^4x \,\Big( 
  \sqrt{ \det \big(\eta_{\mu\nu}+ \tfrac{1}{\Lambda^4}\pa_\mu\phi \pa_\nu\phi \big)} -1\Big)\,,
\ee
where $\Lambda^4$ is the brane tension.  
The action trivially has a constant shift symmetry $\phi \to \phi +c$ which implies that the DBI amplitudes have vanishing single-soft limits. In particular, when one of its momentum lines is taken soft,
\be 
  \label{psoft}
  p_\text{soft}^\mu \to \epsilon\, p_\text{soft}^\mu ~~~\text{with}~~~ \eps \to 0\,,
\ee  
the Feynman vertex it sits on goes to zero as $O(\eps)$.  
 There are no cubic interactions, so propagators remain finite. Hence, every tree-level Feynman diagram goes to zero as $O(\eps)$. What may be surprising is that a cancellation occurs between Feynman diagrams such that the soft behavior of any tree-level DBI $n$-point amplitude is enhanced to $O(\eps^2)$. For example for the 6-point amplitude, the $O(\eps)$-contributions of the pole diagrams cancel against those of the 6-point contact term, leaving an overall $O(\eps^2)$ soft behavior:
\be
  {\begin{tikzpicture}
     \node at (-0.3,0) {$\mathcal{A}_6 ~~=$};
      \node at (1,0) {\LARGE $\sum$};
      \draw[line width = 1.05] (1.5,0)--(3.5,0);
      \draw[line width = 1.05] (2.1,0)--(1.7,0.4);
      \draw[line width = 1.05] (2.1,0)--(1.7,-0.4);
      \draw[line width = 1.05] (2.9,0)--(3.3,0.4);
      \draw[line width = 1.05] (2.9,0)--(3.3,-0.4);
      \node at (4.3,0) {$+$};
      \draw[line width = 1.05] (5,0)--(6,0);
      \draw[line width = 1.05] (5.2,0.4)--(5.8,-0.4);
      \draw[line width = 1.05] (5.2,-0.4)--(5.8,0.4);
      \node at (2.5,-0.7) {$\underbrace{\hspace{20mm}}_{\text{$O(\eps)$}}$};
      \node at (5.5,-0.7) {$\underbrace{\hspace{12mm}}_{\text{$O(\eps)$}}$};
      \node at (3.4,-1.25) {$\underbrace{\hspace{56mm}}_{\text{$O(\eps^2)$}}$};
    \end{tikzpicture}}
\ee
The cancellation of the $O(\eps)$-contributions requires the coefficients of the 4- and 6-particle interactions $(\pa \phi)^4$ and $(\pa \phi)^6$ to be uniquely related. Interestingly we can invert the logic of this argument. Begin with the most general effective action constructed from the operators present in the DBI action, but now with \textit{a priori} independent Wilson coefficients $c_i$, schematically

\begin{equation}
  S_{\text{eff}} \sim \int \text{d}^4x \; \left[(\partial\phi)^2+ \frac{c_1}{\Lambda^4}\partial^4\phi^4 + \frac{c_2}{\Lambda^8}\partial^6\phi^6+...\right].
\end{equation}

Imposing that the amplitudes of this model satsify $O(\epsilon^2)$ low-energy theorems generates an infinite set of relations among the $c_i$. Up to non-physical ambiguities related to field redefinitions, the unique solution to these constraints is the DBI action. In that sense, DBI is the unique leading-order 4d  real single-scalar theory with  $O(\eps^2)$ low-energy theorems \cite{Cheung:2014dqa}.

The cancellation of the $O(\eps)$-terms in the DBI amplitudes is a manifestation of a less obvious symmetry of the action. The broken Lorentz transformations transverse to the brane induce an enhanced shift symmetry on the brane action of the form $\phi\to \phi +  c_\mu x^\mu + \ldots$, where the ``$+\ldots$" stand for field-dependent terms. A theory with interaction terms built from scalar fields with at least two derivatives on every field would trivially  have 
the enhanced shift symmetry that leads to the $O(\eps^2)$ soft behavior, but this is not the case for DBI. Therefore DBI is in a class of EFTs that have been described in previous work as \textit{exceptional}\cite{Cheung:2014dqa}. This example illustrates the Lagrangian-based description of what is meant by an exceptional EFT: a local field theory of massless particles with shift symmetries that lead to an enhanced soft behavior of the scattering amplitudes beyond what is obvious from simple counting of derivatives on the fields.\footnote{This definition is a little imprecise. In standard usage, an EFT is defined by some physical data including the spectrum of particles and associated symmetries and corresponds to an effective action with operators at all orders in the derivative expansion. The defining property of an exceptional EFT however is typically only valid at leading or next-to-leading order. The equivalent on-shell statement is that the scattering amplitudes of the EFT are only recursively constructible at the same order in the expansion.} 

The on-shell significance of the exceptional EFTs was first described in \cite{Cheung:2015ota,Cheung:2016drk}. It was shown, for the case of scalar effective field theories, that the class of exceptional EFTs as defined above coincides precisely with the class of EFTs for which there exists a valid method of on-shell recursion. On-shell recursion for scattering amplitudes in the form of BCFW \cite{Britto:2004ap,Britto:2005fq} or those based on various types of multi-line shifts \cite{Risager:2005vk,Elvang:2008vz,Cohen:2010mi,Cheung:2015cba} have been around for several years now, but they are often not valid in EFTs. Technically,  this is because  higher-derivative interactions tend to give `bad' large-$z$ behavior of the amplitudes under the complex momentum shifts and as a result there are non-factorizable contributions from a pole at $z=\infty$. A more physical reason is that in order for a recursive approach to have a chance, it has to be given information about how higher-point terms are possibly connected to the lower-point interactions. Standard recursion relations basically only `know' gauge-invariance, so in the DBI example they have no opportunity to know about any relation between the couplings of $(\pa \phi)^4$ and $(\pa \phi)^6$. So, naturally, a recursive approach to calculate amplitudes in exceptional EFTs needs to know about the low-energy theorems, since --- as illustrated for DBI --- this is what ties the higher-point interactions to the lower-point ones. This is exactly the additional input introduced to define  the {\em soft subtracted recursion relations} presented in \cite{Cheung:2015ota}; they provide a tool to calculate the leading (and possibly next-to-leading) order contribution to the S-matrix of an exceptional EFT without explicit reference to the action. 

The existence of valid recursion relations gives us our sought-after on-shell characterization of the relation among the Wilson coefficients of Goldstone EFTs. The infinite set of \textit{a priori} independent local operators at leading order in the derivative expansion determine the leading-order part of the S-matrix. For a generic EFT, the presence of independent operators of valence $n$ corresponds  to  the appearance of independent coefficients on contact contributions for amplitudes with $n$ external particles. If the scattering amplitudes are recursively constructible at a given order, then no such  {\em independent} coefficients can appear since the entire amplitude must be determined by factorization into amplitudes with fewer external particles.  Furthermore, the recursion must take as its input a finite set of seed amplitudes that depend on only a finite number of parameters.

Beyond being an efficient method for calculating explicit scattering amplitudes in known models, the subtracted recursion relations can be implemented as a numerical algorithm to explore and classify the landscape of possible EFTs. We term this program the \textit{soft bootstrap} due to the structural similarity of the method with the conformal bootstrap \cite{Simmons-Duffin:2016gjk,Poland:2018epd}. The method is described in detail in Section \ref{consistency}, here we give a simplified description. We consider EFTs as defined by a set of on-shell \textit{soft data}: a spectrum of massless states, linearly realized symmetries and low-energy theorems. We use general ans\"atze for scattering amplitudes of low valence and low mass dimension, consistent with the assumed spectrum and linear symmetries, as input for subtracted recursion. If the ans\"atze satisfy 
a certain criterion guaranteeing the validity of the subtracted recursion relations and if the assumed soft data corresponds to a valid EFT, then the output of the recursion should correspond to a physical scattering amplitude. Here \textit{valid EFT} means the existence of the assumed EFT as a local, unitary, Poincar\'e invariant quantum field theory. 

For tree-level scattering amplitudes this includes the requirement that the only singularities of the amplitude correspond to factorization on a momentum channel. Conversely if no such valid EFT exists, or equivalently if the assumed soft data is inconsistent, then the output of the recursion generically  will not correspond to a physical scattering amplitude and this may be detected through the presence of non-physical or spurious singularities. In practice, the ans\"atze are parametrized by a finite number of coefficients, and the removal of spurious singularities often places constraints on these coefficients. 

The soft bootstrap program was initiated in \cite{Cheung:2016drk}, where it was used to explore the landscape of real scalar EFTs with vanishing low-energy theorems. The results are reviewed and extended  in Section \ref{s:softboot}. This paper should be understood as a continuation and generalization of this program, incorporating richer soft data including spinning particles and linearly realized supersymmetry.  In Section \ref{s:overview}  we provide a brief overview of exceptional EFTs studied in this paper before summarizing our main results in Section \ref{outline} that also provides  an outline of the paper. 

\subsection{Overview of EFTs}
\label{s:overview}

In this paper, we extend the application of the  soft bootstrap from real scalars to any massless helicity-$h$ particle and  we  derive a precise criterion for the validity of the soft subtracted recursion relations. By the new validity criterion, the on-shell characterization of an  exceptional EFT will precisely be that its amplitudes are constructible using soft recursion. 

Our work requires a precise definition of  the degree of softness of the amplitude. This is given in Section  \ref{s:soft}. For now, let us simply introduce the  {\em soft weight} $\sigma$ as 
\be
  \mathcal{A}_n (\eps p_1 , p_2, \ldots ) = \eps^\sigma \, \mathcal{S}_n^{(0)}  + O(\eps^{\sigma+1})
  ~~~~\text{as} ~~\eps \to 0\,,
\ee
where $\mathcal{S}_n^{(0)} \ne 0$. 
Table \ref{Tablesigma} summarizes the soft weights for various known cases of spontaneous symmetry breaking. The earlier example of DBI corresponds to the case of  spontaneously broken higher-dimensional Poincar\'e symmetry; only the breaking of the translational symmetry actually gives rise to a Goldstone mode \cite{Low:2001bw} and it will have $\sigma=2$.
\begin{table}[t!]
\begin{center}
  \begin{tabular}{ | c | c | c |}
    \hline
    Soft degree $\sigma$ & Spin $s$ & Type of symmetry breaking \\ \hline\hline
    1 & 0 &  Internal symmetry (symmetric coset)\\ \hline
    0 & 0 &  Internal symmetry (non-symmetric coset)\\ \hline
    1 & 1/2 & Supersymmetry \\ \hline
    0 & 0 & Conformal symmetry \\ \hline
    0 & 1/2 & Superconformal  symmetry\\ \hline
    2 & 0 & Higher-dimensional Poincar\'e symmetry \\ \hline
    0 & 0 & Higher-dimensional AdS symmetry  \\ \hline
    3 & 0 & Special Galileon symmetry  \\ \hline    
  \end{tabular}
\end{center}
\label{Tablesigma}
\caption{\small The table lists soft weights $\sigma$ associated with the soft theorems $\mathcal{A}_n \to O(\eps^\sigma)$ as $\eps \to 0$ for several known cases. The soft limit is taken holomorphically in 4d spinor helicity, see Section \ref{s:soft} for a precise definition. Conformal and superconformal breaking is discussed in Section \ref{sec:scft}. }
\end{table}
 
Here follows a brief overview of exceptional EFTs that appear in this paper. We include the connection between their soft behavior and Lagrangian shift symmetries:
\begin{itemize}
	\item {\bf DBI} can be extended to a complex scalar Dirac-Born-Infeld theory and coupled supersymmetrically to a fermion sector described by the  \textbf{Akulov-Volkov} action of Goldstinos from spontaneous breaking of supersymmetry. In extended supersymmetric DBI, the vector sector is \textbf{Born-Infeld (BI)} theory. The soft weights are  $\sigma_Z=2$ for the complex scalars $Z$ of DBI, $\sigma_\psi=1$ for the fermions of Akulov-Volkov, and $\sigma_\gamma=0$ for the BI photon. The soft behaviors  can be associated with shift symmetries $Z\to Z+ c+ v_\mu x^\mu$ and $\psi\to\psi+\xi$, where $\xi$ is a constant Grassmann-number.\footnote{We leave out field-dependent terms for simplicity when stating the shift symmetries.} $\mathcal{N}=1$ {\bf supersymmetric Born-Infeld}  couples the BI vector to the Goldstino of Akulov-Volkov. 
	\item {\bf Nonlinear sigma models (NLSM)} describe the Goldstone modes of sponteneously broken internal symmetries and have scalars with constant shift symmetries that give $\sigma=1$ soft weights in the low-energy theorems. A common example of an NLSM is {\bf chiral perturbation theory} in which the scalars live in a coset space $U(N)\times U(N)/U(N)$. 

The complex scalar $\mathbb{CP}^1$ NLSM can be supersymmetrized with a fermion sector that is  \textbf{Nambu-Jona-Lasinio (NJL)} model. The complex scalars have shift symmetry $Z\to Z+c$ and $\sigma_Z=1$ while the fermions have no shift symmetry and $\sigma_\psi=0$. We study both the 
$\mathcal{N}=1$ and $2$ supersymmetric $\mathbb{CP}^1$ NLSM.\footnote{In Section \ref{s:N2NLSM} we show that the $\mathcal N = 2$ $\mathbb{CP}^1$ NLSM requires the presence of 3-point interactions and the soft weight of the scalar is reduced to $\sigma_Z = 0$.}
	
\item A NLSM can have a non-trivial subleading operator that respects the shift symmetry and hence also the low-energy theorems with $\sigma=1$. This operator is known as the {\bf Wess-Zumino-Witten (WZW)} term and has a leading 5-point interaction.  
	
\item {\bf Galileon} scalar EFTs  arise in various contexts and have the extended shift symmetry $\phi\to\phi+c+v_\mu x^\mu$ that gives low-energy theorems with $\sigma=2$. As such they can be thought of as subleading operators of the DBI action, and are called {\bf DBI-Galileons}.  They can also be decoupled from DBI (at the cost of having no UV completion). 

In 4d there are two independent Galileon operators: the quartic and quintic Galileon. (By a field redefinition, the cubic Galileon is not independent from the quartic and quintic.) When decoupled from DBI, the quartic Galileon has an even further enhanced shift symmetry $\phi\to\phi+c+v_\mu x^\mu + s_{\mu\nu}x^\mu x^\nu$ that gives low-energy theorems with soft weight $\sigma=3$ and is then called the {\bf Special Galileon} \cite{Cheung:2016drk,Hinterbichler:2015pqa}.
	
\item The quartic Galileon has a complex scalar version with $\sigma_Z=2$ (but it cannot have $\sigma_Z=3$). It has an ${\mathcal{N}=1}$ supersymmetrization \cite{Farakos:2013zya,Elvang:2017mdq} in which the fermion sector  trivially realizes  a constant shift symmetry that gives $\sigma_\psi=1$.

\item There is evidence \cite{Elvang:2017mdq} that the quintic Galileon may have an ${\mathcal{N}=1}$ supersymmetrization. This involves a complex scalar whose real part is a Galileon  with 
$\sigma=2$ and imaginary part is an R-axion with $\sigma=1$.
\end{itemize}
We now summarize the main results obtained in this paper.

\subsection{Outline of Results}
\label{outline}

\begin{itemize}
    \setlength\itemsep{0.1em}

  \item In Section \ref{s:EFT} a brief review is given of the Wilsonian effective action. The notion of the \textit{reduced dimension} of an operator is defined and the relevance to power-counting in the derivative expansion is explained. 

  \item In Section \ref{s:softrec} we present a review and elaboration on the method of soft subtracted recursion. The asymptotic (large-$z$) behavior of a scattering amplitude under the momentum deformation is determined using a novel method exploiting the properties of tree amplitudes of massless particles under complex scale transformations. This result is then used to formulate a precise \textit{constructibility criterion} (\ref{criterion}) for the applicability of the method. The failure of an EFT (at some order in the derivative expansion) to satisfy the criterion is shown to be equivalent to the existence of independent local operators which are ``trivially'' invariant under an extended shift symmetry. The systematics of the soft bootstrap algorithm for constraining EFTs is described.

\item In Section \ref{s:softboot} several numerical applications of the soft bootstrap are presented. The landscape of constructible EFTs with simple spectra consisting of a single massless complex scalar, Weyl fermion, or vector boson is exhaustively explored. In particular, our analysis shows that there can be no vector Goldstone bosons with vanishing soft theorems. A similar result follows from an algebraic analysis that appeared around the same time as this paper \cite{Klein:2018ylk}.

\item In Section \ref{softsusy} we describe the interplay between soft behavior and supersymmetry. From the supersymmetry Ward identities we show that the soft weights of the states in an $\mathcal{N}=1$ multiplet can differ by at most one. Implications for superconformal symmetry breaking and constraints on low-energy theorems in extended supergravity are presented as examples. 

\item In Section \ref{sec:SUSY_NLSM}, we apply recursion to  construct the scattering amplitudes of the $\mathcal{N}=1$, $2$ $\mathbb{CP}^1$ nonlinear sigma models at leading (two-derivative) order. For the $\mathcal{N}=1$ case, it is shown that recursive constructibility together with the conservation of $U(1)$ charges by the seed amplitudes implies that (at two-derivative order) all tree amplitudes of this model conserve an additional accidental $U(1)$ charge. 
For the $\mathcal{N}=2$ model, recursive constructibility is non-trivial due to the presence of 3-point interactions and non-vanishing scalar soft limits, but can be achieved using the supersymmetry Ward identities (see Appendix \ref{WIN2}). Using this, we show that all tree amplitudes satisfy the Ward identities of $SU(2)_R$ and conserve an additional $U(1)_R$ under which the vector bosons are charged. (A detailed inductive proof of the $SU(2)_R$ Ward identities is given in Appendix \ref{RecWardApp}.) The connection between the existence of such chiral charges for vector bosons and known results about 
special K\"ahler geometry are described, in particular we highlight the emergence of electric-magnetic duality.  Finally, an explicit form of the singular low-energy theorem for the vector bosons of the $\mathcal{N}=2$ model is presented. 

\item Section \ref{sec:sDBI} contains brief comments on supersymmetrizations of DBI and Born-Infeld. 

\item In Section \ref{sec:Galileon} various applications of the soft bootstrap algorithm to Galileon-like models are presented. Previous results on the $\mathcal{N}=1$ supersymmetrization of the quartic- and quintic-Galileon are elaborated upon, in particular the various possible soft weight assignments to the states in the multiplet are described in detail. 

The existence of an extension of the special Galileon with non-trivial couplings to a massless vector is considered and evidence is given in favor of the existence of such a model. The soft bootstrap algorithm is applied to the problem of classifying higher-derivative corrections to the special Galileon effective action  that  
preserve the low-energy theorem via the associated on-shell matrix elements. Compatible amplitudes are classified up to couplings of dimension $-12$ for quartic interactions and $-17$ for quintic interactions. These results are compared with the output of the double-copy in the form of the field theory KLT relations as applied to chiral perturbation theory. These two constructions
are found to agree for quartic interactions but not for quintic.

\item In Appendix \ref{sec:AmplitudeExpressions} many explicit forms of calculated amplitudes for various models considered in this paper are presented.

\end{itemize}

\section{Structure of the Effective Action}
\label{s:EFT}

The low-energy dynamics of a physical system can be described by a Wilsonian effective action containing a set of local quantum fields for each of the on-shell asymptotic states with \textit{all possible} local interactions allowed by the assumed symmetries:
\begin{equation}
  \label{Seff}
  S_{\text{effective}} = S_0 + \sum_{\mathcal{O}} \frac{c_{\mathcal{O}}}{\Lambda^{\Delta[\mathcal{O}]-4}} \int \text{d}^4x\, \mathcal{O}(x)\,.
\end{equation}
Here $S_0$ denotes the free theory, i.e.~the kinetic terms, $\Lambda$ is a characteristic scale of the problem, and $c_{\mathcal{O}}$ are dimensionless constants. The sum is over all local Lorentz invariant operators $\mathcal{O}(x)$ of the schematic form
\begin{equation} \label{operator}
  \mathcal{O}(x) \sim \partial^A \phi(x)^B \psi(x)^C F(x)^D\,,
\end{equation}
where $A,\ldots,D$ are integer exponents. In this paper we focus on EFTs in which the operators $\mathcal{O}$ are manifestly gauge invariant.\footnote{This need not be the case in more general scenarios (though of course we insist on overall gauge invariance). For example in Yang-Mills theory, the gauge invariant operator $\tr F^2$ has a quadratic term which we group into the free part $S_0$ of the action while the interaction terms would be accounted for in the sum of all operators $\mathcal{O}$ in \reef{Seff}. Similarly, for massless spin-2 fields when $\sqrt{-g}R$ is expanded around flat space. }

We assign the following quantities to a local operator

\begin{itemize}
  \item \textit{Dimension:} $\Delta[\mathcal{O}]$ defined as the engineering dimension with bosonic fields of dimension 1 and fermionic fields of dimension 3/2.
  \item \textit{Valence:} $N[\mathcal{O}]$ defined as the sum of the total number of field operators appearing. Equivalently, this is the valence of the Feynman vertex derived from such an interaction.
  \end{itemize}

The schematic operator in \reef{operator} has 
$\Delta[\mathcal{O}]=A+B+\tfrac{3}{2}C + 2D$ and 
$N[\mathcal{O}] = B + C + D$.
    
  In standard EFT lore, operators of lowest dimension dominate in the IR. In many cases this means the marginal and relevant interactions dominate and the irrelevant interactions are sub-dominant and suppressed by powers of the UV scale $\Lambda$. In other cases, such as effective field theories describing the dynamics of Goldstone modes, there are only irrelevant interactions and it may be less clear which operators dominate. It is therefore useful to introduce  the \textit{reduced} dimension 
  \begin{equation}
    \label{redDim}
    \tilde{\Delta}[\mathcal{O}] = \frac{\Delta[\mathcal{O}]-4}{N[\mathcal{O}]-2}\,
  \end{equation}
for the operator basis \reef{Seff}. Operators that minimize $\tilde{\Delta}$ dominate in the IR. 

The authors of \cite{Cheung:2014dqa,Cheung:2015ota,Cheung:2016drk} consider only scalar EFTs and therefore operators of the form $\mathcal{O} \sim \partial^m \phi^n$. They define a quantity 
  \begin{equation}
    \rho \equiv \frac{m-2}{n-2} = \tilde{\Delta}[\mathcal{O}] - 1\,,
  \end{equation}
to determine when two operators of this form produce tree-level diagrams with couplings of the same mass dimension. Morally $\rho$ is the same as the reduced dimension $\tilde{\Delta}[\mathcal{O}]$. The latter is the natural generalization of $\rho$ to operators containing particles of all spins.

The quantity $\tilde{\Delta}$ is useful for clarifying the notion of what it means for an interaction to be leading order in an EFT with only irrelevant interactions. In the deep IR, the relative size of the dimensionless Wilson coefficients in the effective action is unimportant since lower dimension operators will \textit{always} dominate over higher dimension operators. It is therefore only necessary to isolate the contributions that are leading in a power series expansion of the amplitudes in the inverse UV cutoff scale $\Lambda^{-1}$. The dominant interactions in the deep IR are generated by operators that \textit{minimize} this quantity. As an illustrative example, consider an effective action for scalars with interaction terms of the form
\begin{equation}
  S_{\text{effective}} \supset \int \text{d}^4 x \left[\frac{c_4}{\Lambda^4}\partial^4 \phi^4 + \frac{c_5}{\Lambda^5} \partial^4 \phi^5\right].
\end{equation}
The reduced dimensions $\tilde{\Delta}$ are $2$ and $5/3$ for the quartic and quintic interactions respectively. The quintic interaction should therefore dominate over the quartic in the deep IR. To see this explicitly we have to compare amplitudes with the {\em same} number of external states, so  we compare the contributions from tree-level Feynman diagrams to the 8-point amplitude:

\begin{center}
  \begin{tikzpicture}
    \draw (0.3,0.7)--(1,0);
    \draw (0,0)--(4,0);
    \draw (0.3,-0.7)--(1,0);
    \draw (2,0)--(1.4,0.7);
    \draw (2,0)--(2.6,0.7);
    \draw (3,0)--(3.7,0.7);
    \draw (3,0)--(3.7,-0.7);
    \node at (5,0) {$\sim$};
    \node at (6,0) {\Large$\frac{1}{\Lambda^{12}}$};
    \draw (8,0.3)--(9,0);
    \draw (8,-0.3)--(9,0);
    \draw (8.3,0.8)--(9,0);
    \draw (8.3,-0.8)--(9,0);
    \draw (9,0)--(10,0);
    \draw (10,0)--(11,0.3);
    \draw (10,0)--(11,-0.3);
    \draw (10,0)--(10.7,0.8);
    \draw (10,0)--(10.7,-0.8);
    \node at (12,0) {$\sim$};
    \node at (13,0) {\Large$\frac{1}{\Lambda^{10}}$};
  \end{tikzpicture}
\end{center}
This confirms that the diagrams arising from the quintic interaction dominate the 8-point amplitude.

It is useful to introduce the notion of  {\em fundamental interactions} (or {\em fundamental operators}) in an EFT. These are the lowest dimension operator(s) whose on-shell matrix elements can be recursed to define all matrix elements of the theory at leading order in the low-energy expansion.

Consider the DBI action. The leading interaction comes from an operator of the form 
$\frac{1}{\Lambda^4}\partial^4 \phi^4$ and as discussed in the introduction, with the associated 4-point amplitude as input, all other $n$-point amplitudes in DBI can be constructed with soft subtracted recursion relations.  If the action had contained an interaction term of the form $\frac{c_5}{\Lambda^5}\partial^5\phi^4$, then $\frac{1}{\Lambda^4}\partial^4 \phi^4$ would not be sufficient to determine dominating contributions at $n$-point order, i.e.~both interactions would need to be considered fundamental for soft recursion.

The operators immediately subleading to DBI in the brane-effective action are encoded in the DBI-Galileon. In 4d, there are two such independent couplings,\footnote{The cubic Galileon interaction is equivalent to a particular linear combination of the quartic and quintic Galileon after a field redefinition.} namely for a quartic interaction of the schematic form $\frac{b_4}{\Lambda^6}\partial^6 \phi^4$ and a quintic interaction of the form $\frac{b_5}{\Lambda^9} \partial^8 \phi^5$; these both have $\tilde{\Delta} = 3$ whereas DBI has  $\tilde{\Delta} = 2$. Thus the DBI-Galileon has a total of three fundamental operators: the 4-point DBI interaction and the 4- and 5-point Galileon interactions.

\section{Subtracted Recursion Relations}
\label{s:softrec}
We review on-shell subtracted recursion relations for scattering amplitudes of Goldstone modes \cite{Kampf:2013vha,Cheung:2014dqa,Cheung:2015ota,Cheung:2016drk,Luo:2015tat} and derive a new precise criterion for their validity. 

\subsection{Holomorphic Soft Limits and Low-Energy Theorems}
\label{s:soft}

We rely on the 4d spinor helicity formalism (for reviews, see \cite{Elvang:2013cua,Elvang:2015rqa,Henn:2014yza,Dixon:2013uaa}) in which a massless on-shell momentum is written $p = - |p\> [ p|$. This presents an ambiguity in how to take the soft limit \reef{psoft}: it could for example be taken democratically as $\{|p\>,|p]\} \to \{ \eps^{1/2} |p\>, \eps^{1/2} |p]\}$, holomorphically $\{|p\>,|p]\} \to \{ \eps |p\>,  |p]\}$, or anti-holomorphically $\{|p\>,|p]\} \to \{  |p\>,  \eps |p]\}$. These are all equivalent choices, because the momentum $p$ is invariant under {\em little group scaling} $\{|p\>,|p]\} \to \{ t |p\>, t^{-1} |p]\}$.  Amplitudes scale homogeneously under the little group,  
\be 
  \label{lilgrp}
  \mathcal{A}_n\big(\{ |1\>, |1] \} \ldots \{ t |i\>, t^{-1}|i] \}_+ \ldots \big) 
  =  t^{-2h_i} \mathcal{A}_n\big(\{ |1\>, |1] \} \ldots \{ |i\>, |i] \}_+ \ldots \big) \,,.
\ee
so the choice of soft limit is simply reflected in a helicity-dependent overall scaling factor. 
We choose to minimize the power of $\eps$ in the soft limit by letting the choice depend on the sign of the helicity of the particle: specfically, we take $p_\text{soft} \to \epsilon\, p_\text{soft} = -\epsilon |s\> [s|$ holomorphically for any state with non-negative helicity:\footnote{Taking the soft limit as simply  as in \reef{softlimit} is not compatible with overall momentum conservation. To stay on the algebraic locus of momentum conservation in momentum space, we take the limit with appropriate shifts in a subset of the $n-1$ other momentum variables. The precise prescription can be found in equation (6) of \cite{Elvang:2016qvq}. The details will not affect the main line of the discussion in this paper, but we note that all  calculations are done manifestly on-shell, including the soft limits. }
\be
  \label{softlimit}
  |s\> \to \eps |s\>~~~~~\text{for}~~h_s \ge 0 \,.
\ee
For a negative-helicity particle, we use the anti-holomorphic  prescription $|s] \to \eps |s]$. For scalars, it makes no difference which choice is made. 

We  characterize the soft behavior of amplitudes of massless particles in terms of a {\em holomorphic soft weight} $\sigma$ (or, for brevity, just {\em soft weight}). It is defined in terms of the holomorphic soft limit \reef{softlimit} as
\be
  \label{softbe}
  \mathcal{A}_n\big(\{ |1\>, |1] \} \ldots \{ \eps|s\>, |s] \}_+ \ldots \big) 
  =\eps^{\sigma} \,  \mathcal{S}_n^{(0)} +O(\eps^{\sigma+1})
  ~~~~\text{as}~~\eps \to 0\,,
\ee
where $\mathcal{S}_n^{(0)} \ne 0$. This way of taking the soft limit is closely correlated with the shifts introduced for the soft subtracted recursion relations in the following.

\subsection{Review of Soft Subtracted Recursion Relations}

We consider complex momentum deformations of the form 
\be
  \label{pishift}
   p_i \to \hat{p}_i = (1- a_i z) p_i\,
   ~~~~\text{with}~~~~
   \sum_{i=1}^n a_i p_i = 0 \,.
\ee  
The label $i=1,2,\dots,n$ runs over the $n$ massless particles in the scattering amplitude. 
The shifted momenta $\hat{p}_i$ are on-shell by virtue of $p_i^2 = 0$ and satisfy momentum conservation when the shift coefficients $a_i$ satisfy the condition  in \reef{pishift}. (We discuss the solutions to this condition in Section \ref{consistency}.)
When evaluated on the shifted momenta $\hat{p}_i$, an $n$-point amplitude becomes a function of $z$ and we write it as $\hat{\mathcal{A}}_n(z)$.

The {\em subtracted recursion relations} for an $n$-point tree-level amplitude $\mathcal{A}_n$ are derived from the Cauchy integral
\be
  \label{contourA}
  \oint \frac{dz}z \frac{\hat{\mathcal{A}}_n(z)}{F(z)} ~=~ 0\,,
\ee
where the contour surrounds all the poles at finite $z$ and the function $F$ is defined as
\begin{equation}
 \label{Fdef}
  F(z) = \prod_{i=1}^n (1-a_iz)^{\sigma_i}\,.
\end{equation}
The vanishing of the integral in \reef{contourA} requires absence of a simple pole at $z=\infty$. We derive a sufficient criterion for this behavior in Section \ref{s:validity}.

The shift \reef{pishift} is implemented on the spinor helicity variables according to the sign of the helicity $h_i$ of particle $i$ as
\be
\label{softshift}
\begin{array} {lll}
 h_i \ge 0\!: &~~~|i\> \to  (1- a_i z)  |i\> \,,~~~& |i] \to |i]\,,
 \\[1mm]
  h_i < 0\!:  &~~~|i\> \to  |i\> \,,~~~& |i] \to (1- a_i z) |i]\,.
\end{array}
\ee

The  limit $z \to 1/a_i$ is then precisely the soft limit $\hat{p}_i \to 0$ of the $i$th particle in the deformed amplitude. Hence, {\em if the amplitude satisfies low-energy theorems of the form \reef{softbe} with weights $\sigma_i$ for each particle $i$}, the integral \reef{contourA} will not pick up any non-zero residues from poles arising from the function $F$ when it is chosen as in \reef{Fdef}.  
Therefore the only simple poles in \reef{contourA} arise from $z = 0$ and factorization channels in the deformed tree amplitude. They occur where internal momenta go on-shell, $\hat{P}_I^2 = 0$.  The residue theorem then states that the residue at $z=0$ equals minus the sum of all such residues, and factorization on these poles gives 
\begin{equation} \label{softrecursion}
  \mathcal{A}_n 
  = \hat{\mathcal{A}}_n(z=0)
    = \sum_I \sum_{|\psi^{(I)}\rangle} \sum_{\pm}
       \frac{\hat{\mathcal{A}}_L^{(I)}(z_I^\pm)\hat{\mathcal{A}}_R^{(I)}(z_I^\pm)}
       {F(z_I^\pm)P_I^2(1-z_I^\pm/z_I^\mp)}\,.
\end{equation}
The sums are over all factorization channels $I$, the two solutions  $z_I^\pm$ to $\hat{P}_I^2 = 0$, and all possible particle types $|\psi^{(I)}\rangle$ that can be exchanged in channel $I$. These recursion relations are called {\em soft subtracted recursion relations}. When $F=1$, the recursion is called {\em unsubtracted}.

The expression for the solutions $z_I^\pm$ to the quadratic equation $\hat{P}_I^2=0$ involves square roots, but those must cancel since the tree amplitude is a rational function of the kinematic variables. On channels where the amplitude factorizes into two {\em local} lower-point amplitudes (meaning that they have no poles), the cancellations of the square roots can be made manifest. This is done by a second application of Cauchy's theorem, which for each channel $I$ converts the sum of residues at $z=z_I^\pm$ to the sum of the residues at $z=0$ and $z=1/a_i$ for all $i$. Details are provided in Appendix \ref{app:recrel}, here we simply state the result:   {\em if $\mathcal{A}_L^{(I)}$ and $\mathcal{A}_R^{(I)}$  are local for all factorization channels},  the soft recursion relations take the form
\begin{equation} \label{softrecursion2}
  \mathcal{A}_n = \sum_I \sum_{|\psi^{(I)}\rangle}  
  \bigg( \frac{\hat{\mathcal{A}}_L^{(I)}(0)\hat{\mathcal{A}}_R^{(I)}(0)}{P_I^2}
  ~+~
  \sum_{i=1}^n
  \text{Res}_{z=\frac{1}{a_i}}\,\frac{\hat{\mathcal{A}}_L^{(I)}(z)\hat{\mathcal{A}}_R^{(I)}(z)}{z\,F(z)\,\hat{P}_I^2}
  \bigg) \,.
\end{equation}
This form of the recursion relations is manifestly rational in the kinematic variables, and we will be using \reef{softrecursion2} for the applications in this paper. Note that only the first term in \reef{softrecursion2} has pole terms. Therefore the sum of the $1/a_i$ residues over all channels must be a local polynomial in the momenta.

\subsection{Validity Criterion}
\label{s:validity}
The purpose of including $F(z)$ in \reef{contourA} is to improve the large-$z$ behavior of the integrand so that one can avoid a pole at $z=\infty$. This is necessary in EFTs, where the large-$z$ behavior of the amplitude typically does not allow for unsubtracted recursion relations with $F(z)=1$ to be valid without a boundary term from $z=\infty$. A sufficient condition for absence of a simple pole at infinity is that the deformed amplitude vanishes as $z \to \infty$. Below we show that for a theory with a single fundamental interaction (see Section \ref{s:EFT}) of valence $v$ and coupling of mass-dimension $[g_v]$ the criterion for validity of the subtracted recursion relations is
\begin{equation} \label{criterion}
\boxed{ ~~~4-n-\frac{n-2}{v-2}\,[g_v]-\sum_{i=1}^ns_i -\sum_{i=1}^n\sigma_i < 0\,.~~}
\end{equation}
Here $s_i$ is the spin (not helicity) of particle $i$ and $\sigma_i$ is its soft behavior \reef{softbe}. Alternatively, one can write the constructibility criterion in terms of the reduced dimension $\tilde\Delta$, introduced in \reef{redDim}, as
\begin{equation} \label{criterion2}
\boxed{ ~~~4-n+(n-2)\tilde\Delta-\sum_{i=1}^ns_i -\sum_{i=1}^n\sigma_i < 0\,.~~}
\end{equation}

The criterion generalizes to theories with more than one fundamental coupling by replacing $\frac{n-2}{v-2}\,[g_v]$ in \reef{criterion} by the sum over all couplings contributing to the diagrammatic expansion of the amplitude in question; the precise criterion is given in  \reef{morethanonecoupling}.

\vspace{3mm}
\noindent {\bf Proof of the criterion \reef{criterion}}\\
To avoid a pole at infinity in the Cauchy integral \reef{contourA}, it is sufficient to require  
${\hat{\mathcal{A}}_n(z)}/{F(z)} \to 0$ as $z \to \infty$. To start with, we determine the large-$z$ behavior of the deformed amplitude $\hat{\mathcal{A}_n}(z)$. 

Generically, in a theory of massless particles with couplings $g_k$, a tree-level amplitude takes the form
\be
  \label{AnMj}
 \mathcal{A}_n = \sum_j \bigg( \prod_k g_k^{n_{jk}} \bigg) M_j\,,
\ee 
where $\prod_k g_k^{n_{jk}}$ is a product of coupling constants and $M_j$ is a function of spinor brackets only. Since there can be no other dimensionful quantities entering $M_j$, 
the mass dimension $[M_j]$ can be determined via a homogenous scaling of all spinors:
\be
  \label{scaleuni}
   |i\> \to \lambda^{1/2} |i\>  ~~~\text{and}~~~ |i] \to \lambda^{1/2} |i]
   ~~~~\implies~~~~
   M_j \to \lambda^{[M_j]}M_j\,.
\ee
The mass dimension is also fixed by simple dimensional analysis to be 
\begin{equation}
  [M_j] = 4-n -\sum_k n_{jk}[g_k]\,,
\end{equation}
since an $n$-point scattering amplitude in 4d has to have mass-dimension $4-n$. 

It is useful to consider a modified scale transformation defined as
\be
\label{hscale}
\begin{array} {lll}
 h_i \ge 0\!: &~~~|i\> \to  \lambda  |i\> \,,~~~& |i] \to |i]\,,
 \\[1mm]
  h_i < 0\!:  &~~~|i\> \to  |i\> \,,~~~& |i] \to \lambda |i]\,.
\end{array}
\ee
The effect of this scaling can be obtained from the  uniform scaling \reef{scaleuni} via a little group transformation \reef{lilgrp} on all momenta with $t = \lambda^{1/2}$. Therefore under \reef{hscale}, $M_j$ scales as $M_j \to \lambda^{[M_j] - \sum_i s_i}M_j$, where $s_i$ is the spin ({\em not} helicity) of particle $i$. 

For the case of a theory with a single fundamental interaction of valence $v$ with coupling $g_v$, the number of couplings appearing in an $n$-point amplitude is $\frac{n-2}{v-2}$, and therefore we have 
\be
  \label{AnD}
  \mathcal{A}_n \to \lambda^D \mathcal{A}_n\,,~~~~
  D= 4-n -\frac{n-2}{v-2} [g_v]- \sum_i s_i 
\ee 
under the modified scale transformation \reef{hscale}.

Under the momentum shift \reef{softshift},  the deformed tree amplitude $\hat{\mathcal{A}}_n(z)$ can be written 
\be 
  \label{shiftedscaledAn}
  \begin{split}
   \hat{\mathcal{A}}_n(z) 
   =&~\hat{\mathcal{A}}_n\big( \ldots \{ (1-a_i z) |i\> , |i] \}_+ 
   \ldots  \{  |j\> , (1-a_j z) |j] \}_- \big)\\
   =&~\hat{\mathcal{A}}_n\big( \ldots \{ z(1/z-a_i) |i\> , |i] \}_+ 
   \ldots  \{  |j\> , z(1/z-a_j) |j] \}_- \big)\\
   =&~z^D \, \hat{\mathcal{A}}_n\big( \ldots \{ (1/z-a_i) |i\> , |i] \}_+ 
   \ldots  \{  |j\> , (1/z-a_j) |j] \}_- \big)\,,
   \end{split}
\ee
where the subscripts $\pm$ refer to the sign of the helicity of each particle. In the last line we used the behavior \reef{AnD} under the modified scaling  \reef{hscale}.

At large $z$, the amplitude in the last line of \reef{shiftedscaledAn} is the original unshifted amplitude evaluated at a momentum configuration with $q_i = -a_i p_i$. These momenta are all on-shell and satisfy, via \reef{pishift}, momentum conservation. The only way the tree amplitude could have a singularity at this momentum configuration would be if an internal line went on-shell. This can always be avoided for generic momenta.\footnote{The condition \reef{pishift} has a trivial solution with all $a_i$ equal. Therefore any solution to \reef{pishift} can be shifted uniformly $a_i \to a_i + a$ for any real number $a$. Hence, we can always avoid the discrete set of momentum configurations for which an internal line in $\mathcal{A}_n$ goes on-shell.} Thus we conclude from \reef{shiftedscaledAn} that for large $z$, the deformed amplitude behaves as
\be
  \hat{\mathcal{A}}_n(z) \to z^{N} 
  ~~~~\text{with}~~~~N\le D\,,
\ee
where $D$ is given in \reef{AnD}. The inequality allows for the possibility that ${\mathcal{A}}_n$ could have a zero at $q_i = -a_i p_i$.   

Our mission was to find a criterion for ${\hat{\mathcal{A}}_n(z)}/{F(z)} \to 0$ as $z \to \infty$. By the definition \reef{Fdef}, we have  $F(z) \to z^{\sum_i \sigma_i}$ for large $z$. From our analysis of the large-$z$ behavior of ${\hat{\mathcal{A}}_n(z)}$, we can therefore conclude that, at worst, ${\hat{\mathcal{A}}_n(z)}/{F(z)} \to z^{D-\sum_i \sigma_i}$. The sufficient criterion for absence of a pole at infinity, and hence for validity of the subtracted recursion relation, is then $D -\sum_i \sigma_i < 0$. This is precisely the condition \reef{criterion}. This concludes the proof.

It is straightforward to generalize the 
constructibility criterion to EFTs with more than one fundamental interaction, 
\be
 \label{morethanonecoupling}
4-n- \text{min}_j\Big( \sum_k n_{jk}[g_k]\Big) -\sum_{i=1}^ns_i -\sum_{i=1}^n\sigma_i < 0\,.
\ee 
Recall that in effective field theories, the couplings have negative mass-dimension. This means that the constructibility criterion tends to be dominated by the fundamental interactions associated with operators of the highest mass-dimension that can contribute to the $n$-point amplitude. 

\vspace{3mm}
\noindent {\bf Example 1}\\
Let us once again return to the example of DBI. 
The action has a fundamental quartic vertex $g_4 (\partial \phi)^4$ with a coupling of mass-dimension $[g_4] = -4$. The constructibility criterion  \reef{criterion} for the $n$-scalar amplitude is $n(1- \sigma_S) < 0$, where $\sigma_S$ is the soft behavior of the scalar $\phi$. Since $\sigma_S=2$ in DBI, all DBI tree amplitudes are constructible via the subtracted soft recursion relations, as claimed in the introduction. 

The failure of the constructibility criterion for $\sigma_S=1$ is simply the statement that an EFT whose interactions are built from powers of $(\partial \phi)^2$ trivially has a constant shift symmetry and hence $\sigma_S=1$, so there are no constraints from shift symmetry on the coefficients of $(\partial \phi)^{2k}$ in terms of that of $(\partial \phi)^4$ and then one has no chance of recursing $\mathcal{A}_4$ to get all-point amplitudes.

\vspace{3mm}
\noindent {\bf Example 2}\\
Consider a theory of massless fermions with quartic coupling of mass-dimension $[g_4] = -2$. The  criterion \reef{criterion} says that the $n$-fermion amplitudes are constructible when 
\mbox{$4 < n (1+2 \sigma_\psi)$.} 
 Thus all $n>4$ point tree-amplitudes are constructible by \reef{softrecursion} for any soft weight $\sigma_\psi \ge 0$. No such theory exists for $\sigma_\psi>0$ (as we prove in Section \ref{s:fermionEFTs}), but for $\sigma_\psi = 0$ this is exactly the Nambu--Jona-Lasinio (NJL) model, which consists of  the simple 4-fermion interaction $\psi^2 \bar{\psi}^2$ \cite{Nambu:1961}.

\subsection{Non-Constructibility = Triviality}
\label{non-constructibility}
We have derived a constructibility criterion, but what does it mean? The answer is quite simple: if an $n$-point amplitude can be constructed recursively from lower-point on-shell amplitudes, there cannot exist a local gauge-invariant $n$-field operator that contributes to the  amplitude without modifying its soft behavior. We define a {\em trivial operator} to be one with at least 4 fields whose matrix elements  manifestly have a given soft weight $\sigma$. Let us now assess what it takes to make an operator of scalar, fermion, and vector fields trivial. 

\noindent {\bf Triviality.}

{\em Scalars.} Operators with at  least $m$ derivatives on each scalar field will trivially have single-soft scalar limits with $\sigma_S=m$.  

{\em Fermions.} We have chosen the soft limit \reef{softlimit} according to the helicity such that the fermion wavefunctions do not generate any soft factors of $\epsilon$. Thus a trivial soft behavior must come from derivatives on each fermion field in the Lagrangian. We conclude that the trivial soft behavior $\sigma_F=$ smallest number of derivatives on each fermion field.

{\em Photons.}  Gauge invariance tells us that we should construct the interaction terms using the field strength $F_{\mu\nu}$.\footnote{Or covariant derivatives $D_\mu = \pa_\mu + ig A_\mu$. In this paper, we focus on scalars and fermions that do not transform under any gauge-$U(1)$, therefore photons must couple via $F_{\mu\nu}$.} When associated with an external photon, the Feynman rule for $F_{\mu\nu}$ 
gives $p_\mu \eps_\nu - p_\nu \eps_\mu$. Naively, it may seem to be linear in the soft momentum, but under the holomorphic soft shift \reef{softshift} it is actually $O(\eps^0)$. Recall that in spinor helicity formalism, a positive helicity vector polarization takes the form $\eps^\mu_+ \bar{\sigma}_\mu^{\dot{a}b}  =\eps_+^{\dot{a}b} = |q\>^{\dot{a}}[p|^b/\<pq\>$, where $q$ is a reference spinor. Hence, for a positive helicity photon we have
\be
 \label{Ffeynrule}
  (F_+)_a{}^b \equiv (\sigma^{\mu\nu})_a{}^b F_{\mu\nu}
   \longrightarrow
   (\sigma^{\mu\nu})_a{}^b (p_\mu \eps_{+\nu} - p_\nu \eps_{+\mu})
   \sim 
   |p]_a \<p|_{\dot{c}} \frac{|q\>^{\dot{c}}[p|^b}{\<pq\>}
   = |p]_a [p|^b \,.
\ee
This is explicitly independent of the reference spinor $q$ because $F_{\mu\nu}$ is gauge invariant. For a positive helicity particle, we take the soft limit holomorphically as $|p\> \to \eps |p\>$ (while $|p] \to  |p]$), so we explicitly see that $F_{\mu\nu} \longrightarrow |p] [p|$ is $O(\eps^0)$ when $p$ is taken soft. Likewise, for a negative helicity photon, $(F_-)^{\dot a}{}_{\dot b} \longrightarrow |p\> \<p|$. We conclude that an operator with photons has trivial soft behavior that is determined by the smallest number of derivatives on each field strength $F_{\mu\nu}$. 

In an EFT where photon interactions are built only from the field strengths, the matrix elements   are $O(1)$ when a photon is taken soft. This, for example, is exactly the case for Born-Infeld theory in which the photons have $\sigma=0$. 

\noindent {\bf Constructibility.} Suppose we study an $n$-particle amplitude with $n_s$ scalars,  $n_f$ fermions, and $n_\gamma$ photons in an EFT whose fundamental $v$-particle interactions all have couplings of the same mass-dimension $[g_v]$.  The  criterion \reef{criterion} for constructibility via subtracted soft recursion relations can be written as 
\be
\label{crit2}
 4-n-n_v [g_v]
 -\frac{1}{2} n_f - n_\gamma 
 - n_s\sigma_s-  n_f\sigma_f -  n_\gamma\sigma_\gamma < 0\,,
\ee 
where $n_v = (n-2)/(v-2)$ is the number of vertices needed at $n$-point. 

\noindent {\bf Non-constructibility = Triviality.}
Let us assess if there can be a local contact term for an $n$-particle amplitude with $n_s$ scalars,  $n_f$ fermions, and $n_\gamma$ photons and soft behaviors $\sigma_s$, $\sigma_f$, and $\sigma_\gamma$, respectively. As discussed above, a contact term that has such trivial soft behavior takes the form 
\be
   g_n \underbrace{ (\partial^{\sigma_s} \phi) \cdots (\partial^{\sigma_s} \phi) }_{n_s}
   \underbrace{(\partial^{\sigma_f} \psi) \cdots (\partial^{\sigma_f} \psi)}_{n_f}
   \underbrace{(\partial^{\sigma_\gamma} F) \cdots (\partial^{\sigma_\gamma} F)}_{n_\gamma}
\ee
(for brevity we have not distinguished $\psi$ and $\bar\psi$). 
In 4d, the mass-dimension of the coupling $g_n$ is easily computed as
\be
  [g_n] = 
  4 
  - \big(n_s + n_s \sigma_s\big) 
  - \big(\tfrac{3}{2}n_f + n_f \sigma_f\big) 
  - \big(2 n_\gamma + n_\gamma \sigma_\gamma\big)\,.
\ee
Using  $n=n_s + n_f + n_\gamma$, we can rewrite this as 
\be
  4 - n - [g_n] - \frac{1}{2}n_f -n_\gamma
  - n_s \sigma_s - n_f \sigma_f -  n_\gamma \sigma_\gamma
  = 0 \,.
\ee
Compare this with \reef{crit2}; we note that the constructibility criterion is simply that  
$n_v [g_v] >  [g_n]$, or maybe more intuitively, that $g_n$ has more negative mass-dimension than $n_v$ $g_v$-vertices. So, when constructibility holds, the $n$-particle amplitude constructed from the $n_v$ $v$-valent vertices {\em cannot} be influenced by a contact term that trivially has the soft behavior: such a contact term would be too high order in the EFT due to all the derivatives needed to trivialize the soft behavior. That of course makes sense; were there such an independent local contact term, it could be added to the result of recursion with any coefficient without changing any of the properties of the amplitude. Hence recursion cannot possibly work in that case.  (This is analogous to the example in \cite{Elvang:2013cua,Elvang:2015rqa} for constructibility in scalar-QED via BCFW; the difference here is that the subtracted soft recursion relations ``know" about the soft behavior in addition to gauge-invariance.) 

The argument is easily extended to the case where the theory has fundamental vertices of different valences and mass-dimensions. We conclude that the constructibility criterion \reef{criterion}
 is equivalent to the non-existence of local $n$-particle operators with couplings of the same mass-dimension and trivial soft behavior: Non-constructibility = Triviality.
 
\subsection{Implementation of the Subtracted Recursion Relations}
\label{consistency}

Here we present details relevant for the practical implementation of the soft subtracted recursion relations. 

{\bf Solving the shift constraints.} Conservation of the momentum for the shifted momenta $\hat{p}_i$ \reef{pishift} requires the shift variables $a_i$ to satisfy
\be
  \label{momconstraint}
  \sum_i a_i p_i^\mu = 0.
\ee
In 4d, the LHS can be viewed as a $4 \times n$ matrix $p_i^\mu$ of rank $4$ (if $n \ge 5$) multiplying a $n$-component vector $a_i$. Hence the valid choices of parameters $a_i$ form a vector space given by the kernel of the matrix $p_i^\mu$. For $n \ge 5$ any subset of four momenta are generically linearly independent, so the $p_i^\mu$-matrix has full rank. By the rank-nullity theorem,  the dimension of the kernel is therefore $n-4$. However, there is always a trivial solution which consists of all $a_i$'s equal, hence non-trivial solutions to \reef{momconstraint} exist only when $n \ge 6$. 

Practically, the linear system of equations is solved by dotting in $p_j$, i.e.~we have 
\be
  \label{stimesa}
  \sum_i s_{ji} \, a_i  = 0 ~~~\text{for}~~~j=1,2,\ldots,n\,.
\ee
The symmetric $n\times n$-matrix with entries $s_{ji}$ has rank 4, so the linear system \reef{stimesa}
can be solved for say $a_1$, $a_2$, $a_3$, and $a_4$ in terms of the $n-4$ other $a_i$'s. 

{\bf Soft bootstrap.}
Subtracted recursion relations can be used to calculate tree amplitudes in EFTs of Goldstone modes in theories we already know well, such as DBI, Akulov-Volkov etc. However, the soft  subtracted recursion relations can also be used as a tool to {\em classify} and assess the existence of exceptional EFTs  with a given spectrum of massless particles and low-energy theorems with given weights $\sigma$. 

The approach to the classification of special EFTs is as follows:
\begin{enumerate}
\item[(1)] Model input:  the spectrum of massless particles and the coupling dimensions of the fundamental interactions in the model. 
\item[(2)] Symmetry assumptions: the $n$-particle amplitudes have soft behavior with weight $\sigma_i$ for the $i$th particle.
\end{enumerate}
If the constructibility criterion \reef{criterion} is {\em not} satisfied, the assumptions (1) and (2) are trivially satisfied and we cannot constrain the couplings in the EFTs; it is not exceptional.

If the constructibility criterion \reef{criterion} {\em is} satisfied for input (1) and (2), one can use the soft subtracted recursion relations to test whether a theory can exist with the above assumptions. One proceeds as follows. 

The fundamental vertices give rise to local amplitudes which must be polynomials\footnote{This is true at 4-point and higher; for 3-point, massless particle amplitudes are uniquely fixed by the little group scaling.} in the spinor helicity brackets, and it is simple to construct the most general such  ansatz for the local input amplitudes. One can further restrict this ansatz by imposing on it the soft behaviors associated with the assumed symmetries.  The result of recursing this input from the fundamental vertices is supposed to be a physical amplitude and therefore it must necessarily be independent of the $n-4$ parameters $a_i$ that are unfixed by \reef{momconstraint}. If that is {\em not} the case for any ansatz of the fundamental input amplitudes (vertices), we learn that there cannot exist a theory with the properties (1) and (2) above. On the other hand, an $a_i$-independent result is evidence (but not proof) of the existence of such a theory. It may well be that $a_i$-independence requires some of the free parameters in the input amplitudes to be fixed in certain ways and this can teach us important lessons about the underlying theory. The test of $a_i$-independence can be done efficiently numerically, and  this way one can scan through theory-space to test which symmetries are compatible with a given model input. 

Additionally, one can impose further constraints from unbroken global symmetries, for example, one can restrict the input from the fundamental amplitudes by imposing the supersymmetry Ward identities. We shall see examples of this in later sections.

{\bf 4d and 3d consistency checks.}
There is a subtlety that must be addressed for $n=6$. In that case, the solution space is 2-dimensional, but one solution is the trivial one with all $a_i$ equal. Furthermore,  one can  rescale all $a_i$. This means that if the recursed result for the amplitude depends on the $a_i$ only through ratios of the form
\be
\label{ratioofai}
\frac{(a_i-a_j)}{(a_k-a_l)}\,,
\ee
it will appear to be $a_i$-independent numerically, but the result will nonetheless have spurious poles. To detect this problem numerically, we dimensionally reduce the recursed result to 3d.\footnote{ The dimensional reduction from 4d to 3d is carried out by simply replacing all square spinors by  angle spinors.}. Then the space of solutions to \reef{momconstraint} is $(n-3)$-dimensional, so there are non-trivial solutions and a numerical 3d test will reveal dependence on ratios such as \reef{ratioofai} for $n=6$.

We refer to the consistency checks of $a_i$-independence as {\em 4d and 3d consistency checks}, respectively, or simply as {\em $n$-point tests} when applied to construction of $n$-point amplitudes. In this paper, we use 6-, 7- and 8-point tests.  In  Section \ref{s:softboot}, we present an overview of the resulting space of exceptional pure real and complex scalar, fermion, and vector EFTs.

{\bf Special requirements for non-trivial 5-point interactions.}
Consider 5-particle interactions which are non-trivial with respect to a given soft behavior. This could for example be the Wess-Zumino-Witten (WZW) term, which with 4 derivatives on 5 scalars has a non-trivial $\sigma=1$ soft behavior. Or the 5-point Galileon, which with 8 derivatives on 5 scalars has a non-trivial $\sigma=2$. Constructibility  tells us that one must be able to calculate such 5-point amplitudes from soft recursion relations via factorization, i.e.~
\be
\mathcal{A}_5 = \sum_I \frac{\hat{\mathcal{A}}_3 \hat{\mathcal{A}}_4}{P_I^2}\,.
\ee
However, there are no 3-point amplitudes available that could possibly make this work. The reason is that the only 3-scalar interaction with a non-zero on-shell amplitude is $\phi^3$, which  gives rise to amplitudes with $\sigma=-1$ \cite{Elvang:2016qvq}. So we appear to have a contradiction: the 
constructibility criterion tells us that these $5$-particle amplitudes are recursively constructible, but it is obviously impossible to construct them from lower-point input. 

What goes wrong is that at 5-points, there are no non-trivial choices of the $a_i$ parameters that  give valid recursion relations in 4d. So we have to go to 3d kinematics to resolve this issue. The above contradiction persists in 3d, so the only resolution is that these non-trivial constructible 5-point amplitudes must vanish in 3d kinematics. 

Indeed they do: for WZW term and the quintic Galileon, the 5-point matrix elements are
\be
  \label{A5LC}
  A_5^\text{WZW} = g_5 \, \eps_{\mu\nu\rho\sigma} p_1^\mu p_2^\nu p_3^\rho p_4^\sigma \,,
  ~~~~~~
  A_5^\text{Gal} = g_5' \,(\eps_{\mu\nu\rho\sigma} p_1^\mu p_2^\nu p_3^\rho p_4^\sigma)^2\,.
\ee
 The Levi-Civita contraction makes it manifest that these amplitudes vanish in 3d.

We conclude that {\em any} non-trivial (in the sense of soft behavior) 5-particle interaction must vanish in 3d. Thus, it is no coincidence that the WZW and quintic Galileon 5-point amplitudes are  proportional to Levi-Civita contractions.

\section{Soft Bootstrap}
\label{s:softboot}
We now turn to examples of how the soft recursion relations can be used to examine the existence of exceptional EFTs. The landscape of real scalar theories was previously  studied in \cite{Cheung:2014dqa,Cheung:2015cba,Cheung:2015ota,Cheung:2016drk}. We outline it briefly below for completeness, but otherwise focus on new results, in particular for complex scalars, fermions, and vectors. This section considers only theories with one kind of massless particle. One can of course also couple scalars, fermions, and vectors in EFTs, and this is discussed in Sections \ref{sec:SUSY_NLSM}, \ref{sec:sDBI}, and \ref{sec:Galileon}.

\subsection{Pure Scalar EFTs}
Consider an EFT with a single real scalar field $\phi$. There can only be non-vanishing 3-point amplitudes in $\phi^3$-theory and this gives amplitudes with soft weight $\sigma=-1$. Focusing on EFTs with soft weights $\sigma \ge 0$, the lowest-point amplitude is 4-point. 

The on-shell factorization diagrams that contribute in the recursion relations \reef{softrecursion2} for $\mathcal{A}_6(1_\phi\, 2_{{\phi}}\, 3_\phi\, 4_{{\phi}}\, 5_\phi\, 6_{{\phi}})$ are composed of a product of two 4-point amplitudes, for example the 123-channel diagram is
\begin{equation*}
	\mathcal{A}_6^{(123)}= 
	\raisebox{-1.1cm}{
	\begin{tikzpicture}
	\draw (7.9,0)--(9,0);
	\node at (7.6,0) {$2_{{\phi}}$};
	\draw (8.3,0.8)--(9,0);
	\node at (8,0.8) {$1_\phi$};
	\draw (8.3,-0.8)--(9,0);
	\node at (8,-0.8) {$3_\phi$};
	\node at(9.3,0.3) {$-P_{{\phi}}$};
	\draw (9,0)--(11,0);
	\node at (10.8,0.3) {$P_{\phi}$};
	\draw (11,0)--(12.1,0);
	\node at (12.5,0) {$5_\phi$};
	\draw (11,0)--(11.7,0.8);
	\node at (12.1,0.8) {$4_{{\phi}}$};
	\draw (11,0)--(11.7,-0.8);
	\node at (12.1,-0.8) {$6_{{\phi}}$};
	\end{tikzpicture}
	}
	=~
	\frac{\hat{\mathcal{A}}_L(0)\hat{\mathcal{A}}_R(0)}{P_{123}^2}
	+\sum_{i=1}^6
	    \text{Res}_{z=\frac{1}{a_i}}\,
	    \frac{\hat{\mathcal{A}}_L(z)\hat{\mathcal{A}}_R(z)}{z\,F(z)\,\hat{P}_{123}^2}\,,~~~
\end{equation*}
where $\hat{\mathcal{A}}_L = \hat{\mathcal{A}}_4(1_\phi\,2_\phi\,3_\phi\,-P_\phi)$ and $\hat{\mathcal{A}}_R= \hat{\mathcal{A}}_4(P_\phi\, 4_\phi\,5_\phi\,6_\phi)$.\footnote{The momenta in the hatted amplitudes are shifted; for simplicity, we do not write the hats on the momentum variables explicitly. Note that in particular $P_\phi$ should really be understood as $\hat{P}_\phi$ with $\hat{P}_\phi^2=0$.}
One sums over the 10 independent permutations corresponding to the 10 distinct factorization channels.\footnote{We do not consider color-ordering in this section. With color-ordering, one only includes the factorization diagrams from cyclic permutations of the external lines. }

For complex scalars, we assume that the input 4-point amplitudes are of the form 
$\mathcal{A}_4(1_Z\,2_{\bar{Z}}\,3_Z\,4_{\bar{Z}})$;\footnote{There is no color-ordering implied in any of the amplitudes here. We simply alternate $Z$ and $\bar{Z}$ states as odd/even numbered momentum lines. In later sections, other helicity states are grouped similarly, in particular for supersymmetric cases, states that belong to the positive helicity sector sit on odd-numbered lines and negative helicity sector states on even-numbered lines. This is convenient for the practical implementation but should not be misunderstood as an indication of color-ordering.} one can also consider more general input but it would not be compatible with supersymmetry, so in the present paper we do not discuss such options. 
At 6-point, there is only one type of amplitude that can arise from such 4-point input via recursion, and that is
 $\mathcal{A}_6(1_Z\,2_{\bar{Z}}\,3_Z\,4_{\bar{Z}}\,5_Z\,6_{\bar{Z}})$. The 123-channel diagram is \begin{equation}
     \label{123channelscalar}
	\mathcal{A}_6^{(123)}= 
	\raisebox{-1.1cm}{
	\begin{tikzpicture}
	\draw (7.9,0)--(9,0);
	\node at (7.6,0) {$2_{\bar{Z}}$};
	\draw (8.3,0.8)--(9,0);
	\node at (8,0.8) {$1_Z$};
	\draw (8.3,-0.8)--(9,0);
	\node at (8,-0.8) {$3_Z$};
	\node at(9.3,0.3) {$-P_{\bar{Z}}$};
	\draw (9,0)--(11,0);
	\node at (10.8,0.3) {$P_{Z}$};
	\draw (11,0)--(12.1,0);
	\node at (12.5,0) {$5_Z$};
	\draw (11,0)--(11.7,0.8);
	\node at (12.1,0.8) {$4_{\bar{Z}}$};
	\draw (11,0)--(11.7,-0.8);
	\node at (12.1,-0.8) {$6_{\bar{Z}}$};
	\end{tikzpicture}
	}
\end{equation}
To get the full amplitude, one must sum over all factorization channels:
\be
  \label{A6scalar}
\mathcal{A}_6(1_Z\,2_{\bar{Z}}\,3_Z\,4_{\bar{Z}}\,5_Z\,6_{\bar{Z}}) = 
\left(\mathcal{A}_6^{(123)}
+ \left(2\leftrightarrow4\right)+ \left(2\leftrightarrow6\right) \right)+ \left(1\leftrightarrow5\right)+ \left(3\leftrightarrow5\right).
\ee
In the following we consider real and complex scalar theories with 4- and 5-point fundamental vertices.

\subsubsection{Fundamental 4-point Interactions}

Consider  a theory of a single real scalar with fundamental 4-point interactions. We parameterize $ \mathcal{A}^\text{ansatz}_4$ as the most general polynomial in the Mandelstam variables $s, t, u$ (with $s+t+u=0$) and full Bose symmetry. We subject the recursed result for $\mathcal{A}_6$ to the test of $a_i$-independence, as described in Section \ref{consistency}. The result is
\bea
  \label{4ptrealscalarin}
  &&\pa^{2m} \phi^4 
  \\[2mm]
   \nonumber
  &&\begin{array}{|c|c|l|ccccc|}
  \hline
 \text{- [g]} & m & 
 \mathcal{A}^\text{ansatz}_4(1_\phi\,2_{{\phi}}\,3_\phi\,4_{{\phi}}) & \sigma=0 & 1 &2 & 3 & 4~~~ \\
  \hline
  0 & 0  & g &  \phi^4\text{-theory} & \text{F} & \text{F} & \text{F} & \text{F} \\
  2 & 1  & 0 &  - & \text{F} & \text{F}  & \text{F} & \text{F} \\ 
  4&  2  & g (s^2+t^2+u^2) &  -&  - &  \text{DBI} &\text{F} & \text{F}\\
  6&  3  & g \, s t u  &  -&  -& \text{Gal$_4$}&\text{Spec Gal$_4$} & \text{F}\\
  8&  4  & g (s^4 + t^4 + u^4)  &  -&  -&  -& \text{F} & \text{F}\\
  \hline
  \end{array}
\eea
In the table, we list the coupling dimension $[g]$ of the fundamental quartic couplings along with the most general ansatz for the corresponding 4-point amplitude. The dash, $-$, indicates that the constructibility criterion \reef{criterion} fails; this means ``triviality" in the sense described in Section \ref{non-constructibility}). ``F" indicates that the soft recursion fails to give an $a_i$-independent result,  and hence no such theory can exist with the given assumptions. When a case passes the 6-point test, we are able to uniquely identify which theory it is.  In the above table, the non-trivial theories that pass the 6-point test are: $\phi^4$-theory, DBI, and the quartic Galileon. The latter automatically has $\sigma=3$ (which is called the Special Galileon) and passes 6-point test for both 
$\sigma=2$ and $\sigma=3$.

The analysis for complex scalars proceeds similarly and the results are
\bea
  \label{4ptscalarin}
  &&\pa^{2m} Z^2 \bar{Z}^2
  \\[2mm]
   \nonumber
  &&\begin{array}{|c|c|l|cccc|}
  \hline
 \text{- [g]} & m & \mathcal{A}^\text{ansatz}_4(1_Z,2_{\bar{Z}},3_Z,4_{\bar{Z}}) & \sigma=0 & 1 &2 & 3 \\
  \hline
  0 & 0  & g &  |Z|^4\text{-theory} & \text{F} & \text{F} & \text{F} \\
  2 & 1  & g t &  - & \text{$\mathbb{CP}^1$ NLSM} & \text{F}  & \text{F} \\ 
  4&  2  & g t^2 + g' s u &  -&  - &  \text{$g'=0$ cmplx DBI} &\text{F}\\
  6&  3  & g t^3 + g' s t u  &  -&  -& \text{$g=0$ cmplx Gal$_4$}& \text{F}\\
  8&  4  & g t^4 + g' t^2 s u + g'' s^2 u^2  &  -&  -&  -& \text{F}\\
  \hline
  \end{array}
\eea
The non-trivial theories are $|Z|^4\text{-theory}$, the $\mathbb{CP}^1$ NLSM (which is studied in further detail in Section \ref{sec:SUSY_NLSM}), and the complex scalar versions of DBI and the quartic Galileon. Note that there does not exist a complex scalar version of the Special Galileon with $\sigma=3$. The results for the 6-point amplitudes of each of the theories with $\sigma>0$ can be found in Appendix \ref{sec:AmplitudeExpressions}.

\subsubsection{Fundamental 5-point Interactions}
\label{s:5ptsc}
At 5-point, the input amplitudes are constructed as polynomials of Mandelstam variables $s_{ij}$ and  Levi-Civita contractions of momenta. They must obey  (1) momentum conservation, (2) Bose symmetry,  and (3) assumed soft behavior $\sigma$. In many cases, these constraints on the 5-point input amplitudes are sufficient to rule out such theories (assuming no other interactions) without even applying soft recursion. 

As discussed at the end of Section \ref{consistency}, non-trivial 5-point
 amplitudes must vanish in 3d kinematics, so they are naturally written using the Levi-Civita tensor, 
as in the two cases of WZW and the quintic Galileon \reef{A5LC}.

We can summarize the results in the following:
\begin{itemize}
\item {\bf 1 real scalar.} There are only two non-trivial theories based on a fundamental 5-point interaction, namely $\phi^5$-theory, which has $[g_5]=-1$ and $\sigma=0$, and the quintic Galileon, which has $[g_5]=-9$ and $\sigma=2$.

\item {\bf 1 complex scalar.}
We assume input amplitudes of the form $\mathcal A_5 ( 1_Z 2_{\bar Z} 3_Z 4_{\bar Z} 5_Z )$. Two cases pass the 8-point test:

The quintic $g_5 (Z^3 \bar{Z}^2+Z^2 \bar{Z}^3)$-theory with $[ g_5 ] = -1$ has $\sigma_Z=0$.

The  complex-scalar version of the quintic Galileon with $[ g_5 ] = - 9$ and $\sigma_Z = 2$.
The 5-point amplitude is
\begin{equation} \label{Gal5Z}
    \mathcal A_5 ( 1_Z 2_{\bar Z} 3_Z 4_{\bar Z} 5_Z ) = g_5 (\eps_{\mu\nu\rho\sigma} p_1^\mu p_2^\nu p_3^\rho p_4^\sigma)^2\,,
\end{equation}
same as for the real-scalar quintic Galileon. The fact that it passes the 8-point test is somewhat trivial: because of the two explicit factors of momentum for 4 out of  5 particles, the residues at $1/a_i$ vanish identically for each factorization channel. The same is true for the real Galileon, so the 8-point test is not really effective as an indicator of whether such a theory may exist. 

Suppose the putative complex-scalar quintic Galileon is coupled to the complex scalar DBI. Then we can conduct a 7-point test based on factorization into a quantic Galileon  and a quartic DBI subamplitude. The test of $a_i$-independence requires the coupling constant $g_5$ to vanish. This means that the DBI-Galileon with a complex scalar cannot have a 5-point interaction.

At $[ g_5 ] = - 9$, there is a 6-parameter family of 5-point amplitudes with $\sigma_Z = 1$. The EFT with such amplitudes is generally non-constructible. However, a 1-parameter sub-family is compatible with the constraints of supersymmetry. As discussed in  \cite{Elvang:2017mdq} and further in Section \ref{s:galsusy} this may be a candidate for a supersymmetric quintic Galileon with a limited sector of constructible amplitudes. 

\end{itemize}

\subsection{Pure Fermion EFTs}
\label{s:fermionEFTs}
Let us now consider EFTs with only fermions and fundamental interactions of the form 
$\pa^{2m} \psi^2 \bar{\psi}^2$. This is not the only choice, but it is the option compatible with supersymmetry. Moreover, we have found that couplings of ``helicity violating" 4-point interactions in the fermion sector must vanish by the 6-point test in all pure-fermion cases we tested. The calculations proceed much the same way as for scalars, except that one must be  more careful with signs when inserting fermionic states on the internal line. The diagrams needed for the recursive calculation of the 6-fermion amplitude 
$A_6(1^+_\psi\,2^-_\psi\,3^+_\psi\,4^-_\psi\,5^+_\psi\,6^-_\psi)$ are just like those in the scalar case \reef{123channelscalar}, but now the permutations have to be taken with a sign:
\be
\mathcal{A}_6(1^+_\psi\,2^-_\psi\,3^+_\psi\,4^-_\psi\,5^+_\psi\,6^-_\psi)=\left(\mathcal{A}_6^{(123)}-(1\leftrightarrow 5)-(3\leftrightarrow 5)\right)-(2\leftrightarrow 4)-(2\leftrightarrow 6).
\ee

The input 4-point amplitudes $\mathcal{A}_4(1_\psi^+\,2_{{\psi}}^-\,3_\psi^+\,4_{{\psi}}^-)$ are fixed by little group scaling to be $\<24\>[13]$ times a Mandelstam polynomial of degree $m-1$ that must be symmetric under $s \lra u$ to ensure Fermi antisymmetry for identical fermions. The most general input amplitudes for low values of $m$ are summarized in the table below that also shows the result of the recursive 6-point test:
\bea
  \label{4ptfermionin}
  &&\pa^{2m} \psi^2 \bar{\psi}^2\\ \nonumber
  &&\begin{array}{|c|c|l|cccc|}
  \hline
 \text{- [g]} & m & \mathcal{A}_4(1_\psi^+\,2_{{\psi}}^-\,3_\psi^+\,4_{{\psi}}^-) = \<24\>[13] \times & \sigma = 0 & 1 & 2 & 3 \\
  \hline
  2 & 0  & g & \text{NJL} & \text{F} & \text{F} & \text{F}\\
  4&  1  & g t &  - & \text{A-V} & \text{F} & \text{F}\\
  6&  2  & g t^2 + g' s u & - &- & \text{F} & \text{F} \\
  8&  3  & g t^3 + g' s t u &  - & - & g=0~\text{new} & \text{F}\\
  \hline
  \end{array}
\eea

We  comment briefly on these results:
\begin{itemize}
\item The {\bf NJL model} has the fundamental 4-fermion interaction $\bar\psi^2 \psi^2$ and the result of recursing it to 6-point is given in Appendix \ref{a:NLSMamp}.  The relevance of this model will for our purposes be as part of the supersymmetrization of the NLSM (see Section \ref{sec:SUSY_NLSM}).

\item  {\bf Akulov-Volkov theory} of Goldstinos is the only non-trivial EFT with coupling of mass-dimension $-4$. The Goldstinos in this theory have low-energy theorems with $\sigma=1$. The 6-fermion amplitude is given in \eqref{eq:DBI_A6_4} in Appendix \ref{a:susydbi}.

\item There are no constructible purely fermionic EFTs with fundamental quartic coupling  $[g_4] =-6$. Nonetheless, as was shown in \cite{Elvang:2017mdq}, the quartic Galileon has a supersymmetrization with a 4-fermion fundamental interaction, however, the fermion has $\sigma=1$, so the all-fermion amplitudes  in that theory are not constructible by soft recursion: one needs additional input from supersymmetry. We refer the reader to \cite{Elvang:2017mdq} and present some further details in Section \ref{s:galsusy}. 

\item For  $[g]=-8$ and $\sigma=2$, the 6-point numerical test is passed in  4d kinematics without constraints on $g$ and $g'$; that is because the recursed result depends only on ratios \reef{ratioofai}. When the 3d consistency check is employed, we learn that we must set $g=0$ to ensure $a_i$-independence.  (This is not a strong test since the particular form of the interaction, $stu$, ensures that all $1/a_i$-poles cancel in each factorization individual diagram.) 
Hence, the theory that passes the 6-point test with $\sigma=2$ has  $\mathcal{A}_4(1_\psi,2_{\bar{\psi}},3_\psi,4_{\bar{\psi}}) = g' \<24\>[13] s t u$. The subtracted recursion relations fail at $n > 6$, which means that at 8-point and higher, this model is not uniquely determined by its symmetries. The Lagrangian construction of this theory has been studied as a fermionic generalization of the scalar Galileon \cite{CliffUnpub}.
\end{itemize}

\subsection{Pure Vector EFTs}
Pure abelian vector EFTs consist of interaction terms built from $F_{\mu\nu}$-contractions, possibly dressed with extra derivatives.  In 4d, the Cayley-Hamilton relations imply that theories built from just field strengths $F_{\mu\nu}$ can be constructed from two types of index-contractions, namely  (see for example \cite{Cheung:2018oki})
\be
  f = -\frac{1}{4} F_{\mu\nu} F^{\mu\nu}
  ~~~~\text{and}~~~~
  g = -\frac{1}{4} F_{\mu\nu} \tilde{F}^{\mu\nu}\,, 
\ee
where $\tilde{F}^{\mu\nu} = \frac{1}{2} \eps^{\mu\nu\rho\sigma} F_{\rho\sigma}$.
If one assumes parity, the Lagrangian can only contain even powers of $g$. One can then write an ansatz for the Lagrangian as 
\be
  \label{Lvector}
  \mathcal{L} = f + \frac{b_1}{\Lambda^4} f^2 + \frac{b_2}{\Lambda^4} g^2 
  + \frac{b_3}{\Lambda^8} f^3 + \frac{b_4}{\Lambda^8}  f g^2 + \ldots 
\ee 
As established in Section \ref{non-constructibility}, a model with photon interactions built of $F_{\mu\nu}$-contractions only have soft behavior $\sigma=0$. The simplest 4-photon interactions may naively look like the vector equivalent of the constructible $\phi^4$ scalar EFT. However, that is not the case. For the scalar, the 6-particle operator $\frac{1}{\Lambda^2} \phi^6$ is subleading to the pole contributions with two $\phi^4$-vertices. However, for photons the pole terms with two $\frac{1}{\Lambda^4}F^4$-vertices are exactly the same order as 
$\frac{1}{\Lambda^8} F^6$. Therefore amplitudes in a theory with $F^n$ interactions and $\sigma=0$ are non-constructible, in other words it is trivial to have $\sigma=0$ for any choice of coefficients $b_i$. One may ask if it is possible to choose the parameters $b_i$ in \reef{Lvector} such that the amplitudes have enhanced soft behavior $\sigma>0$. The  6-point soft recursive test shows that this is impossible, i.e.~no models exist with Lagrangians of the form \reef{Lvector} and $\sigma>0$.

Nonetheless, the class of theories with pure $F^n$-interactions do include one particularly interesting case, namely Born-Infeld (BI) theory. The BI Lagrangian can be written in 4d as
\be
   \mathcal{L}_\text{BI} = 
   \Lambda^4 \bigg(1 - \sqrt{-\det\big(\eta_{\mu\nu} + F_{\mu\nu}/\Lambda^2 \big)} \bigg)\,.
\ee
Upon expansion, the Lagrangian will take the form \reef{Lvector} with some particular coefficients $b_i$. As noted, those particular coefficients do not change the single-soft behavior of amplitudes, the BI photon also has $\sigma=0$. Nonetheless, BI theory does have the distinguishing feature of being the vector part of a supersymmetric EFT. In particular,  $\mathcal{N}=1$  supersymmetric Born-Infeld theory couples the BI vector to a Goldstino mode whose self-interactions are described by the Akulov-Volkov action. One can also view Born-Infeld as the vector part of the $\mathcal{N}=2$  or $\mathcal{N}=4$ supersymmetrization of DBI.
 It was argued recently \cite{Cheung:2018oki} that supersymmetry ensures  BI amplitudes to vanish in certain multi-soft limits. Based on that, the BI amplitudes can be calculated unambiguously using on-shell techniques \cite{Cheung:2018oki}. Alternatively, one can show that the $\mathcal{N}=1$ supersymmetry Ward identities uniquely fix the BI amplitudes in terms of amplitudes with Goldstinos; we  discuss this briefly in Section \ref{sec:sDBI} and in further  detail in the context of partial breaking of supersymmetry in a forthcoming paper.

Next, one can consider EFTs in which the field strengths are dressed with derivatives, for example 
\be
   \mathcal{L} = - \frac{1}{4}F^2 + \frac{c_1}{\Lambda^6}\partial^2 F^4  + \frac{c_1}{\Lambda^{12}}\partial^4 F^6 +\ldots 
\ee 
Theories with fundamental 4-point interactions are non-constructible for $\sigma=0$ and fail the soft recursion $a_i$-independence 6-point test for $\sigma > 0$. One implication of this is that there can be no vector Goldstone bosons with vanishing low-energy theorems. This conclusion was also reached in \cite{Klein:2018ylk}, but from a very different algebraically-based analysis.   A second implication is that the pure vector sector of an $\mathcal{N}\ge2$ Galileon model is  non-constructible with the basic soft recursion, and other properties (such as supersymmetry)  have to be specified in order to determine those amplitudes recursively.

There are other interesting vector EFTs: we study in detail the $\mathcal{N}=2$ supersymmetric NLSM in  Section \ref{sec:SUSY_NLSM}. Furthermore, massive gravity \cite{deRham:2010gu,deRham:2010ik,Hinterbichler:2011tt} motivates the existence of a vector-scalar theory coupling Galileons to a vector field; we explore this in Section \ref{s:galvec}.

\section{Soft Limits and Supersymmetry}
\label{softsusy}

For models with unbroken supersymmetry, the on-shell amplitudes satisfy a set of  linear 
relations known as the \textit{supersymmetry Ward identities} \cite{Grisaru:1976vm,Grisaru:1977px}. (For recent reviews and results, see  \cite{Elvang:2009wd,Elvang:2013cua,Elvang:2015rqa}.)  In this section, we use $\mathcal{N}=1$ supersymmetry to derive general consequences for the soft behavior for massless particles in the same supermultiplet. It is not assumed that these particles are Goldstone or quasi-Goldstone modes; the results apply to all $\mathcal{N}=1$ supermultiplets of massless particles. The consequences for extended supersymmetry are directly inferred from the $\mathcal{N}=1$ constraints. 

\subsection{\texorpdfstring{$\mathcal{N}=1$}{N=1} Supersymmetry Ward Identities}

We consider $\mathcal{N}=1$ \textit{chiral} and \textit{vector}  supermultiplets. We use the following shorthand for the action of the supercharges on individual particles with momentum label $i$: for chiral multiplets
\begin{equation}
  \label{susychiral}
  \begin{array}{|c||c|c||c|c|}
  \hline
  \text{state}~i & \mathcal{Q} \cdot i & \mathcal{A}_n~\text{prefactor} & \mathcal{Q}^\dagger \cdot i & \mathcal{A}_n~\text{prefactor}\\
  \hline
  ~\psi^+ & Z & |i] & 0 & 0 \\
  Z & 0 & 0 & ~\psi^+ &-|i\> \\
  \overline{Z} & ~\psi^- & |i] & 0 & 0 \\
  ~\psi^- &0 & 0 &   \overline{Z} & -|i\> \\
  \hline
  \end{array}
\end{equation}
where $Z$ is a complex scalar and $\psi$ is a Weyl fermion. The superscripts $\pm$ refer to the helicity of the particle.  $\mathcal{Q}^\dagger$ raises helicity by $1/2$ while $\mathcal{Q}$ lowers it by $1/2$. The prefactor is what goes outside the amplitude when the supercharge acts on it, e.g.~
\be
  \begin{split}
   \mathcal{Q} \cdot \mathcal{A}_n \big(1_Z\, 2_\psi^+ \, 3_\psi^+ \, 4_{\overline{Z}} \ldots \big) 
    = & ~~0~+~|2] \mathcal{A}_n \big(1_Z\, 2_Z  \, 3_\psi^+ \, 4_{\overline{Z}} \ldots \big) 
      ~-~ |3] \mathcal{A}_n \big(1_Z\, 2_\psi^+ \, 3_Z \, 4_{\overline{Z}} \ldots \big) \\
   &   
      ~+ ~|4] \mathcal{A}_n \big(1_Z\, 2_\psi^+ \, 3_\psi^+ \, 4_\psi^- \ldots \big)
      ~+ \ldots
   \end{split} 
\ee
Due to the Grassmann nature of the supercharges, there is a minus sign for each fermion that the supercharge has to move past to get to the $i$th state. 

Similarly for a vector multiplet:
\begin{equation}
  \label{susyvector}
  \begin{array}{|c||c|c||c|c|}
  \hline
  \text{state}~i & \mathcal{Q} \cdot i & \mathcal{A}_n~\text{prefactor} & \mathcal{Q}^\dagger \cdot i & \mathcal{A}_n~\text{prefactor}\\
  \hline
  \gamma^+ & \psi^+ & |i] & 0 & 0 \\
  \psi^+  & 0 & 0 & \gamma^+ &-|i\> \\
  \psi^- & \gamma^- & -|i] & 0 & 0 \\
  \gamma^- &0 & 0 &   \psi^- & |i\> \\
  \hline
  \end{array}
\end{equation}
where $\psi$ is a Weyl fermion and $\gamma$ is a vector boson. 

 In this notation, the supersymmetry Ward identities are equivalent to the statement that the following action of the supercharges annihilates the amplitude \cite{Elvang:2013cua,Elvang:2015rqa,Elvang:2009wd}
\be
\label{SWI}
\begin{split}
  0= \mathcal{Q}\cdot \mathcal{A}_n \left(1,\ldots,n\right) &=  \sum_{i=1}^n(-1)^{L_i+P_i}|i] \mathcal{A}_n\left(1,\ldots,\mathcal{Q}\cdot i, \ldots ,n\right), \\
  0 = \mathcal{Q}^\dagger \cdot \mathcal{A}_n \left(1,\ldots,n\right) &=  \sum_{i=1}^n(-1)^{L_i+P_i}|i\rangle \mathcal{A}_n\left(1,\ldots,\mathcal{Q}^\dagger\cdot i, \ldots ,n\right),
\end{split}
\ee
where $L_i$ is equal to the number of fermions to the \textit{left} of $\mathcal{Q}^{(\dagger)}\cdot i$ and the factors $P_i=0$ or $1$ correspond to the additional minus signs associated with the spinor prefactors as described in Tables \ref{susychiral} and \ref{susyvector}. Note that the action of the supercharges always changes the number of fermions by $\pm 1$, but that amplitudes are non-vanishing only if the number of fermions is even. So to get an interesting relation among amplitudes on the right-hand-side, the amplitude on the left-hand-side must vanish identically. 

\subsection{Soft Limits and Supermultiplets}
\label{sec:softsusy}
We consider the chiral multiplet and vector multiplet separately and then extend the results to enhanced supersymmetry.

{\bf Chiral multiplet.} Define the \textit{soft factors} $\mathcal{S}_n^{(i)}$ as the momentum dependent coefficients in the holomorphic soft expansion taken here for simplicity on the first particle
\be \label{softexp}
  \begin{split}
    \mathcal{A}_n\left(\{\epsilon|1\rangle,|1]\}_Z,\ldots\right) 
  &~\rightarrow~ 
  \mathcal{S}_n^{(0)}(1_Z,\ldots)\,\epsilon^{\sigma_Z} + \mathcal{S}_n^{(1)}(1_Z,\ldots)\,\epsilon^{\sigma_Z+1} + \mathcal{O}\left(\epsilon^{\sigma_Z+2}\right)\,, \\[1.5mm]
  \mathcal{A}_n\left(\{\epsilon|1\rangle,|1]\}_\psi^+,\ldots\right) 
  &~\rightarrow~ \mathcal{S}_n^{(0)}(1_\psi^+,\ldots)\,\epsilon^{\sigma_\psi}+ \mathcal{S}_n^{(1)}(1_\psi^+,\ldots)\,\epsilon^{\sigma_\psi+1} + \mathcal{O}\left(\epsilon^{\sigma_\psi+2}\right).
    \end{split}
\ee
The soft weights are $\sigma_Z$  and $\sigma_\psi$ for the scalar and fermion, respectively. To see how supersymmetry forces relations among the soft weights and soft factors we use (\ref{SWI}) to write
\be \label{softSWI}
  \begin{split}
   \mathcal{A}_n\left(1_Z,\ldots,n\right) &= \sum_{i=2}^n(-1)^{L_i+P_i+1} \frac{[X i ]}{[ X 1 ]} 
   \,\mathcal{A}_n\left(1_\psi^+,\ldots,\mathcal{Q} \cdot i , \ldots,n\right),\\
  \mathcal{A}_n\left(1_\psi^+,\ldots,n\right) &= \sum_{i=2}^n(-1)^{L_i+P_i+1} \frac{\langle X i \rangle}{\langle X 1 \rangle} 
  \,\mathcal{A}_n\left(1_Z,\ldots,\mathcal{Q}^\dagger \cdot i , \ldots,n\right),
    \end{split}
\ee
where the arbitrary $X$-spinor cannot be proportional to $|1\>$ or $|1]$.

Taking the holomorphic soft expansion on the right-hand-side of these expressions, in the second line only, an extra power of $\epsilon$ appears in the denominator and we find
\be
  \begin{split}
  &\mathcal{S}_n^{(0)}(1_Z,\ldots)\,\epsilon^{\sigma_Z} 
  + \mathcal{O}\left(\epsilon^{\sigma_Z+1}\right) 
  = \sum_{i=2}^n(-1)^{L_i+P_i+1} \frac{[X i ]}{[ X 1 ]} 
  \mathcal{S}_n^{(0)}(1_\psi^+,\ldots,\mathcal{Q}\cdot i,\ldots)\,\epsilon^{\sigma_\psi} 
  + \mathcal{O}\left(\epsilon^{\sigma_\psi+1}\right),\nonumber\\[1.5mm]
   &
   \mathcal{S}_n^{(0)}(1_\psi^+,\ldots)\,\epsilon^{\sigma_\psi} 
   + \mathcal{O}\left(\epsilon^{\sigma_\psi+1}\right) 
  = \sum_{i=2}^n(-1)^{L_i+P_i+1} \frac{\langle X i \rangle}{\langle X 1 \rangle} 
  \mathcal{S}_n^{(0)}(1_Z,\ldots,\mathcal{Q}^\dagger\cdot i,\ldots)\,\epsilon^{\sigma_Z-1} 
  + \mathcal{O}\left(\epsilon^{\sigma_Z}\right).
  \end{split}
\ee
The leading power of $\epsilon$ on the right-hand-side must match the leading power on the left. It is possible that cancellations among the terms on the right-hand-side may effectively increase the leading power but never decrease it. This then gives the following inequalities
\begin{equation}
  \sigma_Z \geq \sigma_\psi \;\;\; \text{and} \;\;\; \sigma_\psi \geq \sigma_Z -1\,,
\end{equation}
for which there are only two solutions
\begin{equation}
\label{susysigma1}
\boxed{ ~ \sigma_Z = \sigma_\psi+1 \;\;\; \text{or} \;\;\; \sigma_Z = \sigma_\psi \,.~}
\end{equation}
These two options have different consequences for the soft factors.
For $\sigma_Z = \sigma_\psi +1$, we have
\be
  \begin{split}
  0~=~ &\sum_{i=2}^n(-1)^{L_i+P_i}[Xi]\,\mathcal{S}^{(0)}_n\left(1_\psi^+,\ldots,\mathcal{Q}\cdot i,\ldots\right) ,\\
  \mathcal{S}_n^{(0)}\left(1_\psi^+,\ldots\right) ~=~ &\sum_{i=2}^n(-1)^{L_i+P_i+1}\frac{\langle X i\rangle}{\langle X 1 \rangle}\,\mathcal{S}^{(0)}_n\left(1_Z,\ldots,\mathcal{Q}^\dagger\cdot i,\ldots\right),
  \end{split}
\ee
while for $\sigma_\phi = \sigma_\psi$, we have
\be
  \begin{split}
  0~=~&\sum_{i=2}^n(-1)^{L_i+P_i}\langle Xi\rangle\mathcal{S}^{(0)}_n\left(1_Z,\ldots,\mathcal{Q}^\dagger\cdot i,\ldots\right) , \\
  \mathcal{S}_n^{(0)}\left(1_Z,\ldots\right) ~=~& \sum_{i=2}^n(-1)^{L_i+P_i+1}\frac{[X i]}{[X 1]}\mathcal{S}^{(0)}_n\left(1_\psi^+,\ldots,\mathcal{Q}\cdot i,\ldots\right). 
  \end{split}
\ee
In addition there will be an infinite number of similar relations which come from matching higher powers in $\epsilon$. 

{\bf Vector multiplet.} We define the soft factors as
\begin{equation}
 \mathcal{A}_n\left(\{\epsilon|1\rangle,|1]\}_\gamma^+,\ldots\right) 
 ~\rightarrow~ 
 \mathcal{S}_n^{(0)}(1_\gamma^+,\ldots)\,\epsilon^{\sigma_\gamma} + \mathcal{S}_n^{(1)}(1_\gamma^+,\ldots)\,\epsilon^{\sigma_\gamma+1} + \mathcal{O}\left(\epsilon^{\sigma_\gamma+2}\right).
\end{equation}
The analysis of the supersymmetry Ward identities proceeds similarly to that of the chiral multiplet and results in only two options for the soft weights:
\begin{equation}
\label{susysigma2}
  \boxed{ ~ \sigma_\psi = \sigma_\gamma + 1, \;\;\; \text{or} \;\;\; \sigma_\psi = \sigma_\gamma\,.~}
\end{equation}
The consequences for the soft factors are for  $\sigma_\psi = \sigma_\gamma +1$ 
\be
  \label{bonusWID}
  \begin{split}
  0~=~&\sum_{i=2}^n(-1)^{L_i+P_i}[Xi]\mathcal{S}^{(0)}_n\left(1_\gamma^+,\ldots,\mathcal{Q}\cdot i,\ldots\right) , \\
  \mathcal{S}_n^{(0)}\left(1_\gamma^+,\ldots\right) ~=~& \sum_{i=2}^n(-1)^{L_i+P_i+1}\frac{\langle X i\rangle}{\langle X 1 \rangle}\mathcal{S}^{(0)}_n\left(1_\psi^+,\ldots,\mathcal{Q}^\dagger\cdot i,\ldots\right), 
  \end{split}
\ee
and  for $\sigma_\gamma = \sigma_\psi$ 
\be
  \begin{split}
  0~=~&\sum_{i=2}^n(-1)^{L_i+P_i}\langle Xi\rangle\mathcal{S}^{(0)}_n\left(1_\psi^+,\ldots,\mathcal{Q}^\dagger\cdot i,\ldots\right) ,\\
  \mathcal{S}_n^{(0)}\left(1_\psi^+,\ldots\right) ~=~& \sum_{i=2}^n(-1)^{L_i+P_i+1}\frac{[X i]}{[X 1]}\mathcal{S}^{(0)}_n\left(1_\gamma^+,\ldots,\mathcal{Q}\cdot i,\ldots\right). 
  \end{split}
\ee

Note that we have made no assumptions about the sign of $\sigma$, so the relations derived here are totally general. Also, the supersymmetry Ward identities hold at all orders in perturbation theory, so the relations among the soft behaviors remain true at loop-level. 

{\bf Extended supersymmetry.} Relations between the soft weights of particles in the same massless supermultiplets in extended supersymmetry follow directly from the $\mathcal{N}=1$ results above, since the supersymmetry Ward identities take the same form for each pair of $(s,s+\tfrac{1}{2})$-multiplets. In particular, the soft weights of the boson ($\sigma_B$) and fermion ($\sigma_F$) in a $(s,s+\tfrac{1}{2})$-multiplet are related as
\be
  \label{sigmaBFsusy}
   \left\{
   \begin{array}{ll}
        \sigma_B = \sigma_F+1 \;\;\; \text{or} \;\;\; \sigma_B = \sigma_F 
        & ~\text{for $s$ integer}\,, \\[1.5mm]
        \sigma_B = \sigma_F-1 \;\;\; \text{or} \;\;\; \sigma_B = \sigma_F 
        & ~\text{for $s$ half-integer}    \,.
   \end{array}
   \right.
\ee
These relations will be useful in later applications in this paper. For now, we make a small aside and demonstrate the application of \reef{sigmaBFsusy} to the case of spontaneously broken superconformal symmetry and for unbroken extended supergravity. 

\subsection{Application to Superconformal Symmetry Breaking}
\label{sec:scft}
The breaking of conformal symmetry gives rise to a single Goldstone mode \cite{Low:2001bw}, often called the dilaton. 
It has been established in the literature \cite{Boels:2015pta,DiVecchia:2015jaq,Bianchi:2016viy} that this dilaton obeys low-energy theorems with $\sigma=0$. In a superconformal theory, breaking of conformal invariance must be accompanied by breaking of the superconformal symmetries. This follows from the algebra: $\{\mathcal{S},\mathcal{S}^\dagger\} = \mathcal{K}$, $[\mathcal{Q},\mathcal{K}] = \mathcal{S}^\dagger$ and $[\mathcal{Q}^\dagger,\mathcal{K}] = \mathcal{S}$, where $\mathcal{K}$ are the generators of conformal boosts, $\mathcal{S}$ and $\mathcal{S}^\dagger$ are the superconformal fermionic generators, and $\mathcal{Q}$ and $\mathcal{Q}^\dagger$ are the regular supercharges with $\{\mathcal{Q},\mathcal{Q}^\dagger\} = \mathcal{P}$. 

Assuming $\mathcal{Q}$-supersymmetry to be unbroken, the dilaton will be joined by a Goldstone mode from the broken R-symmetry to form a complex scalar $Z$ with $\sigma_Z=0$.\footnote{An example of the bosonic part of an $\mathcal{N}=1$ effective action of the dilaton and a $U(1)_R$ Goldstone boson can be found in \cite{Bobev:2013vta}.} It follows from our general analysis that the fermionic partner of $Z$ will have $\sigma=0$ or $\sigma=-1$. For the latter, Yukawa-interactions are necessary  \cite{Elvang:2016qvq} and supersymmetry then requires cubic scalar interactions $Z |Z|^2 + \text{h.c.}$ which would imply $\sigma=-1$ for the dilaton. Since $\sigma_Z=0$, $\sigma=-1$  is not possible for the dilaton and we conclude that the Goldstino mode associated with the breaking of the superconformal fermionic symmetries generated by $\mathcal{S}$ and $\mathcal{S}^\dagger$ must have low-energy theorems with soft weight $\sigma=0$.

An example is $\mathcal{N}=4$ SYM on the Coulomb branch with the simplest breaking pattern.\footnote{See \cite{Craig:2011ws,Bianchi:2016viy} for explicit amplitudes on the Coulomb branch of $\mathcal{N}=4$ SYM.} The $R$-symmetry is broken from $SO(6)$ to $SO(5)$ and the five broken generators give rise to five Goldstone modes which join the dilaton of the conformal breaking to be the 6 real scalars of an $\mathcal{N}=4$ massless multiplet. The supermultiplet also contains the 4 Goldstinos associated with the four broken superconformal generators. The supermultiplet is capped off by a $U(1)$ vector whose soft weight, by the above analysis, must be either $\sigma=0$ or $-1$. The states that are charged under this $U(1)$ are the massive $W$-multiplets and in their presence, one can have $\sigma=-1$, otherwise $\sigma=0$ for the vector.

\subsection{Application to Supergravity}
\label{sec:extsusy}

It is well-known that gravitons have a universal soft behavior \cite{Weinberg:1965shj}: when the soft limit \reef{softlimit} is applied to a single graviton, the amplitude diverges as $1/\eps^3$, i.e.~the soft weight is $\sigma_2 =-3$. (In this section, we use a subscript on the soft weight to indicate the spin of the particle.) 
Applying \reef{sigmaBFsusy} shows that the gravitino can have $\sigma_{3/2} =-2$ or $-3$. However, unitarity and locality constraints show \cite{Elvang:2016qvq} that amplitudes cannot be more singular than $1/\eps^2$ for a single soft gravitino, so it must be that $\sigma_{3/2} = -2$. This must be true in any supergravity theory.

Consider now a graviphoton in $\mathcal{N} \ge 2$ supergravity. Its supersymmetry Ward identities with the gravitino imply $\sigma_1 = -2$ or $\sigma_1 = -1$. The $\sigma_1 = -2$ behavior requires the graviphoton, and by supersymmetry also the gravitino, to interact with a pair of electrically charged particles via a dimensionless coupling; however, for the gravitino such a coupling is inconsistent with unitarity and locality  \cite{Elvang:2016qvq}. 
So there is only one option, namely $\sigma_1 = -1$. 

In pure $\mathcal{N} \ge 3$ supergravity, we also have spin-$\tfrac{1}{2}$ fermions in the graviton supermultiplet. By \reef{sigmaBFsusy} and the previous results, they can have either $\sigma_{1/2} = -1$ or $0$. The analysis in \cite{Elvang:2016qvq} shows that $\sigma_{1/2} = -1$ requires a dimensionless coupling of the spin-$\tfrac{1}{2}$ particle with two other particles, for example via a Yukawa coupling. Since there are no dimensionless couplings in pure supergravity, it follows from  \cite{Elvang:2016qvq} that the amplitude has to be $O(\eps^0)$ or softer. This leaves only one option, namely that $\sigma_{1/2} = 0$ in pure supergravity. 

In pure $\mathcal{N} \ge 4$ supergravity, the scalars in the supermultiplet can have $\sigma_{0} = 0$ or $\sigma_0=1$. If we focus on the MHV sector, the supersymmetry Ward identities give 
\be
  \mathcal{A}_n(1_ Z\, 2_{\bar{Z}}\, 3_h^-\,  4_h^+  \ldots n_h^+) = 
  \frac{\<13\>^4}{\<23\>^4} \,\mathcal{A}_n(1_ h^+\, 2_h^-\, 3_h^- \, 4_h^+  \ldots n_h^+)\,,
\ee
where $Z$ and $\bar{Z}$ denote any pair of conjugate scalars and $h$ are gravitons. Taking line 1 soft holomorphically, $|1\> \to \eps |1\>$, the graviton amplitude on the RHS diverges as $1/\eps^3$ but the prefactor vanishes as $\eps^4$. It follows that the MHV amplitude vanishes as  $O(\eps)$ in the single soft-scalar limit. In other words, for MHV amplitudes $\sigma_0 = 1$. It is tempting to conclude that one must have $\sigma_0 = 1$ for all amplitudes, but that is too glib, as we now explain.

\begin{table}[t]
\begin{center}
\begin{tabular}{|c|c|}
\hline
helicity state & $\sigma$ \\
\hline
+2 graviton & $-3$ \\
+3/2 gravitino & $-2$ \\
+1 graviphoton & $-1$ \\
+1/2 fermion & $0$ \\
0 scalar & $0$ or $+1$ \\
-1/2 fermion & $+1$ \\
-1 graviphoton & $+1$ \\
-3/2 gravitino & $+1$\\
-2 graviton & $+1$\\
\hline
\end{tabular}
\caption{\small Holomorphic soft weights $\sigma$  for the $\mathcal{N}=8$ supermultiplet. Note that the soft weights in this table follow from taking the soft limit holomorphically, $|i\> \to \eps |i\>$ for all states, independently of the sign of their helicity. At each step in the spectrum, the soft weight either changes by 1 or not at all. Note that one could also have used the anti-holomorphic definition $|i] \to \eps |i]$ of taking the soft limit; in that case the soft weights would just have reversed, to start with $\sigma=-3$ for the negative helicity graviton, but no new constraints would have been obtained on the scalar soft weights.  In $\mathcal{N}=8$ supergravity, the 70 scalars are Goldstone bosons of the coset $E_{7(7)}/SU(8)$ and hence $\sigma=1$. Including higher-derivative  corrections  may change this behavior to $\sigma=0$ depending on whether the added terms are compatible with the coset structure.\label{fig:N=8}}
\end{center}
\end{table}

It is known that the scalar cosets of $\mathcal{N} \ge 4$ pure supergravity theories  in 4d are  symmetric, and therefore lead to $\sigma_0 = 1$ vanishing low-energy theorems. But at the level of the on-shell amplitudes, this conclusion does not follow from the supersymmetry Ward identities alone: as we have seen, they give $\sigma_0 = 1$ or $\sigma_0 = 0$. That analysis has to remain true at all loop-orders. In $\mathcal{N} = 4$ supergravity, for example, the anomaly of the $U(1)$ R-symmetry can be expected to affect the soft behavior at some order. Our arguments show that it cannot happen in the MHV sector, but does not rule it out beyond MHV; this is what the  $\sigma_0 = 0$ accounts for. Furthermore,   one can add higher-derivative operators to the supergravity action such that supersymmetry is preserved but the low-energy theorems are not. Indeed, string theory does this in the $\alpha'$-expansion by adding to the $\mathcal{N} = 8$ tree-level action a supersymmetrizable operator $\alpha'^3 e^{-6\phi}R^4$. This operator does not affect the soft behavior of MHV amplitudes, but it is known that it does result  in non-vanishing single soft scalar limits for 6-particle NMHV amplitudes at order $\alpha'^3$ \cite{Elvang:2010kc,Brodel:2009hu}.

The results for $\mathcal{N}=8$ supersymmetry are summarized in Table \ref{fig:N=8}.

\subsection{MHV Classification and Examples of Supersymmetry Ward Identities}
\label{s:MHVclass}
For later convenience, we state here the explicit form of the supersymmetry Ward identities \reef{SWI} for a few particularly useful cases. We focus on the chiral multiplet, but similar results apply to the vector multiplet. 

First we make the simple observation that amplitudes with all $Z$'s or only one $\bar{Z}$ and rest $Z$'s vanish:
\be
  \label{MHVviol}
   \mathcal{A}_n \big(1_Z\ 2_{Z}\ 3_Z\ 4_{Z}\ldots  n_{Z} \big) = 0\,
   ~~~~\text{and}~~~~
   \mathcal{A}_n \big(1_Z\ 2_{\bar{Z}}\ 3_Z\ 4_{Z}\ldots \big) = 0\,.
\ee
This follows from the  supersymmetry Ward identities such as 
\be \nonumber
  \begin{split}
 0 &= \mathcal{Q}\cdot \mathcal{A}_n \big(1_\psi^+\ 2_{{Z}}\ 3_Z\ 4_{{Z}} \ldots  n_{Z}\big) 
 = |1] \,\mathcal{A}_n \big(1_Z\ 2_{{Z}}\ 3_Z\ 4_{Z}\ldots  n_{{Z}} \big) \,,
 \\
 0 &= \mathcal{Q}\cdot \mathcal{A}_n \big(1_\psi^+\ 2_{\bar{Z}}\ 3_Z\ 4_{{Z}} \ldots  n_{Z}\big)
  = |1] \,\mathcal{A}_n \big(1_Z\ 2_{\bar{Z}}\ 3_Z\ 4_{Z}\ldots n_{Z} \big) 
       - |2] \,\mathcal{A}_n(1_\psi^+\ 2_\psi^-\ 3_Z\ 4_{Z} \ldots n_{Z}\big) \,.
   \end{split}
\ee
Dotting in $[2|$ gives  \reef{MHVviol}.
Similarly $\mathcal{A}_n(1_\psi^+\ 2_\psi^-\ 3_Z\ 4_{Z} \ldots n_{Z}\big) = 0$ and so on. 
In the context of gluon scattering, the equivalent statements are that amplitudes with helicity structure $+++\ldots+$ or $-++\ldots+$ vanish. These helicity configurations are often called ``helicity violating".

The simplest non-vanishing amplitudes are often denoted MHV (Maximally Helicity Violating) in the context of gluon scattering and we adapt the same nomenclature here. MHV amplitudes obey the simplest supersymmetry Ward identities in that they are just linear proportionality relations. For example, it follows from 
\be
  \begin{split}
 0 &= \mathcal{Q}\cdot \mathcal{A}_n \big(1_\psi^+\ 2_{\bar{Z}}\ 3_Z\ 4_{\bar{Z}}\ 5_{{Z}} \ldots  n_{{Z}}\big)
 \\
 &= |1] \,\mathcal{A}_n \big(1_Z\ 2_{\bar{Z}}\ 3_Z\ 4_{\bar{Z}}\ldots \big) 
     - |2] \,\mathcal{A}_n(1_\psi^+\ 2_\psi^-\ 3_Z\ 4_{\bar{Z}} \ldots \big) 
     - |4] \,\mathcal{A}_n(1_\psi^+\ 2_{\bar{Z}}\ 3_Z\ 4_\psi^- \ldots \big) 
     \end{split}
\ee
upon dotting in $[4|$ that
\be
\label{WI4phi}
\mathcal{A}_n(1_\psi^+\ 2_\psi^-\ 3_Z\ 4_{\bar{Z}}\ 5_{{Z}} \ldots  n_{{Z}})=\frac{\sq{14}}{\sq{24}}\ \mathcal{A}_n(1_Z\ 2_{\bar{Z}}\ 3_Z\ 4_{\bar{Z}}\ 5_{{Z}} \ldots  n_{{Z}}).\\
\ee
Similarly, one finds that the MHV amplitude with four fermions is proportional to the one with two fermions. To summarize, MHV amplitudes satisfy
\be
\label{WI2phi2psi}
\begin{split}
\mathcal{A}_n(1_\psi^+\ 2_\psi^-\ 3_\psi^+\ 4_\psi^- \ 5_{{Z}} \ldots  n_{{Z}})
&=\frac{\sq{13}}{\sq{14}}\ \mathcal{A}_n(1_\psi^+\ 2_\psi^-\ 3_Z\ 4_{\bar{Z}} \ 5_{{Z}} \ldots  n_{{Z}})
\\
&=\frac{\sq{13}}{\sq{24}}\ \mathcal{A}_n(1_Z\ 2_{\bar{Z}}\ 3_Z\ 4_{\bar{Z}} \ 5_{{Z}} \ldots  n_{{Z}})\,.
\end{split}
\ee
The second-simplest class of supersymmetric Ward identities relate amplitudes in the NMHV class. 
In this paper, the 6-particle amplitudes play a central role, so we write down the 6-point NMHV supersymmetry Ward identities explicitly:
\begin{align}
\begin{split}
\label{WI6-1}
|1]\mathcal{A}_6(1_Z\ 2_{\bar{Z}}\ 3_Z\ 4_{\bar{Z}}\ 5_Z\ 6_{\bar{Z}})&-|2]\mathcal{A}_6(1_\psi^+\ 2_\psi^-\ 3_Z\ 4_{\bar{Z}}\ 5_Z\ 6_{\bar{Z}})\\
&
\hspace{-3cm}
-|4]\mathcal{A}_6(1_\psi^+\ 2_{\bar{Z}}\ 3_Z\ 4_\psi^-\ 5_Z\ 6_{\bar{Z}})-|6]\mathcal{A}_6(1_\psi^+\ 2_{\bar{Z}}\ 3_Z\ 4_{\bar{Z}}\ 5_Z\ 6_\psi^-)=0\,,
\end{split}
\\[2.5mm]
\begin{split}
\label{WI6-2}
|1]\mathcal{A}_6(1_Z\ 2_\psi^-\ 3_\psi^+\ 4_{\bar{Z}}\ 5_Z\ 6_{\bar{Z}})&+|3]\mathcal{A}_6(1_\psi^+\ 2_\psi^-\ 3_Z\ 4_{\bar{Z}}\ 5_Z\ 6_{\bar{Z}})\\
&
\hspace{-3cm}
-|4]\mathcal{A}_6(1_\psi^+\ 2_\psi^-\ 3_\psi^+\ 4_\psi^-\ 5_Z\ 6_{\bar{Z}})-|6]\mathcal{A}_6(1_\psi^+\ 2_\psi^-\ 3_\psi^+\ 4_{\bar{Z}}\ 5_Z\ 6_\psi^-)=0\,,
\end{split}
\\[2.5mm]
\begin{split}
\label{WI6-3}
|1]\mathcal{A}_6(1_Z\ 2_\psi^-\ 3_\psi^+\ 4_\psi^-\ 5_\psi^+\ 6_{\bar{Z}})&+|3]\mathcal{A}_6(1_\psi^+\ 2_\psi^-\ 3_Z\ 4_\psi^-\ 5_\psi^+\ 6_{\bar{Z}})\\
&
\hspace{-3cm}
+|5]\mathcal{A}_6(1_\psi^+\ 2_\psi^-\ 3_\psi^+\ 4_\psi^-\ 5_Z\ 6_{\bar{Z}})-|6]\mathcal{A}_6(1_\psi^+\ 2_\psi^-\ 3_\psi^+\ 4_\psi^-\ 5_\psi^+\ 6_\psi^-)=0\,.
\end{split}
\end{align}
We now turn to applications of these results.

\section{Supersymmetric Non-linear Sigma Model}
\label{sec:SUSY_NLSM}

Perhaps the simplest and most familiar class of models that exhibit both linearly realized supersymmetry and interesting low-energy theorems are the \textit{supersymmetric non-linear sigma models}. Of particular interest are the \textit{coset} sigma models for which the target manifold is a homogeneous space $G/H$. At lowest order, the coset sigma model captures the universal low-energy behavior of the scalar Goldstone modes of a spontaneous symmetry breaking pattern $G\rightarrow H$, where $G$ and $H$ are the \textit{isometry} and \textit{isotropy} groups of the target manifold respectively. If the target manifold is additionally a \textit{symmetric} space and there are no 3-point interactions, then the off-shell Ward-Takahashi identities for the spontaneously broken currents imply $\sigma=1$ vanishing low-energy theorems for the Goldstone scalars. An interesting recent perspective on coset sigma models can be found in \cite{Low:2018acv}.

At leading order it is fairly straightforward to calculate the on-shell scattering amplitudes for such a model from the (two-derivative) non-linear sigma model effective action. Using the methods of on-shell recursion, the use of an effective action is unnecessary. Instead, we may assume low-energy theorems and on-shell Ward identities of the isotropy group $H$ as the on-shell data that defines the model. Using the procedure of the soft bootstrap described in Section \ref{consistency}, we may apply subtracted recursion to construct the contributions to the S-matrix at leading order.

 A particularly simple and well-studied example of such a construction has previously been given for the $\frac{U(N)\times U(N)}{U(N)}$ coset sigma model \cite{Cheung:2015ota,Kampf:2013vha}. There are several nice features of this model which make it an appealing toy-model to study on-shell. As will be discussed in Section \ref{s:doublecopy}, at leading order ($\tilde{\Delta} =1$ or equivalently two-derivative)  the isotropy $U(N)$ symmetry allows for the construction of \textit{flavor}-ordered partial amplitudes with only $(n-3)!$ independent amplitudes for the scattering of $n$ Goldstone scalars.  

The situation is somewhat less straightforward for models describing the low-energy dyna\-mics of the Goldstone modes of internal symmetry breaking with some amount of linearly realized supersymmetry.\footnote{In this more general context \textit{internal} symmetry includes R-symmetry. For our purposes the relevant property is that the conserved charges are Lorentz scalars and so correspond to a spectrum of spin-0 Goldstone modes.} There are several interesting consequences of this combination of symmetries. The states must form mass degenerate multiplets of the supersymmetry algebra, which in this case means that the Goldstone scalars must always transform together with additional massless spinning states. As discussed in Section \ref{sec:softsusy}, the low-energy theorems of each of the particles in these \textit{Goldstone} multiplets are not independent.

It is well-known in the literature of supersymmetric field theories that to construct a supersymmetric action, the massless scalar modes must parametrize a target space manifold with \textit{K\"ahler} structure for $\mathcal{N}=1$ supersymmetry \cite{Zumino:1979et}. For $\mathcal{N}=2$ supersymmetry the target space manifold must have the structure
\begin{equation}
  \mathcal{M}_{\mathcal{N}=2} = \mathcal{M}_{\text{V}}\times \mathcal{M}_{\text{H}},
\end{equation}
where the scalars of the \textit{vector} multiplets parametrize the \textit{special-K\"ahler} manifold $\mathcal{M}_{\text{V}}$ while the scalars belonging to \textit{hyper} multiplets parametrize the \textit{hyper-K\"ahler} manifold $\mathcal{M}_{\text{H}}$ \cite{Freedman:2012zz}. As a consequence, despite the obvious virtues of a flavor ordered representation, this makes studying the supersymmetrization of the $\frac{U(N)\times U(N)}{U(N)}$ coset sigma model using subtracted recursion more difficult, since even in the $\mathcal{N}=1$ case the target manifold is not K\"ahler. This does not mean that the internal symmetry breaking pattern $U(N)\times U(N) \rightarrow U(N)$ is impossible in an $\mathcal{N}=1$ supersymmetric model. Rather it means that the target space contains $\frac{U(N)\times U(N)}{U(N)}$ as a non-K\"ahler submanifold and includes additional directions in field space or equivalently includes additional massless \textit{quasi-Goldstone} scalars \cite{Bando:1983ab}. In general there is no unique way to extend the symmetry breaking coset to a K\"ahler manifold, because in any given example the spectrum of quasi-Goldstone modes depends on the details of the UV physics. 
Correspondingly, the quasi-Goldstone scalars do not satisfy the kind of universal low-energy theorems necessary for us to construct the scattering amplitudes recursively. 

Instead, in this section we will study the interplay of low-energy theorems and supersymmetry by considering the simplest symmetric coset that is both K\"ahler and special-K\"ahler
\begin{equation}
  \frac{SU(2)}{U(1)} \cong \mathbb{CP}^1\,,
\end{equation}
and therefore should admit both an $\mathcal{N}=1$ and $\mathcal{N}=2$ supersymmetrization. Our assumption here is that the target manifold \textit{is} the coset manifold and therefore the massless spectrum should contain only two real scalar degrees of freedom, both Goldstone modes. They form a single complex scalar field $Z,\overline{Z}$ which carries a conserved charge associated with the isotropy $U(1)$. These properties uniquely determine the Goldstone multiplets as an $\mathcal{N}=1$ chiral and $\mathcal{N}=2$ vector multiplet respectively.

The main results of this section are (1) the demonstration that both the $\mathcal{N}=1$ and $\mathcal{N}=2$ $\mathbb{CP}^1$ non-linear sigma models are constructible on-shell using recursion without the need to explicitly construct an effective action. And (2) this construction gives a new \textit{on-shell} perspective on the relationship between the linearly realized target space isotropies of $\mathcal{M}_{\text{V}}$ and electric-magnetic duality transformations of the associated vector bosons. 

\subsection{$\mathcal{N}=1$ $\mathbb{CP}^1$ NLSM}

The $\mathcal{N}=1$ $\mathbb{CP}^1$ non-linear sigma model is defined by the following on-shell data:
\begin{itemize}  
\item A spectrum consisting of a massless $\mathcal{N}=1$ chiral multiplet $(Z, \bar Z,\psi^+,\psi^-)$.
\item Scattering amplitudes satisfy $\mathcal{N}=1$ supersymmetry Ward identities.
\item Scattering amplitudes satisfy isotropy $U(1)$ Ward identities under which $Z,\bar Z$ are charged.
\item $\sigma_Z=\sigma_{\bar Z} = 1$ soft weight for the scalars.
\end{itemize}
Using the approach of the soft bootstrap, we begin by constructing the most general on-shell amplitudes at lowest valence that are consistent with the above data and minimize $\tilde{\Delta}$. There are no possible 3-point amplitudes consistent with the assumptions and so we must begin at 4-point. A $|Z|^4$ interaction, corresponding to $\tilde{\Delta}=0$, is consistent with $U(1)$ conservation but violates the assumed low-energy theorem. The next-to-lowest reduced dimension interactions correspond to $\tilde{\Delta}=1$ and have a unique 4-point amplitude  consistent with the assumptions
\begin{equation}
  \label{NLSMA4}
  \mathcal{A}_4(1_Z\ 2_{\bar{Z}}\ 3_Z\ 4_{\bar{Z}})= \frac{1}{\Lambda^2} s_{13}.
\end{equation}
Note that at 4-point, the conservation of the $U(1)$-charge for the complex scalar is automatically enforced as a consequence of the supersymmetry Ward identitites. We will see that this implies the conservation of the $U(1)$ charge for amplitudes with arbitrary number of external particles corresponding to $\tilde{\Delta}=1$. Note that this is not automatic for higher order ($\tilde{\Delta}>1$) corrections and must be imposed as a separate constraint. Using (\ref{WI2phi2psi}) the remaining 4-point amplitudes are completely determined by supersymmetry; it is convenient to summarize the component amplitudes in a single \textit{superamplitude} 
\cite{Elvang:2011fx} 
\begin{equation} \label{supamp}
  \mathcal{A}_4(1_{\Phi^+}2_{\Phi^-}3_{\Phi^+}4_{\Phi^-}) = \frac{1}{\Lambda^2}\sq{13}\delta^{(2)}(\tilde{Q}) = \frac{1}{2\Lambda^2}\sq{13}\sum_{i,j=1}^4 \ang{ij} \eta_{i}\eta_{j}\,.
\end{equation}
Here we have introduced two chiral superfields $\Phi^+$ and $\Phi^-$ that contain the positive and negative helicity fields of the $\mathcal{N}=1$ chiral multiplet as
\be
  \Phi^+=\psi^+ +\eta Z\ \,,~~~~~~
  \Phi^-=\bar Z -\eta \,\psi^-.
\ee
$\eta$ is the Grassmann coordinate of $\mathcal{N}=1$ on-shell superspace and $\eta_i$ denotes the $\eta$-coordinate of the $i^{\text{th}}$ superfield. We can obtain all the component amplitudes by projecting out components of the superfield. For example, the all-fermion amplitude can be derived as follows
\begin{equation}
  \mathcal{A}_4(1^+_{\psi}2^-_{\psi}3^+_{\psi}4^-_{\psi})=\frac{\partial}{\partial\eta_2}\frac{\partial}{\partial\eta_4} \mathcal{A}_4(1_{\Phi^+}2_{\Phi^-}3_{\Phi^+}4_{\Phi^-}) =-\frac{1}{\Lambda^2}\sq{13}\ang{24}.
\end{equation} 
It is useful to note that the expression \reef{supamp} is manifestly local. It follows that all component amplitudes are free of factorization singularities, indicating the absence of 3-point interactions in this theory. Note also that the pure fermion sector is exactly the NJL model detected by the soft bootstrap in Section \ref{s:fermionEFTs}.

Next, we use these 4-point amplitudes to recursively construct $n$-point amplitudes. Following the discussion in Section \ref{softsusy}, we note that the soft weight of the fermion must be either $\sigma_\psi=0$ or $\sigma_\psi=1$. Making the conservative choice $\sigma_\psi=0$, we evaluate the constructibility criterion on the above on-shell data,
\begin{equation}
4<2n_s +n_f,
\end{equation}
where $n_f$ is the number of external fermion states of the $n$-point amplitude and $n_s=n-n_f$ is the number of external scalar states. For $n>4$, this condition is satisfied for all $n$-point amplitudes. We find that recursively constructing the 6-point amplitudes yields an $a_i$-independent expression. All the 6-point amplitudes can be found in Appendix \ref{a:NLSMamp}. Since our input 4-point amplitudes are MHV, the only non-zero constructible amplitudes at 6-point are NMHV and can be verified to satisfy the NMHV 6-point Ward identities \eqref{WI6-1}, \eqref{WI6-2}, \eqref{WI6-3}.

If however we make the stronger assumption $\sigma_\psi=1$, the recursively constructed 6-point amplitude is $a_i$-dependent and therefore fails the consistency checks. As a result we conclude that the true soft weight of the fermion of our theory is $\sigma_\psi=0$ and this is sufficient to construct the S-matrix at leading order from the 4-point seed amplitudes (\ref{supamp}).

The recursive constructibility of the S-matrix has non-trivial consequences for the possible conserved additive quantum numbers. In a recursive model the only non-zero amplitudes are those which can be constructed by gluing together lower-point on-shell amplitudes
\begin{center}
  {\begin{tikzpicture}[line width=1 pt, scale=0.6]
      \draw[-, line width=0.75 pt] (-10.3,1.7)--(-9,0);
      \draw[-, line width=0.75 pt] (-10.3,-1.7)--(-9,0);
      \draw[-, line width=0.75 pt] (-10.7,1.3)--(-9,0);
      \draw[-, line width=0.75 pt] (-10.7,-1.3)--(-9,0);
      \node at (-10.7,0) {$\vdots$};
      \node at (-7.3,0) {$\vdots$};
      \draw[-, line width=0.75 pt] (-9,0)--(-7.7,1.7);
      \draw[-, line width=0.75 pt] (-9,0)--(-7.7,-1.7);
      \draw[-, line width=0.75 pt] (-9,0)--(-7.3,1.3);
      \draw[-, line width=0.75 pt] (-9,0)--(-7.3,-1.3);
      \draw[black,fill=lightgray] (-9,0) circle (4 ex);
      \node at (-9,0) {$\mathcal{A}_n$};
      \node at (-5.6,0) {\Large$\cong$};
      \node at (-3.5,0) {\Huge $\sum$}; 
      \node at (-3.6,-1.25) {$I, X$};
      \draw[-, line width=0.75 pt] (-1.3,1.7)--(0,0);
      \draw[-, line width=0.75 pt] (-1.3,-1.7)--(0,0);
      \draw[-, line width=0.75 pt] (-1.7,1.3)--(0,0);
      \draw[-, line width=0.75 pt] (-1.7,-1.3)--(0,0);
      \draw[-, line width=0.75 pt] (0,0)--(5,0);  
      \node at (2.5,-0.6) {$P_I$};
      \draw[black,fill=lightgray] (0,0) circle (4 ex);
      \node at (0,0) {$\mathcal{A}_{L}$};
      \node at (-1.7,0) {$\vdots$};
      \draw[-, line width=0.75 pt] (5,0)--(6.3,1.7);
      \draw[-, line width=0.75 pt] (5,0)--(6.3,-1.7);
      \draw[-, line width=0.75 pt] (5,0)--(6.7,1.3);
      \draw[-, line width=0.75 pt] (5,0)--(6.7,-1.3);
      \draw[black,fill=lightgray] (5,0) circle (4 ex);
      \node at (5,0) {$\mathcal{A}_{R}$};
      \node at (6.7,0) {$\vdots$};
      \node at (1.4,0.63) {$X$};          
      \node at (3.6,0.7) {$\overline{X}$};
    \end{tikzpicture}}
\end{center}
where the states $X,\bar X$ on either side of the factorization channel $I$ have CP conjugate quantum numbers. As discussed further in Appendix \ref{RecWardApp}, if an additive quantum number is conserved by all seed amplitudes then it must be conserved by all recursively constructible amplitudes. 

For example, in the present context the seed amplitudes conserve two \textit{independent} $U(1)$ charges:
\begin{center}
\begin{tabular}{|c|c|c|}
  \hline
  & $U(1)_A$& $U(1)_B$\\
  \hline
  \hline
  $Z$ & $q_A$ & 0\\
  $\bar Z$  & $-q_A$& 0\\
  $\psi^+$ & 0 & $q_B$\\
  $\psi^-$ & 0 & $-q_B$\\
  \hline
  $\eta$ & $-q_A$ & $q_B$\\
  $\Phi^+$ & 0 & $q_B$\\
  $\Phi^-$ & $-q_A$ & $0$ \\
  \hline
\end{tabular}
\end{center}
We know to expect the existence of an isotropy $U(1)$ under which the scalars are charged, but from our on-shell construction it is unclear whether this should be $U(1)_A$ or a combination of $U(1)_A$ and $U(1)_B$. We have presented the charges as two independent R-symmetries but more correctly we should consider them as a single \textit{global} $U(1)$ and a $U(1)_R$. The presence of a second conserved quantum number is not part of the definition of the $\mathbb{CP}^1$ non-linear sigma model but is instead an \textit{emergent} or \textit{accidental} symmetry at lowest order in the EFT. In general one would expect $U(1)_A\times U(1)_B$ to be explicitly broken to the isotropy $U(1)$ by higher dimension operators.

\subsection{$\mathcal{N}=2$ $\mathbb{CP}^1$ NLSM} 
\label{s:N2NLSM}
The $\mathcal{N}=2$ $\mathbb{CP}^1$ NLSM is defined by the following on-shell data:
\begin{itemize}  
\item A spectrum consisting of a massless $\mathcal{N}=2$ vector multiplet $(Z, \bar Z,\psi^{a+},\psi^-_a, \gamma^+, \gamma^-)$, where $a=1,2$.
\item Scattering amplitudes satisfy $\mathcal{N}=2$ supersymmetry Ward identities.
\item Scattering amplitudes satisfy isotropy $U(1)$ Ward identities under which $Z,\bar Z$ are charged.
\end{itemize}  
Note that, importantly, we do {\em not} impose the the soft weight of the scalars $\sigma_Z =\sigma_{\bar{Z}} =1$. As we will explain further below, no model with the above properties \textit{and} vanishing scalar soft limits exists. 

To proceed, interactions with reduced dimension $\tilde{\Delta}=0$ (such as Yukawa interactions) are incompatible with $\mathcal{N}=2$ supersymmetry for a single vector multiplet. Thus, the minimal value is $\tilde{\Delta}=1$; that is of course also the value for the  $\mathcal{N}=1$ model. It is curious to note that $\mathcal{N}=2$ supersymmetry is sufficient to uniquely construct the S-matrix at this order in $\tilde{\Delta}$. As we show in the following, \textit{without} assuming vanishing scalar soft limits, the restriction of the external states to a single chiral multiplet $(Z,\bar{Z},\psi^{1+},\psi_1^-)$ reproduces the $\mathcal{N}=1$ $\mathds{C}\mathds{P}^1$ sigma model. 

As in the previous section, for $\tilde{\Delta}=1$ the 4-point scalar amplitude takes the form \eqref{NLSMA4}. All 4-point component amplitudes are uniquely fixed by the 4-scalar amplitudes by the $\mathcal{N}=2$ supersymmetry Ward identities and they can be encoded compactly into superamplitudes using two chiral superfields \cite{Elvang:2011fx}
\be
   \begin{split}
	\Phi^+&=\gamma^+ +\eta_1 \psi^{1+} +\eta_2\psi^{2+}-\eta_1\eta_2 Z,\\
	\Phi^-&=\bar{Z} +\eta_1 \psi^-_2 -\eta_2\psi^-_1-\eta_1\eta_2 \gamma^-\,.
   \end{split}
\ee
Here $\eta_1$ and $\eta_2$ are the Grassmann coordinates of $\mathcal{N}=2$ on-shell superspace. The $R$-indices on $\psi^a$ are raised and lowered using $\epsilon_{ab}$, so $\psi^-_2=\epsilon_{21}\psi^{1-}=\psi^{1-}$ and $\psi^-_1=\epsilon_{12}\psi^{2-}=-\psi^{2-}$. In terms of 
the superfields, the 4-point superamplitude can be expressed as
\be
  \label{supamp2}
  \mathcal{A}_4(1_{\Phi^+}2_{\Phi^-}3_{\Phi^+}4_{\Phi^-})=\frac{1}{\Lambda^2}\frac{\sq{13}}{\ang{13}} \delta^{(4)}(\tilde{Q})=\frac{1}{4\Lambda^2}\frac{\sq{13}}{\ang{13}} \prod_{a=1}^2 \sum_{i,j=1}^4 \ang{ij} \eta_{ia}\eta_{ja}.
\ee
We use $\eta_{ia}$ to denote the $a^\text{th}$ Grassmann coordinate of the $i^{\text{th}}$ external superfield. In contrast to \eqref{supamp}, the superamplitude \eqref{supamp2} generates component amplitudes that are not local due  to the factorization singularity at $P_{13}^2\to 0$. 
For example, consider the following component amplitude
\begin{equation}
\label{NLSMA4_ggff}
\mathcal{A}_4(1^+_{\gamma} 2^-_{\gamma} 3^+_{\psi^1} 4^-_{\psi_1})=-\frac{\partial}{\partial\eta_{21}}\frac{\partial}{\partial\eta_{22}}\frac{\partial}{\partial\eta_{31}}\frac{\partial}{\partial\eta_{42}}\mathcal{A}_4(1_{\Phi^+}2_{\Phi^-}3_{\Phi^+}4_{\Phi^-})=-\frac{1}{\Lambda^2}\frac{\sq{13}\sq{14}\ang{24}}{\sq{24}}.
\end{equation}
Locality and unitarity imply that this 4-point amplitude must factorize into 3-point amplitudes on the singularity at $P_{13}^2\to 0$. Denoting the helicity of the exchanged particle $h$,
 the amplitude factorizes as
\begin{center}	
	\begin{tikzpicture}
	\draw (8.3,0.8)--(9,0);
	\node at (8,0.8) {$1^+_{\gamma}$};
	\draw (8.3,-0.8)--(9,0);
	\node at (8,-0.8) {$3^+_{\psi^1}$};
	\node at(9.4,0.3) {$P^h_{13}$};
	\draw (9,0)--(10.25,0);
	\draw (10.75,0) -- (12,0);
	\node at (11.2,0.3) {$-P^{-h}_{13}$};
	\draw (12,0)--(12.7,0.8);
	\node at (13.1,0.8) {$ 2^-_{\gamma}$};
	\draw (12,0)--(12.7,-0.8);
	\node at (13.1,-0.8) {$4^-_{\psi_1}$};
	\end{tikzpicture}
      \end{center}
The contribution to the residue on the singularity takes the form 
 \begin{align}
	&P_{13}^2\mathcal{A}_4(1^+_{\gamma} 2^-_{\gamma} 3^+_{\psi^1} 4^-_{\psi_1})\biggr\vert_{P_{13}^2=0}
	\,=\,\mathcal{A}_3\big(1^+_{\gamma} 3^+_{\psi^1}(P_{13})_h\big)
	\mathcal{A}_3\big((-P_{13})_{-h}2^-_{\gamma} 4^-_{\psi_1}\big)\nonumber\\
        &\hspace{5mm}= \left(\frac{g_1}{\Lambda}[13]^{3/2-h}[1P_{13}]^{1/2+h}[3P_{13}]^{-1/2+h}\right)\left(\frac{g_2}{\Lambda}\langle 24\rangle^{3/2-h}\langle 2P_{13}\rangle^{1/2+h}\langle 4P_{13}\rangle^{-1/2+h}\right)\nonumber\\
        &\hspace{5mm}=\frac{g_1g_2}{\Lambda^2}(-1)^{2h}\sq{13}^{3/2-h}\ang{24}^{3/2+h}\sq{23}^{1/2-h}\sq{14}^{1/2+h} \,,
\end{align}
with the 3-point amplitudes completely determined by Poincar\'e invariance and little group scaling. Comparing with the explicit form of the residue calculated from (\ref{NLSMA4_ggff})
\begin{equation}
  P_{13}^2\mathcal{A}_4(1^+_{\gamma} 2^-_{\gamma} 3^+_{\psi^1} 4^-_{\psi_1})\biggr\vert_{P_{13}^2=0}=\frac{1}{\Lambda^2} \sq{13}\sq{14}\ang{24}^2,
\end{equation}
we find that $h=1/2$ and $g_1 g_2=-1$. The exchanged particle of helicity $h=1/2$ can be either $\psi^{1+}$ or $\psi^{2+}$. The locality of the $\mathcal{A}_4(1^+_{\psi^1} 2^-_{\psi_1} 3^+_{\psi^1} 4^-_{\psi_1})$ and $\mathcal{A}_4(1^+_{\psi^2} 2^-_{\psi_2} 3^+_{\psi^2} 4^-_{\psi_2})$ tells us that they do not factorize on the $(P_{13})^2\to 0$ pole. We conclude that 
$\mathcal{A}_3(1^+_\gamma 2^+_{\psi_1} 3^+_{\psi_1})=\mathcal{A}_3(1^+_\gamma 2^+_{\psi_2} 3^+_{\psi_2})=0$,  while 
\be \label{gpp}
  \mathcal{A}_3(1^+_\gamma 2^+_{\psi_1} 3^+_{\psi_2})=\frac{g_1}{\Lambda}\sq{12}\sq{13}\,,~~~~~
  \mathcal{A}_3(1^-_\gamma 2^-_{\psi_1} 3^-_{\psi_2})=\frac{g_2}{\Lambda}\ang{12}\ang{13}\,.
\ee
We carry out a similar exercise with $\mathcal{A}_4(1^+_{\gamma} 2^-_{\gamma} 3^+_\gamma 4^-_\gamma)$ for a particle of helicity $h$ in the $P_{13}^2\to 0$ factorization channel. Comparing with the 4-point amplitude \eqref{supamp2} fixes $h=0$. 
This could correspond to either $Z$ or $\bar{Z}$ exchange. The absence 
of a $P_{14}^2\to 0$ pole in $\mathcal{A}_4(1^+_{\gamma} 2^-_{\gamma} 3_Z 4_{\bar{Z}})$ shows that $\mathcal{A}_3(1^+_\gamma 2^+_\gamma 3_{\bar{Z}})=0$ and
\be \label{ggz}
 \mathcal{A}_3(1^+_\gamma 2^+_\gamma 3_Z)=\frac{g_3}{\Lambda}\sq{12}^2\,, ~~~~~
  \mathcal{A}_3(1^-_\gamma 2^-_\gamma 3_{\bar{Z}})=\frac{g_4}{\Lambda}\ang{12}^2\,,
\ee
where $g_3 g_4=1$. Demanding that all non-local 4-point amplitudes 
factorize correctly fixes $-g_1=g_2=g_3=g_4=-1$. The 3-point superamplitudes are
\bea
	\mathcal{A}_3(1_{\Phi^-}2_{\Phi^-}3_{\Phi^-})\!\!&=&\!\!\delta^{(4)}(\tilde{Q})=\frac{1}{4\Lambda}\prod_{a=1}^2\ \sum_{i,j=1}^3 \ang{ij}\eta_{ia}\eta_{ja}\,,\\
	\nonumber
	\mathcal{A}_3(1_{\Phi^+}2_{\Phi^+}3_{\Phi^+})\!\!&=&\!\!\frac{1}{\Lambda}\delta^{(2)}(\eta_1 \sq{23}+\eta_2\sq{31}+\eta_3\sq{12}) = \frac{1}{\Lambda}\prod_{a=1}^2(\eta_{1a} \sq{23}+\eta_{2a}\sq{31}+\eta_{3a}\sq{12})\,,
\eea
where $\prod_{a=1}^2 f_a$ is defined as $f_1 f_2$.

 It is interesting to observe that even though the $\mathcal{N}=0,1$ and $2$ $\mathbb{CP}^1$ NLSM have the pure scalar 4-point amplitude in common, in the latter case the extended supersymmetry together with locality require the presence 3-point interactions. 

 We are now in a position to address the constructibility of general $n$-point amplitudes. Since we are not assuming vanishing soft limits as part of our on-shell data, we are not able to make use of subtracted recursion. This is only problematic for a subset of the amplitudes in this model, at least at leading order. The unsubtracted constructibility criterion for this model reads
 \begin{equation} \label{concrit}
   4< n_f + 2n_v,
 \end{equation}
where $n_f$ and $n_v$ are the number of fermions and vector bosons respectively. It turns out that the amplitudes that do not satisfy this criterion can be determined from the $\mathcal{N}=2$ supersymmetry Ward identities in terms of those that do; explicit formulae are given in Appendix \ref{WIN2}. Remarkably, without making any strong assumptions about the structure of low-energy theorems for the scalars, which usually characterize the sigma model coset structure, the $\mathcal{N}=2$ supersymmetry is sufficient at leading order to both construct the entire S-matrix and reproduce the amplitudes of the $\mathcal{N}=1$ and $\mathcal{N}=0$ models as special cases. 

This same statement can be made in the perhaps more familiar language of local field theory. At this order in the EFT expansion, the S-matrix elements should be calculable from some effective action, the bosonic sector of which should be described by a two-derivative Lagrangian of the general form
\begin{equation}
  \mathcal{L}_{\text{eff}} = P\left(|Z|^2\right)|\partial_\mu Z|^2 + Q\left(|Z|^2\right) 
  Z\, F_+^2 + \text{h.c.}
\end{equation}
where $P(|Z|^2)$ and $Q(|Z|^2)$ are some functions analytic around $Z\sim 0$. Insisting that the S-matrix elements satisfy the on-shell $\mathcal{N}=2$ supersymmetry Ward identities is equivalent to requiring the existence of off-shell $\mathcal{N}=2$ supersymmetry transformations under which the effective action is invariant. The on-shell uniqueness result is equivalent to the statement that the off-shell $\mathcal{N}=2$ supersymmetry uniquely (up to field redefinitions) determines the form of the two-derivative effective action. In particular, the function $P(|Z|^2)$ is uniquely determined to be
\begin{equation}
  P\left(|Z|^2\right) = \left(\frac{1}{1+|Z|^2}\right)^2,
\end{equation}
corresponding to the Fubini-Study metric on $\mathds{C}\mathds{P}^1$. 

Since the entire S-matrix is determined, we can explicitly demonstrate how the presence of the vector bosons modifies the structure of the low-energy theorems from the naive vanishing soft limits suggested by the coset structure. Consider the following relation among 5-point amplitudes given by the $\mathcal{N}=2$ supersymmetry Ward identities
\begin{equation}
  \mathcal{A}_5\left(1_\gamma^+,2_\gamma^+,3_Z,4_{Z},5_{\bar Z}\right) = \frac{\langle 34\rangle^2 }{\langle 45 \rangle^2}\mathcal{A}_5\left(1_\gamma^+,2_\gamma^+,3_\gamma^+,4_{Z},5_\gamma^-\right).
\end{equation}

The amplitude on the right-hand-side satisfies \eqref{concrit} and therefore is constructible using unsubtracted recursion. This gives the non-constructible amplitude on the left-hand-side as
\begin{equation}
  \mathcal{A}_5\left(1_\gamma^+,2_\gamma^+,3_Z,4_{Z},5_{\bar Z}\right) = \frac{1}{\Lambda^3}\langle 34 \rangle^2\left(\frac{[12][34]}{\langle 12 \rangle \langle 34\rangle}+\frac{[23][14]}{\langle 23 \rangle \langle 14\rangle}+\frac{[31][24]}{\langle 31 \rangle \langle 24\rangle}\right).
\end{equation}
The soft limits on particles 1, 2, 3 and 4 vanish, as expected. 
The soft limit on particle 5, however, is $\mathcal{O}(1)$, contrary to the expected soft behavior for a Goldstone mode of a symmetric coset. Explicitly 
\begin{equation}
  \mathcal{A}_5\left(1_\gamma^+,2_\gamma^+,3_Z,4_{Z},5_{\bar Z}\right) \xrightarrow{|5]\rightarrow \epsilon|5]} \frac{1}{\Lambda^3}[12]^2 +\mathcal{O}(\epsilon).
\end{equation}
It is interesting that the coupling to the photons, required by $\mathcal{N}=2$ supersymmetry, results in non-vanishing soft scalar limits for a theory with a symmetric coset. In principle, this amplitude could have had a contact contribution of the form $\propto [12]^2$, but our calculation shows that such a term would be incompatible with $\mathcal{N}=2$ supersymmetry.

 The maximal $R$-symmetry group that this model can realize is $U(2)_R=U(1)_R\times SU(2)_R$. We will now verify that the $SU(2)_R$ symmetry Ward identities hold for the seed amplitudes, the $U(1)_R$ we will address separately. To do this we choose a basis for the generators of $SU(2)_R$. The scalars and vectors both transform as $SU(2)$ singlets. The positive helicity fermion species $\psi^{1,2+}$ will transform in the fundamental representation under
\be
\label{eq:Tgenerators}
	\mathcal{T}_0=\begin{pmatrix}
	1 & 0\\
	0 & -1
	\end{pmatrix},~~~~
	 \mathcal{T}_+ =\begin{pmatrix}
	0 & 1\\
	0 & 0
	\end{pmatrix},~~~~
	 \mathcal{T}_- =\begin{pmatrix}
	0 & 0\\
	1 & 0
	\end{pmatrix}.
\ee
The negative helicity fermions transform in the anti-fundamental with 
$\bar{\mathcal{T}}_i=-\mathcal{T}_i^\dagger$. This tells us that the $\mathcal{T}_0$-Ward identity is satisfied as long as the fermion species appear in pairs of (a) different helicity, same species or (b) same helicity, different species. This is true of all the non-zero amplitudes in this model. The action of $\mathcal{T}_+$ and $\mathcal{T}_-$ are
\begin{equation}
  \label{Tplusaction}
  \begin{array}{|c||c|c||c|c||c|c|}
  \hline
  \text{state}~i & \mathcal{T}_+ \cdot i & \mathcal{A}_n~\text{prefactor} & \mathcal{T}_- \cdot i & \mathcal{A}_n~\text{prefactor} & \mathcal{T}_0 \cdot i & \mathcal{A}_n~\text{prefactor}\\
  \hline
  \psi^{1+} & 0 & 0& \psi^{2+} & 1 & \psi^{1+} & 1\\
  \psi^{2+} & \psi^{1+}   &  1 & 0 &0 & \psi^{2+} & -1\\
  \psi^{-}_1 & \psi^{-}_2 & \!\!\!\! -1 & 0 & 0 & \psi^{-}_1 & -1\\
  \psi^{-}_2 &0 & 0 &  \psi^{-}_1 & \!\!\!\! -1 & \psi^{-}_2 & 1\\
  \hline
  \end{array}
\end{equation}

We find that all 3-point and 4-point amplitudes in this model satisfy the $SU(2)_R$ Ward identities, for example
\be
  \begin{split}
	\mathcal{T}_-\cdot \mathcal{A}_4(1_{\psi^1}^+ 2_{\psi_2}^- 3^+_{\psi^1} 4^-_{\psi_1})&=\mathcal{A}_4(1_{\psi^2}^+ 2_{\psi_2}^- 3^+_{\psi^1} 4^-_{\psi_1})-\mathcal{A}_4(1_{\psi^1}^+ 2_{\psi_1}^- 3^+_{\psi^1} 4^-_{\psi_1})
	+\mathcal{A}_4(1_{\psi^1}^+ 2_{\psi_2}^- 3^+_{\psi^2} 4^-_{\psi_1})\\
	&=-\frac{\sq{13}}{\sq{24}}\left(s+t+u\right)=0\,.
  \end{split}
\ee
As discussed above, we conclude that at leading order the $SU(2)_R$ Ward 
identities are satisfied by all amplitudes in the $\mathcal{N}=2$ model. 

Following the same approach as described for the $\mathcal{N}=1$ model, conservation laws satisfied by the seed amplitudes imply that the same quantities are conserved by all leading-order amplitudes if they are recursively constructible (see  Appendix \ref{RecWardApp}). 
This result extends to non-Abelian symmetries, which in the on-shell language correspond to Ward identities for non-diagonal generators; this is shown for $SU(2)$ in Appendix \ref{RecWardApp}. 
The amplitudes that are not constructible using recursion are fixed by supersymmetry in terms of those that are. Therefore, they will also respect the conservation laws and non-Abelian symmetries of the seed amplitudes. 

This model also conserves a separate $U(1)_R$ charge. We know to expect the conservation of the charge associated with the $U(1)$ isotropy group. In the $\mathcal{N}=1$ case we found that the scattering amplitudes conserve an R-charge $U(1)_A$ assigned only to the complex scalar but it was consistent with the existence of $U(1)_B$ that the isotropy $U(1)$ might also assign a charge to the fermion or even to assign equal charges in the form of a global symmetry. In the present context we also have two independent $U(1)$ symmetries. The first is the $U(1)\subset SU(2)_R$ which assigns opposite charges to the fermions $\psi^{1+}$ and $\psi^{2+}$. The second assigns charges to each of the states which, up to overall normalization can be deduced from the 3- and 4-point seed amplitudes and are summarized in the following table:

\begin{center}
\begin{tabular}{|c|c|c|}
  \hline
  & $U(1)_R$& $SU(2)_R$\\
  \hline
  \hline
  $Z$ & $-4$ & $\textbf{1}$ \\
  $\bar Z$  & $4$&  $\textbf{1}$\\
  $\psi^{a+}$ & $-1$ & $\textbf{2}$\\
  $\psi_a^{-}$ & $1$ & $\textbf{2}$\\
  $\gamma^{+}$ & 2 & $\textbf{1}$\\
  $\gamma^{-}$ & --2 & $\textbf{1}$\\
  \hline
  $\eta_a$ & $3$ & $\textbf{2}$\\
  $\Phi^+$ & $2$ & $\textbf{1}$ \\
  $\Phi^-$ & 4 & $\textbf{1}$ \\
  \hline
\end{tabular}
\end{center}

These are the \textit{only} linear symmetries compatible with the seed amplitudes. The isotropy $U(1)$ must therefore be identified with some linear combination of $U(1)_R$ and $U(1)\subset SU(2)_R$. This is perhaps surprising, it tells us that the massless vector boson must also be charged under the isotropy $U(1)$. Just as for the fermions, the vector charges are \textit{chiral} meaning that the positive and negative helicity states have opposite charges. Such charges for vectors are associated with electric-magnetic duality symmetries. 

Such an extra $U(1)_R$ symmetry is possible because the maximal outer-automorphism group of the $\mathcal{N}=2$ supersymmetry algebra is $U(2)_R$. The assignment of the associated charges is, up to normalization, fixed by the charge of the highest helicity state in the multiplet. It
 is interesting to observe that in the present context, knowledge of the non-vanishing 4-point amplitudes is \textit{insufficient} to determine the $U(1)_R$ charge assignments. It is only from considering the 3-point amplitudes that we find the assignment of a non-zero chiral charge for the vector bosons unavoidable. Consider for example the amplitudes (\ref{ggz}). Since the scalar is required to be charged under the isotropy $U(1)$, which in this case must be the $U(1)_R$ since there are no other symmetries under which the scalar is charged, we see that the vector must also be charged and satisfy $2q[\gamma^+] = -q[Z]$. The existence of fundamental 3-point interactions in this model was deduced by demanding that the singularities of the 4-point amplitudes be identified with physical factorization channels. From an on-shell point of view, it is therefore an {\em unavoidable} consequence of locality, unitarity and supersymmetry that the $\mathcal{M}_V$ isotropy group of an $\mathcal{N}=2$ non-linear sigma model acts on the vector bosons as an electric-magnetic duality transformation. 

The necessary existence of the fundamental 3-point amplitudes (\ref{gpp}) and (\ref{ggz}) has a further interesting consequence for the low-energy behavior of the vector boson. In \cite{Elvang:2016qvq} it was shown that \textit{singular} low-energy theorems arise from the presence of certain 3-point amplitudes. In the notation used in \cite{Elvang:2016qvq} the 3-point amplitudes (\ref{gpp}) and (\ref{ggz}) are classified as $\mathtt{a}=1$ in the soft limit of a positive helicity vector boson. Therefore a vector boson present in amplitudes which contain at least one of the following other particles: $Z$, $\psi^{a+}$ or $\gamma^+$ has soft weight $\sigma_\gamma = -1$. Using the general formalism developed in \cite{Elvang:2016qvq},  we can write down the low-energy theorem of the vector bosons in this subclass of amplitudes
\begin{equation}
\label{softphoton}
  \mathcal{A}_{n+1}\left(s_\gamma^+,1,2,...,n\right) ~
  \xrightarrow{p_s \to \eps p_s~\text{as}~\eps \to 0} ~
  \sum_{k=1}^{n} \frac{[sk]}{\epsilon\langle sk\rangle} \mathcal{A}_{n}\left(1,2,...,\mathcal{F}_+ \cdot  k,...,n\right) +\mathcal{O}\left(\epsilon^0\right).
\end{equation}
Here we are using a notation similar to \cite{Laddha:2017vfh} with the introduction of an operator $\mathcal{F}_+$ which acts on the one-particle states as
\begin{equation}
  \label{softphoton}
  \begin{array}{|c||c|c|}
  \hline
  \text{state}~i & \mathcal{F}_+ \cdot i & \mathcal{A}_n~\text{prefactor} \\
  \hline
   Z & \gamma^- &  \phantom{-}1  \\
  \psi^{1+}  & \psi^{-}_2 & -1  \\
  \psi^{2+} & \psi^{-}_1 & -1\\
  \gamma^+ & \overline{Z} & -1  \\
  \hline
  \end{array}
\end{equation}
and annihilates the states of the negative helicity multiplet. A similar operator $\mathcal{F}_-$ can be defined for the soft limit of a negative helicity vector. Using equation \reef{bonusWID} in conjunction with the soft behavior \reef{softphoton} of the $n+1$-point amplitude results in the following identity for the residual $n$-point amplitudes 
\begin{equation}
\label{newID}
  \sum_{i=1}^n\sum_{j=1}^n (-1)^{L_i+P_i}\frac{[Xi][Yj]}{\langle Yj\rangle} \mathcal{A}_n\left(1,2,...,\mathcal{Q}_{1}\cdot i,...,\mathcal{F}_+\cdot j,...,n\right)=0\,,
\end{equation}
where here $P_i = 0$ or $1$ corresponds to the additional signs associated with the prefactors of both the supersymmetry Ward identities and the operator $\mathcal{F}_+$ given in Table \ref{softphoton}. Note that the action of $\mathcal{Q}_{1}$ and $\mathcal{F}_+$ commute on all physical states, so there is no ambiguity when $i=j$ in the sums. Moreover, rearranging the order of the sums, it becomes clear that for each fixed $j$, the sum over $i$ expresses 
a supersymmetry Ward identity for the $n$-point amplitudes. As such, the identity \reef{newID} does not impose further constraints beyond supersymmetry.

\section{Super Dirac-Born-Infeld and Super Born-Infeld}
\label{sec:sDBI}
In the soft bootstrap analysis of Section \ref{s:softboot}, we encountered three theories with a fundamental quartic interaction whose couplings are of mass-dimension $-4$: 
DBI, Akulov-Volkov, and Born-Infeld. These EFTs can all be related by supersymmetry. We will discuss them in further detail in future work, so for now we simply note the following:
\begin{itemize}
\item The $\mathcal{N}=1$ supersymmetric Dirac-Born-Infeld model has as its pure scalar sector the complex scalar DBI theory with $\sigma_Z=2$ and as its pure fermion sector Akulov-Volkov theory with $\sigma_\psi =1$. All amplitudes are constructible with soft subtracted recursion. We present the  expressions for the 4- and 6-point amplitudes in Appendix \ref{a:susydbi}.
\item The $\mathcal{N}=1$ supersymmetric Born-Infeld model combines Akulov-Volkov theory with 
Born-Infeld theory with $\sigma_\gamma=0$. All amplitudes are constructible with the soft subtracted recursion relations of Section \ref{s:softrec}, except the pure vector ones, but they are uniquely fixed by the supersymmetry Ward identities. The 4- and 6-point amplitudes are given in Appendix \ref{a:susybi}.
\item Extended supersymmetry binds BI, Akulov-Volkov, and DBI into one supersymmetric exceptional EFT. For the case with $\mathcal{N}=4$ supersymmetry, the amplitudes can be constructed using the CHY approach \cite{Heydeman:2017yww}.
\end{itemize}

\section{Galileons}
\label{sec:Galileon}
Galileons are scalar effective field theories that arise in a multitude of contexts and as a result can be defined in different ways. In 4d, Galileons  are
\begin{enumerate}
	\item Higher-derivative scalar field theories with second-order equations of motion and absence  of Ostrogradski ghosts. These theories  have three free parameters: the cubic, quartic and quintic interaction coupling constants. A field redefinition removes the cubic interaction in favor of a linear combination of the quartic and quintic. The scattering amplitudes are of course invariant under the field redefinition, so for the purpose of studying perturbative scattering amplitudes, we consider only the quartic and quintic Galileons.

	\item The non-linear realization of the algebra $\mathfrak{Gal}(4,1)$ which is an \.{I}n\"{o}n\"{u}-Wigner contraction of the $\text{ISO}(4,1)$ symmetry algebra \cite{Goon:2012dy}. Truncated to leading order in the reduced dimension $\tilde{\Delta}$, this gives an effective field theory of a real massless scalar $\phi$ with $\sigma=2$ vanishing soft limits and coupling dimensions $[g_4]=-6$ and $[g_5]=-9$ for the quartic and quintic interactions respectively. 
		
	\item Subleading contributions to the low-energy effective action on a 3-brane embedded in a $5d$ Minkowski space. The leading contribution to this EFT is the DBI action and including the Galileon terms, the model is often called the DBI-Galileon. In the limit of infinite brane tension, the Galileons decouple from DBI. The non-$\mathbb{Z}_2$-symmetric cubic and quintic interactions arise from considering the effective action on an end-of-the-world brane.

	\item Scalar effective field theories that arise from 
	the massless decoupling limit of Fierz-Pauli-type massive gravity \cite{deRham:2010ik, deRham:2010gu} and from the decoupling limit of Proca theories. 
\end{enumerate}

It is not obvious if these definitions are equivalent. The equivalence between Definitions 2 and 3 is straightforward since $\text{ISO}(4,1)$ is the Poincar\'e symmetry of the 5d embedding space. In the brane picture of Definition 3, the DBI-Galileon scalar is a Goldstone boson that arises from the spontaneous breaking of translational symmetry transverse to the brane, with the contraction of the 5d Poincar\'e algebra equivalent to the non-relativistic limit of the fluctuations of the brane into the extra dimension \cite{deRham:2010eu}. 

In an approach based on scattering amplitudes, it is natural to use the second definition of Galileon theories, based on their soft weight $\sigma=2$ and fundamental coupling dimension. This is what we do in the following, however, we do comment on the connections to the other definitions. In Section \ref{s:galsusy}, we briefly review our recent results about the supersymmetrization of \mbox{(DBI-)Galileon} theories in 4d and cover some details that were left out in \cite{Elvang:2017mdq}. Motivated by Definition 4,  we investigate  the possibility of a scalar-vector Galileon theory in Section \ref{s:galvec}. In Sections \ref{s:SubleadingSpGal} and \ref{s:doublecopy}, we focus our attention on the Special Galileon. In Section \ref{s:SubleadingSpGal} we address the question of subleading operators respecting the enhanced $\sigma=3$ soft behavior. In Section \ref{s:doublecopy}, we approach the same question from a double-copy construction.  

\subsection{Galileons and Supersymmetry}
\label{s:galsusy}
This section reviews and expands on the results 
of \cite{Elvang:2017mdq}  for $\mathcal{N}=1$ supersymmetrization of Galileon models. Two approaches to forming a complex scalar $Z = \phi + i \chi$ are considered:
\begin{enumerate}
\item[(a)] Both $\phi$ and $\chi$ are Galileons so that the complex scalar $Z$ has soft weight $\sigma_Z=2$, or 
\item[(b)] $\phi$ is a Galileon but $\chi$ only has constant shift symmetry; then $\sigma_\phi =2$ and $\sigma_\chi =1$, and hence  $\sigma_Z=1$. A natural interpretation of $\chi$ is as an R-axion. 
\end{enumerate}
Both options were considered in \cite{Elvang:2017mdq}.

\noindent {\bf Option (a): $\sigma_Z = 2$}\\
Consider first the {\em quartic} Galileon. 
As discussed in Section \ref{s:MHVclass}, to be compatible with supersymmetry, the 4-point complex scalar amplitudes must have two $Z$'s and two $\bar{Z}$'s; such an amplitude is in the MHV class. 
It is also clear from the table of ``soft bootstrap" results in \reef{4ptscalarin} that there is a unique complex scalar quartic Galileon theory\footnote{That analysis also shows that it is impossible for this kind of model to have special Galileon symmetry with $\sigma_Z=3$.} with $\sigma_Z = 2$ based on the 4-point interaction with $\mathcal{A}_4(1_Z\, 2_{\bar{Z}}\, 3_Z\, 4_{\bar{Z}})=g_4 stu$. The other 4-point amplitudes in a supersymmetric theory are fixed by $\mathcal{A}_4(1_Z\, 2_{\bar{Z}}\, 3_Z\, 4_{\bar{Z}})$ using the supersymmetry Ward identity \reef{WI2phi2psi}. 

By \reef{susysigma1}, the soft behavior of the fermion must be either $\sigma_\psi =1$ or $2$.  
The all-fermion amplitudes are constructible when $\sigma_\psi=2$, and our soft bootstrap results for fermion theories \reef{4ptfermionin} show that no such theory exists. Therefore, the fermions in a supersymmetric  Galileon  theory with $\sigma_Z=2$ must have $\sigma_\psi=1$. 

In a supersymmetric quartic Galileon theory with $\sigma_Z=2$ and $\sigma_\psi=1$, the constructibility criterion \reef{criterion} for $n$-point amplitudes with $n_s$ scalars and $n_f$ fermions is $n_f<4$. Thus at 6-point, we can only use soft subtracted recursion to compute the amplitudes with at most two fermions. However, as discussed in \cite{Elvang:2017mdq},  two of the six supersymmetry Ward identities \reef{WI6-1}-\reef{WI6-3} uniquely determine the 4- and 6-fermion amplitudes. The remaining four identities in \reef{WI6-1}-\reef{WI6-3} are used as consistency checks. The expressions for the 6-point amplitudes of the supersymmetric quartic Galileon can be found in Appendix \ref{susygal4amplitudes}. We have checked that the recursively constructed  4- and 6-point amplitudes match those that we calculate from the Lagrangian superspace construction of the quartic Galileon in \cite{Farakos:2013zya}. 

The supersymmetry Ward identities at 8-point and higher do not uniquely determine the non-constructible amplitudes of the supersymmetric quartic Galileon. We therefore suspect that the quartic Galileon fails to be unique at 8-point and higher \cite{Elvang:2017mdq}.

The {\em quintic} 
Galileon does not admit a supersymmetrization with $\sigma_Z = 2$ for the complex scalar. As discussed at the end of Section \ref{s:5ptsc}, there are no obvious obstructions from the soft-recursion tests to a complex scalar decoupled quintic Galileon with $\mathcal{A}_5(1_Z\, 2_{\bar{Z}}\, 3_Z\, 4_{\bar{Z}}\, 5_Z)=\left(\epsilon_{\mu\nu\rho\sigma}p_1^\mu p_2^\nu p_3^\rho p_4^\sigma\right)^2$. 
However, it is not compatible with the 5-point supersymmetry Ward identities. It follows that the cubic Galileon also cannot be supersymmetrized with $\sigma_Z = 2$.

\noindent {\bf Option (b): $\sigma_Z = 1$.}\\
Consider a {\em quartic} complex scalar theory where the real part of the complex scalar $Z$ is the Galileon $\phi$ and the imaginary part is an R-axion $\chi$. The constructibility criterion with 
$\sigma_\phi=2$ and  $\sigma_\chi = \sigma_\psi= 1$ is
$2n_\chi+n_f<4$, so there are only two mixed amplitudes to check; 
they do not restrict the 2-parameter family of input amplitudes \cite{Elvang:2017mdq}. We have checked that the constructible 6-point amplitudes are compatible with DBI. 

For a {\em quintic} Galileon with $\sigma_Z = 1$, we found \cite{Elvang:2017mdq} a unique solution to the supersymmetry Ward identities
\be
	\mathcal{A}_5 ( 1_Z\, 2_{\bar Z} \, 3_Z\, 4_{\bar Z}\, 5_Z ) 
	= - \frac{[ 24 ]}{[ 25 ]} \mathcal{A}_5 ( 1_Z \, 2_{\bar Z} \, 3_Z \, 4_{\bar \psi} \, 5_\psi ) 
	= \frac{[ 24 ]}{[ 35 ]} \mathcal{A}_5 ( 1_Z\, 2_{\bar \psi} \, 3_\psi\, 4_{\bar \psi} \, 5_\psi )\, ,
\ee	
namely
\be	
  \begin{split}
	\mathcal{A}_5(1_Z\, 2_{\bar Z} \, 3_Z\, 4_{\bar Z}\, 5_Z )=&\ s_{24}\left(6 s_{24}s_{25}s_{45}+\left(4s_{12}s_{23}s_{45}+2 s_{12}s_{24}s_{34}+2s_{25}^2s_{45}+s_{24}s_{25}^2+(2\leftrightarrow 4)\right)\right.\nonumber\\
	&\left.+(1\leftrightarrow 5)+(3\leftrightarrow 5)\right)-4s_{24}^2\,.
   \end{split}	
	\label{fivePtSUSY-WI}
\ee
The amplitudes 
$\mathcal{A}_5 ( 1_{\bar Z}\, 2_Z\, 3_{\bar Z} \, 4_Z\, 5_{\bar Z} )$, 
$\mathcal{A}_5 ( 1_{\bar Z} \, 2_Z\,  3_{\bar Z} \, 4_\psi 5_{\bar \psi} )$, 
and $\mathcal{A}_5( 1_{\bar Z}\, 2_\psi \, 3_{\bar \psi}\, 4_\psi\, 5_{\bar \psi})$ follow from conjugation of the above.\footnote{These 5-point amplitudes are not required to vanish in $3d$ kinematics (and they do not) because they do not satisfy the constructibility criterion.} It is interesting  that the fermions in these 5-point amplitudes automatically have $\sigma_\psi=1$.

To test consistency of  a supersymmetric quintic Galileon with $\sigma_\phi=2$, $\sigma_\chi=1$, and $\sigma_\psi=1$, we consider the 7-point and 8-point amplitudes in 
the decoupled Galileon theory. In both cases, the constructibility criterion is $2n_\chi+n_f<4$. The (few) non-trivial constructible amplitudes pass the soft subtraction recursive tests of $a_i$-independence. We have also tested compatibility with the supersymmetric DBI interactions: at 7-point the constructibility criterion is $2n_\chi+n_f<\,8$ and again the constructible 7-point amplitudes pass the test. This indicates that there may indeed be a supersymmetric brane-theory with both quartic and quintic  terms subleading to DBI. The scalar $\phi$ is the Goldstone mode of the broken transverse translational symmetry whereas the scalar $\chi$ is an R-axion. The fermion $\psi$ is a genuine Goldstino of partial broken supersymmetry. We discuss such scenarios further in forthcoming work.

\subsection{Vector-Scalar Special Galileon}
\label{s:galvec}

It is known that scalar Galileon theories arise in certain limits of massive gravity \cite{deRham:2010gu,deRham:2010ik} (for a review, see \cite{Hinterbichler:2011tt}). An on-shell massive graviton in 4d has 5 polarization states and the decoupling limit gives one real massless scalar (the Galileon) and a massless photon in addition to the massless graviton. So we  expect there to be an EFT of a real Galileon scalar coupled to vector.\footnote{The decoupling of these interactions from the graviton is not clear \cite{Hinterbichler:2011tt}.} The vector couples quadratically to the scalar  and was  consistently truncated off in \cite{deRham:2010ik}. 
Some subsequent studies have discussed the photon-scalar coupling of Galileons, see for example \cite{Jimenez:2016isa}. Here, we use soft recursion to give some definitive results about the possible scattering amplitudes in such a theory. 

If the scalar has $\sigma_\phi = 2$, only the scalar amplitudes are constructible, and we are not able to say anything about the vector sector and its couplings to the scalar. If however the couplings are tuned in such a way that the cubic and quintic Galileon interactions are set to zero then in the scalar sector the soft weight of the scalar is enhanced to $\sigma_\phi=3$, the \textit{special Galileon} scenario. At present it is unknown whether this enhancement of symmetry can be understood in some natural way from the decoupling limit of some model of massive gravity. Moreover, it is not a priori clear if the $\sigma_\phi=3$ enhancement can survive coupling to other  particles.

We   use the power of the soft bootstrap to construct the most general amplitudes consistent with the special Galileon low-energy theorem. We use the 6-point test to exclude EFTs with a special Galileon coupled non-trivially to a photon with $\sigma_\gamma>0$. For the model with $\sigma_\phi = 3$ and $\sigma_\gamma=0$, we find that the soft recursion 6-point test reduces the most general  6 real-parameter ansatz for the scalar and scalar-vector interactions to a 3 real-parameter family:
\be 
  \begin{split}
    \mathcal{A}_4 ( 1_\phi\, 2_\phi\, 3_\phi\, 4_\phi ) &= g_1 s t u \,,\\
    \mathcal{A}_4 ( 1_\phi\, 2_\phi\,2_ \gamma^+ \,4_\gamma^+ ) &= g_2 [ 3 4 ]^2 \left ( t^2 + u^2 + 3 t u \right ) \,,\\
  \mathcal{A}_4 ( 1_\gamma^- \,2_\phi \,3_\phi \,4_\gamma^+ ) &= g_1 \langle 1 2 \rangle [ 2 4 ] \langle 1 3 \rangle [ 3 4 ] u\,, \\
  \mathcal{A}_4 ( 1_\phi\, 2_\phi \,3_\gamma^- \,4_\gamma^-  ) &= 
  g_2^* \langle 34 \rangle^2 \left (  t^2 + u^2 + 3 t u \right ) \,.
  \end{split}
\ee
The couplings of the pure vector sector are unconstrained; the most general ansatz is
\be
  \begin{split}
  \mathcal{A}_4 ( 1_\gamma^+\, 2_\gamma^+ \,3_\gamma^+ \,4_\gamma^+ ) 
  &= g_3 \Big ( [ 1 2 ]^2 [ 3 4 ]^2 s + [ 1 3 ]^2 [ 2 4 ]^2 t + [ 1 4 ]^2 [ 2 3 ]^2 u \Big ) \,,\\
  \mathcal{A}_4 ( 1_\gamma^-\, 2_\gamma^-\, 3_\gamma^+\, 4_\gamma^+ ) 
  &= g_4 \langle 1 2 \rangle^2 [ 3 4 ]^2 s\,, \\
  \mathcal{A}_4 ( 1_\gamma^-\, 2_\gamma^-\, 3_\gamma^- \,4_\gamma^- ) 
  &= g_3^* \Big ( \langle 1 2 \rangle^2 \langle 3 4 \rangle^2 s + \langle 1 3 \rangle^2 \langle 2 4 \rangle^2 t + \langle 1 4 \rangle^2 \langle 2 3 \rangle^2 u \Big )\,.
    \end{split}
\ee 
The most interesting feature of the above result is the relation between the coefficients of the amplitudes $\mathcal{A}_4 ( 1_\phi\, 2_\phi\, 3_\phi\, 4_\phi )$ and 
$ \mathcal{A}_4 ( 1_\gamma^- \,2_\phi \,3_\phi \,4_\gamma^+ )$. 
The former is the familiar quartic Galileon, while the latter would arise from an operator of the form
\begin{equation}
  \mathcal{O} \sim g_1 (\partial_{\mu}F_+^{\alpha\beta})(\partial^\mu F_-^{\dot{\alpha}\dot{\beta}})(\sigma^\nu_{\alpha\dot{\alpha}}\partial_\nu\phi)(\sigma^\rho_{\beta\dot{\beta}}\partial_\rho \phi),
\end{equation}
where $F_\pm$ are as defined in and below  \eqref{Ffeynrule} 

The relation between the couplings strongly indicates the existence of a non-linear symmetry which mixes the scalar and vector modes. Describing the action of this symmetry and its consequences is left for future work.

\subsection{Higher Derivative Corrections to the Special Galileon}
\label{s:SubleadingSpGal}

The real quartic Galileon has low-energy theorems with $\sigma = 3$ soft weight. 
Being agnostic about the origin of the special Galileon, from an EFT perspective, one should write a Lagrangian with all possible operators that respect the symmetries of the theory in a derivative expansion.
The authors of \cite{Padilla:2016mno} found that among a specific subclass of Lagrangian operators, namely those with the schematic form $\partial^4 \phi^4$, $\partial^6 \phi^4$ and $\partial^8 \phi^5$, the special Galileon is the unique choice that can give enhanced soft limits with $\sigma = 3$ soft weight. In this section, we investigate much more exhaustively the possible higher-derivative quartic and quintic operators compatible with $\sigma=3$ soft behavior.  This is done using soft-subtracted recursion relations to calculate the 6- and 7-point scattering amplitudes of the model. 

Let us start our discussion with  the 6-point case.
The constructibility criterion \eqref{crit2} implies that recursion relations are valid if the coupling constant $g_6$ of the 6-point amplitude satisfies
\begin{equation}
    [ g_6 ] > - 20\,.
\end{equation}
Given that this coupling is the product of two quartic couplings and that the leading order quartic coupling has mass dimension $-6$ recursion relations can probe contributions to the 4-point amplitude with mass dimension in the range
\begin{equation}
    - 14 <  [ g_4 ] \le - 6\,.
\end{equation}
Taking into account Bose symmetry, the most general ansatz one can write down for the 4-point matrix element of local operators is 
\begin{equation}
\label{ansatz4ptSpGal}
    \begin{aligned}
        \mathcal A_4 ( 1_\phi 2_\phi 3_\phi 4_\phi ) = & \frac{c_0}{\Lambda^6} s t u \\
         + & \frac{c_1}{\Lambda^8} \left ( s^4 + t^4 + u^4 \right ) \\
         + & \frac{c_2}{\Lambda^{10}} \left ( s^5 + t^5 + u^5 \right  ) \\
         + & \frac{1}{\Lambda^{12}} \left ( c_3 \left ( s^6 + t^6 + u^6 \right ) + c_3' s^2 t^2 u^2 \right ) + \mathcal O ( \Lambda^{-14} ) \,.
    \end{aligned}
\end{equation}
The leading term with coupling $c_0/\Lambda^6$ is the usual quartic Galileon. The terms suppressed by higher powers of the the UV cutoff $\Lambda$ encode all possible higher-derivative quartic operators of the scalar field up to order $\Lambda^{-14}$.

We apply the 6-point test with $\sigma=3$ and find that consistency requires $c_1 = c_3 = 0$ in the ansatz \reef{ansatz4ptSpGal}. The 4-point amplitude then becomes
\begin{equation}
    \mathcal A_4 ( 1_\phi \,2_\phi\, 3_\phi\, 4_\phi ) = \frac{c_0}{\Lambda^6} s t u + \frac{c_2}{\Lambda^{10}} \left ( s^5 + t^5 + u^5 \right ) + \frac{c_3'}{\Lambda^{12}} s^2 t^2 u^2 + \mathcal O ( \Lambda^{-14} )\,.
    \label{eq:specialGalileonSubleading}
\end{equation}
From this, we understand that there cannot exist an 8-derivative Lagrangian operator that preserves the special Galileon symmetry. Additionally, at 6-, 10- and 12-derivative order there exist unique 
quartic operators compatible with $\sigma=3$. In Section \ref{s:doublecopy}, we show explicitly that the result \reef{eq:specialGalileonSubleading} can also be obtained from an application of the BCJ double-copy.

Next we examine the possible existence of quintic operators compatible with $\sigma=3$. We  combine input from the quartic Galileon with the most general possible ansatz for the 5-point matrix elements and use the 7-point test to assess compatibility with $\sigma=3$. The soft subtracted recursion relations at 7 points are valid if
\begin{equation}
    [ g_7 ] >  - 24\,.
\end{equation}
Since the 7-point coupling constant is the product of a quartic (with mass dimension $-6$ or lower) and a quintic coupling,  the latter must then  satisfy
\begin{equation}
    [ g_5 ] > - 18\,.
\end{equation}
With Bose symmetry and the requirement that the ansatz for the 5-point amplitude must have soft weight $\sigma = 3$, we are left with
\bea
    \label{eq:5ptResult}
        &&\!\!\!\!\!\!\!\mathcal A_5 ( 1_\phi\, 2_\phi\, 3_\phi\, 4_\phi\, 5_\phi ) 
        = \frac{d_1}{\Lambda^{15}} \epsilon ( 1 2 3 4) \sum_P ( - 1 )^{| P |} s_{P_1 P_2} s_{P_2 P_3} s_{P_3 P_4} s_{P_4 P_5} s_{P_5 P_1} \\
        \nonumber
       && \!\!\!\!\!\!\!+ \frac{1}{\Lambda^{17}} \Bigg [ d_2\ \epsilon ( 1 2 3 4 )^4 + d_3  \epsilon ( 1 2 3 4 ) \sum_P ( - 1 )^{| P |} s_{P_1 P_2} s_{P_2 P_3}^2 \left ( s_{P_2 P_3}^2 s_{P_3 P_4} - s_{P_1 P_2}^2 s_{P_2 P_4} \right )  \\
       \nonumber
       && + d_4 \left ( \frac 4 5 \sum_{i < j} s_{ij}^3 \sum_{i < j} s_{ij}^5 + \sum_{i<j} \sum_{k \neq i, j} \left ( 20 s_{ij}^2 s_{ik}^3 s_{jk}^3 + 9 s_{ij}^4 s_{ik}^2 s_{jk}^2 - 2 s_{ij}^6 s_{ik} s_{jk} \right ) \right )
       \Bigg ] + \mathcal O ( \Lambda^{-19} )\,.
\eea
In the above, $\epsilon ( 1 2 3 4) = \epsilon_{\mu \nu \rho \sigma} p_{1}^\mu p_2^\nu p_3^\rho p_4^\sigma$, the sum $\sum_{i < j}$ means $\sum_{i = 1}^4 \sum_{j = i + 1}^5$, while the sum $\sum_P$ is over all permutations of $\{ 1, 2, 3, 4, 5 \}$, $( - 1 )^{| P |}$ is the signature of the permutation and $P_i$ is its $i$th element. There are no contributions to the amplitude that have less than 14 derivatives. The $1/\Lambda^{14}$-term satisfies the constructibility criterion and vanishes in 3d kinematics, in agreement with the discussion of Section \ref{consistency}.  Two of the $1/\Lambda^{17}$-terms also vanish in 3d kinematics, but this was not a priori expected since they are too high order to satisfy constructibility. 
 
The 7-point test  implies no constraints on the coefficients $d_1$, $d_2$, $d_3$ and $d_4$. This is  evidence in favor of the existence of four 5-point operators that preserve the special Galileon symmetry. Next, in Section \ref{s:doublecopy}, we investigate whether this result can be obtained from a double-copy prescription, similar to the 4-point case.

\subsection{Comparison with the Field Theory KLT Relations}
\label{s:doublecopy}
The significance of the special Galileon extends well beyond the contraction limit of the 3-brane effective field theory and the decoupling limit of massive gravity. The enhancement of the soft behavior to $\sigma=3$ (which degenerates to $\sigma=2$ when the DBI interactions are re-introduced) or correspondingly the extension of the non-linearly realized symmetry algebra suggests that this model has a fundamental significance of its own that is at present only partially understood. Perhaps one of the deepest and least understood aspects of the special Galileon is its role in the (field theory) \textit{KLT algebra} as the product of two copies of the $\frac{U(N)\times U(N)}{U(N)}$ non-linear sigma model. For $N=2,3$ this coset sigma model has been intensively studied as a phenomenological model of the lightest mesons under the name \textit{Chiral Perturbation Theory} ($\chi\text{PT}$). Henceforth we will use this name to avoid confusion with the $\mathbb{CP}^1$ non-linear sigma model discussed in Section \ref{sec:SUSY_NLSM}.

The \textit{double-copy} relation between $\chi \text{PT}$ and the special Galileon was first understood in the CHY auxilliary world-sheet formalism\cite{Cachazo:2014xea}. Specifically, it was shown in the CHY formalism that the \textit{leading order} contribution to scattering in the special Galileon model can be obtained from the KLT product
\begin{equation} \label{KLT}
  \mathcal{A}_n^{\text{sGal}} = \sum_{\alpha,\beta}\mathcal{A}_n^{\chi \text{PT}}[\alpha]S_{\text{KLT}}[\alpha|\beta]\mathcal{A}_n^{\chi \text{PT}}[\beta]\, ,
\end{equation}
where $\alpha,\beta$ index the $(n-3)!$ independent color(flavor)-orderings.\footnote{We use square brackets for the arguments of a color-ordered amplitude.} The KLT kernel $S_{\text{KLT}}[\alpha|\beta]$ is universal in the sense that the explicit form of the relations (\ref{KLT}) are identical to the perhaps more familiar field theory KLT relations giving a double-copy construction of Einstein-dilaton-$B_{\mu\nu}$ gravity from two copies of Yang-Mills theory. Concretely, the first few relations have the form 
\begin{align} \label{KLT2}
  \mathcal{A}^{\text{sGal}}_4\left(1,2,3,4\right) &= -s_{12}\mathcal{A}_4^{\chi\text{PT}}\left[1,2,3,4\right]\mathcal{A}_4^{\chi\text{PT}}\left[1,2,4,3\right]\,, \nonumber\\[1mm]
  \mathcal{A}^{\text{sGal}}_5\left(1,2,3,4,5\right) &= s_{23}s_{45}\mathcal{A}_5^{\chi\text{PT}}\left[1,2,3,4,5\right]\mathcal{A}_5^{\chi\text{PT}}\left[1,3,2,5,4\right]+(3\leftrightarrow 4) \,, \nonumber\\[1mm]
  \mathcal{A}^{\text{sGal}}_6\left(1,2,3,4,5,6\right) &= -s_{12}s_{45}\mathcal{A}_6^{\chi\text{PT}}\left[1,2,3,4,5,6\right]\left(s_{35}\mathcal{A}_6^{\chi\text{PT}}\left[1,5,3,4,6,2\right]\right.\nonumber\\
&\hspace{5mm}\left.+(s_{34}+s_{35})\mathcal{A}_6^{\chi\text{PT}}\left[1,5,4,3,6,2\right]\right) +\mathcal{P}(2,3,4)\,,
\end{align}
where $\mathcal{P}(2,3,4)$ denotes the sum of all permutations of legs 2, 3 and 4. 

For the formulae (\ref{KLT}) and (\ref{KLT2}) to even be well-defined,  the color-ordered amplitudes on the right-hand-side must satisfy a number of non-trivial relations to reduce the number of independent partial amplitudes to $(n-3)!$ for the scattering of $n$ particles. The existence of a color-ordered representation is itself non-trivial and not guaranteed to be satisfied in all models with color structure\cite{Broedel:2012rc}. In all known cases where the double-copy relations (\ref{KLT}) give a sensible, physical output, the reduction to a reduced basis of size $(n-3)!$ is accomplished by two sets of identities among the partial amplitudes, namely the \textit{Kleiss-Kuijf}  and \textit{fundamental Bern-Carrasco-Johansson} relations. That these identites obtain for amplitudes calculated in the leading two-derivative action of $\chi$PT was first established in \cite{Chen:2013fya} using semi-on-shell recursion techniques developed in \cite{Kampf:2012fn}. 

Our goal in this section is to connect two (possibly discrepant) definitions of the special Galileon model: 
\begin{itemize}
  \item[1.] The special Galileon is the most general effective field theory of a real massless scalar with $\sigma=3$ vanishing soft limits.
  \item[2.] The special Galileon is the double-copy of two copies of $\chi$PT.
\end{itemize}
What we have described above is the known fact that these definitions agree at the lowest non-trivial order. In the previous section we used soft subtracted recursion to construct the most general 4- and 5-point amplitudes consistent with the first definition up to order $\Lambda^{-12}$ and $\Lambda^{-17}$ respectively. To determine if these results agree with the second definition we must first construct the most general 4- and 5-point amplitudes in $\chi$PT compatible with the requirements of the double-copy. Here we are following the approach of \cite{Broedel:2012rc} and making the most conservative possible assumptions. Specifically we assume that both the explicit form of the double-copy (\ref{KLT2}) \textit{and} the relations the amplitudes must satisfy to reduce the basis of partial amplitudes to size $(n-3)!$ are \textit{identical} to what is required at leading order. 

Let us begin with the 4-point amplitudes. The relations we impose are cyclicity (C)
\begin{equation}
  \mathcal{A}_4^{\chi \text{PT}}[1,2,3,4] = \mathcal{A}_4^{\chi \text{PT}}[2,3,4,1]\,,
\end{equation}
Kleiss-Kuijf (KK) or $U(1)$-decoupling
\begin{equation}
  \mathcal{A}_4^{\chi \text{PT}}[1,2,3,4]+\mathcal{A}_4^{\chi \text{PT}}[2,1,3,4]+\mathcal{A}_4^{\chi \text{PT}}[2,3,1,4]=0\,,
\end{equation}
and the fundamental BCJ relation
\begin{equation}
  (-s-t)\mathcal{A}_4^{\chi \text{PT}}[1,2,3,4]-t\mathcal{A}_4^{\chi \text{PT}}[1,2,4,3] = 0\,.
\end{equation}
Since there are no additional quantum number labels in the partial amplitudes, at each order the 4-point amplitude is determined by a single polynomial function of the available Lorentz singlets
\begin{equation}
  \mathcal{A}_4^{\chi \text{PT}}[1,2,3,4] = F^{(0)}(s,t)+\frac{1}{\Lambda^2}F^{(2)}(s,t) + \frac{1}{\Lambda^4}F^{(4)}(s,t) +\ldots
\end{equation}
The superscript $k$ counts both the mass dimension of the function and the number of derivatives in the underlying effective operator. In this language, the double-copy-compatibility conditions take the  form 
\be 
\label{con}
\begin{array}{lll}
\text{C:} &F^{(k)}(s,t) = F^{(k)}(-s-t,t)\,,\\[1mm]
\text{KK:} &F^{(k)}(s,t) + F^{(k)}(s,-s-t)+F^{(k)}(-s-t,s)=0 \,,\\[1mm]
\text{BCJ:} &(-s-t)F^{(k)}(s,t) - tF^{(k)}(s,-s-t)=0 \,.
\end{array}
\ee
 We make a  general parametrization of the polynomial functions as
\be
  \begin{split}
  F^{(0)}(s,t) &= c_1^{(0)}, \\
  F^{(2)}(s,t) &= c_1^{(2)}s+c_2^{(2)}t, \\
  F^{(4)}(s,t) &= c_1^{(4)}s^2+c_2^{(4)}st+c_3^{(4)}t^2, \\
  F^{(6)}(s,t) &= c_1^{(6)}s^3+c_2^{(6)}s^2t+c_3^{(6)}st^2+c_4^{(6)}t^3, \\
  F^{(8)}(s,t) &= c_1^{(8)}s^4+c_2^{(8)}s^3t+c_3^{(8)}s^2t^2+c_4^{(8)}st^3+c_5^{(8)}t^4, 
  \end{split}
\ee
and so on. Imposing the conditions (\ref{con}) gives a system of linear relations among the coefficients $c_i^{(k)}$. These are straightforward to solve and give 
\begin{equation} \label{comp}
  \mathcal{A}_4^{\chi\text{PT}}[1,2,3,4] =  \frac{g_2}{\Lambda^2}t + \frac{g_6}{\Lambda^6}t (s^2+t^2+u^2) + \frac{g_8}{\Lambda^8}t (stu)+\ldots
\end{equation}
A few comments about this result. As expected, the leading 2-derivative contribution is compatible with the conditions (\ref{con}). Surprisingly, there are no compatible contributions from 4-derivative operators, but there are unique contributions at 6- and 8-derivative order. Moreover, the structure of the result here agrees with the 4-point amplitude of \textit{Abelian Z-theory} \cite{Carrasco:2016ldy}. 
The Z-theory model is a top-down construction which gives open string scattering amplitudes as the field theory double-copy of Yang-Mills and a higher-derivative extension of $\chi$PT. The Z-amplitudes are by construction guaranteed to satisfy the double-copy-compatibility conditions but with Wilson coefficients $g_i$ having precise values calculated from the known string amplitudes. The method of this section can be understood as the bottom-up converse of the Z-theory construction, and at 4-point we find agreement.

To summarize, we have shown that up to 8-derivative order there is a 3-parameter family of operators that generate 4-point matrix elements compatible with the conditions required for the double-copy to be well-defined. We could continue this to higher order, but our ability to compare with the methods of Section \ref{s:SubleadingSpGal} are bounded above at this order by the constructibility criterion.

To construct the associated amplitudes in the special Galileon model (according to the second definition described above) we use the first relation in (\ref{KLT2}). The result is 
\begin{equation}
  \mathcal{A}_4^{\text{sGal}}(1,2,3,4) = \frac{c_1}{\Lambda^6}stu + \frac{c_2}{\Lambda^{10}}\left(s^5+t^5+u^5\right) + \frac{c_3}{\Lambda^{12}}s^2t^2u^2 +\ldots\,,
  \label{eq:sGal_A4}
\end{equation}
in precise agreement with the special Galileon amplitude \eqref{eq:specialGalileonSubleading}. 

As an additional check to the results obtained above, we calculate the 6-point amplitudes of both $\chi$PT and the special Galileon.
Up to order $\mathcal O ( \Lambda^{-6} )$ the $\chi$PT amplitude can be calculated using soft subtracted recursion with \eqref{comp} as input.
Note that only three factorization channels contribute to this calculation because the rest do not preserve color ordering.
The resulting amplitude,
\begin{equation}
    \mathcal A_6^{\chi\text{PT}} [ 1, 2, 3, 4, 5, 6 ] = \frac{g_2^2}{\Lambda^4} \bigg [ \frac{s_{13} s_{46}}{p_{123}^2} + \frac{s_{24} s_{15}}{p_{234}^2} + \frac{s_{35} s_{26}}{p_{345}^2} - s_{246}  \bigg ] + \mathcal O ( \Lambda^{-8} )\,,
\end{equation}
satisfies all C, KK and BCJ constraints.
Contributions subleading to the ones listed above do not satisfy the constructibility criterion \eqref{crit2} and cannot be calculated using soft subtracted recursion.
However, we were able to uniquely determine them up to order $\mathcal O ( \Lambda^{-10} )$, by demanding that they have the correct pole structure, consistent with unitarity and locality, have $\sigma = 1$ soft weight and satisfy C, KK and BCJ conditions.
The result of this calculation is listed in \eqref{eq:chiPT_A6}.

We are now in position to calculate the 6-point special Galileon amplitude with two different methods.
We can either use the  6-point KLT relation in  \eqref{KLT2} or use soft subtracted recursion with \eqref{eq:sGal_A4} as input. The results of these calculations match perfectly up to order $ \mathcal O ( \Lambda^{-18} )$, which is the furthest the recursive calculation can go.

Shifting our focus to 5-point amplitudes, we find that it is  {\em not}
 possible to reproduce \eqref{eq:5ptResult} as a double-copy of two (identical or non-identical) color-ordered scalar amplitudes, despite the perfect agreement at 4- and 6-points.
Starting from a general ansatz for the scalar color-ordered amplitude, we find that the leading contribution that satisfies all C, KK and BCJ constraints is $\mathcal O ( \Lambda^{-15} )$ corresponding to a valence 5 scalar-field 
 operator with 14 derivatives. The existence of such an operator at all is interesting since there are apparently \textit{no odd point amplitudes in Z-theory} \cite{Carrasco:2016ldy}! At this order we find that the kinematic structure of Z-theory does not coincide with the most general possible double-copy-compatible higher-derivative extension of $\chi$PT. Or perhaps said differently, just like string theory fixes the Wilson coefficients in the 4-point result  \reef{comp} to take particular (non-zero) values, it appears to fix the Wilson coefficients of the odd-point amplitudes to be zero. 

When we use the second relation of \eqref{KLT2} with this result, we obtain a 5-point scalar amplitude of order $\mathcal O ( \Lambda^{-33} )$, which is significantly subleading to the amplitude \eqref{eq:5ptResult} we calculated in the previous section for the special Galileon.

\section{Outlook}

There are several interesting 
questions that remain unanswered in this work. In Section \ref{s:softboot} we applied the soft bootstrap to classes of models with simple spectra consisting of a single particle of a particular spin. Furthermore, we gave a limited examination of classes of models with linearly realized supersymmetry with spectra consisting of a single multiplet. There is a potentially vast landscape of constructible models with more complicated spectra and possible futher interesting linearly realized symmetries. 

We have already seen examples of this; in Section \ref{sec:SUSY_NLSM} further symmetry (in this case electromagnetic duality symmetry) emerges as an unavoidable consequence of the combination of low-energy theorems and linear $\mathcal{N}=2$ supersymmetry. Similarly we should expect the soft bootstrap to reveal models with complicated \textit{non-linear} symmetries. In Section \ref{s:galvec} we have given evidence in favor of the existence of such a symmetry underlying a vector-scalar extension of the special Galileon.

Our results also suggest two additional applications for the soft bootstrap. The first is to the classification of higher-derivative operators. The method applied in Sections \ref{s:SubleadingSpGal} and \ref{s:doublecopy} to the special Galileon and $\chi$PT is generalizable to a large class of EFTs with manifest advantages over traditional methods. 
The second is as a useful cross-check on results concerning exceptional EFTs obtained via the double copy. In Section \ref{s:doublecopy} we found the puzzling result that there exist valence 5 operators invariant under the special Galileon symmetry which apparently cannot be constructed as the double copy of subleading $\chi$PT operators. 

 It would be reasonable to expect further, similarly rich and unexpected, phenomena to be present throughout the landscape of constructible EFTs.

\section*{Acknowledgements}
We would like to thank Clifford Cheung, Kurt Hinterbichler,  Chia-Hsien Shen, and Jaroslav Trnka for useful discussions. The authors are grateful to the Kavli Institute for Theoretical Physics, UC Santa Barbara, for hospitality during the `Scattering Amplitudes and Beyond' program which was supported under Grant No.~NSF PHY17-48958 to the KITP. 
This work was supported in part by the US Department of Energy under Grant No.~DE-SC0007859.
CRTJ was supported by a Leinweber Graduate Fellowship and MH by a Rackham Predoctoral Fellowship from the University of Michigan.

\appendix

\section{Derivation of \reef{softrecursion2}}
\label{app:recrel}
In this appendix, we derive the manifestly local form \reef{softrecursion2} of the subtracted recursion relations. For a given factorization channel, consider from the recursion relations \reef{softrecursion} the  expression 
\be
 \label{recrelexpr1}
  \frac{\hat{\mathcal{A}}_L^{(I)}(z_I^\pm)\hat{\mathcal{A}}_R^{(I)}(z_I^\pm)}{F(z_I^\pm)P_I^2(1-z_I^\pm/z_I^\mp)} 
  = \sum_{z_I=z_I^\pm} \text{Res}_{z=z_I}
  \frac{\hat{\mathcal{A}}_L^{(I)}(z)\hat{\mathcal{A}}_R^{(I)}(z)}{z\,F(z)\,\hat{P}_I^2} 
  = \oint_\mathcal{C} dz\, \frac{\hat{\mathcal{A}}_L^{(I)}(z)\hat{\mathcal{A}}_R^{(I)}(z)}{z\,F(z)\,\hat{P}_I^2}\,,
\ee 
where the contour surrounds only the two poles $z_I^\pm$. 
The second equality is non-trivial and deserves clarification. In the second expression, the  subamplitudes $\hat{\mathcal{A}}_L^{(I)}(z)$ and $\hat{\mathcal{A}}_R^{(I)}(z)$ are only defined precisely on the residue values $z= z_I^\pm$ for which the internal momentum $\hat{P}_I$ is on-shell; in general one cannot just think of $\hat{\mathcal{A}}_{L,R}^{(I)}(z)$ as functions of $z$. However, in the product $\hat{\mathcal{A}}_L^{(I)}(z)\hat{\mathcal{A}}_R^{(I)}(z)$, one can eliminate the internal momentum $\hat{P}_I$ in favor of the $n$ shifted external momenta by using momentum conservation. Then the resulting expression can be analytically continued in $z$ away from the residue value. This is implicitly what has been done in performing the second step  in \reef{recrelexpr1}.

Let us assess the large-$z$ behavior of the integrand in \reef{recrelexpr1}. The L and R subamplitudes have couplings $g_L$ and $g_R$ such that $g_L g_R = g_n$, with $g_n$ the coupling of $\mathcal{A}_n$. Their mass-dimensions are related as $[g_L]+[g_R] = [g_n]$. Hence, using $n_L + n_R = n+2$ and \reef{AnD}, we find that the numerator behaves at large $z$ as 
\be
\hat{\mathcal{A}}_L^{(I)}(z)\hat{\mathcal{A}}_R^{(I)}(z)
\to z^{D_L} z^{D_R}
 = z^{6-n- [g_n] -\sum_{i=1}^n s_i - 2 s_P}
 = z^{D+2 - 2 s_P}\,,
\ee
where $s_P$ denotes the spin of the particle exchanged on the internal line and $D$ is the large $z$ behavior of the $\mathcal{A}_n$ which we know satisfies $D- \sum_{i=1}^n \sigma_i <0$, by the assumption that the amplitude  $\mathcal{A}_n$ is recursively constructible by the criterion \reef{criterion}. We therefore conclude that the integrand in \reef{recrelexpr1} behaves as $z^{D-1- \sum_{i=1}^n \sigma_i - 2 s_P}$, i.e.~it goes to zero as $1/z^2$ or faster. Hence, there is no simple pole at $z \to \infty$. 

If we deform the contour, we get the  sum over all  poles $z \ne z_I^\pm$ in  
$\hat{\mathcal{A}}_L^{(I)}(z)\hat{\mathcal{A}}_R^{(I)}(z)/(z\,F(z)\,\hat{P}_I^2)$. Let us  {\em assume that $\mathcal{A}_L^{(I)}$ and $\mathcal{A}_R^{(I)}$  are both local}: they have no poles and hence we pick up exactly the simple poles at $z=0$ and $z=1/a_i$ for $i=1,2,\dots,n$. We then conclude that the soft recursion relations take the form
\begin{equation} \label{Asoftrecursion2}
  \mathcal{A}_n = \sum_I \sum_{z'=0, \frac{1}{a_1},\dots,\frac{1}{a_n}}\sum_{|\psi^{(I)}\rangle}
  \text{Res}_{z=z'}\,\frac{\hat{\mathcal{A}}_L^{(I)}(z)\hat{\mathcal{A}}_R^{(I)}(z)}{z\,F(z)\,\hat{P}_I^2}\,,
\end{equation}
where $F(z) = \prod_{i=1}^n (1-a_iz)^{\sigma_i}$. This form of the recursion relation is manifestly rational in the momenta. 

Note that only the $z=0$ residues give pole terms in $\mathcal{A}_n$. Therefore the sum of the $1/a_i$ residues over all channels must be a local polynomial in the momenta.


\section{Explicit expressions for amplitudes}
\label{sec:AmplitudeExpressions}
In this appendix, we present expressions for the 4- and 6-point amplitudes of the theories discussed in the main text.
The 6-point amplitudes were reconstructed with the 4-point ones as input, by means of the subtracted recursion relations and the the supersymmetry Ward identities also discussed in the main text.

\subsection{Supersymmetric $\mathbb{CP}^1$ NLSM}
\label{a:NLSMamp}
Below, we list the amplitudes for the $\mathbb{CP}^1$ $\mathcal{N}=1$ supersymmetric NLSM.
This model is discussed in Section \ref{sec:SUSY_NLSM} as an illustration of our methods.

The 4-point amplitudes are:
\begin{align}
    \mathcal A_4 ( 1_Z 2_{\bar Z} 3_Z 4_{\bar Z}) & = \frac{1}{\Lambda^2} s_{13}\,,
    \label{eq:NLSM_A4_1} \\
    \mathcal A_4 ( 1_Z 2_{\bar Z} 3_\psi^+ 4_\psi^- ) & = - \frac{1}{\Lambda^2} \sq{23} \ang{24} = \frac{1}{2 \Lambda^2} \la 4 | p_1 - p_2 | 3 ]\,,
    \label{eq:NLSM_A4_2} \\
    \mathcal A_4 ( 1_\psi^+ 2_\psi^- 3_\psi^+ 4_\psi^- ) & = - \frac{1}{\Lambda^2} \sq{13} \ang{24}\,.
    \label{eq:NLSM_A4_3}
\end{align}
They serve as the input for computing the 6-point amplitudes recursively:
\begin{align}
    \mathclap{\mathcal A_6 ( 1_Z 2_{\bar Z} 3_Z 4_{\bar Z} 5_Z 6_{\bar Z} )} & \nonumber \\
    = \frac{1}{\Lambda^4} \bigg [ & \left ( \frac{s_{13}s_{46}}{p_{123}^2} + ( 1 \leftrightarrow 5 ) + ( 3 \leftrightarrow 5 ) \right ) + ( 2 \leftrightarrow 4 ) + ( 2 \leftrightarrow 6 ) + 3 p_{135}^2 \bigg ]\,,
    \label{eq:NLSM_A6_1} \\
   \mathclap{\mathcal A_6 ( 1_Z 2_{\bar Z} 3_Z 4_{\bar Z} 5_\psi^+ 6_\psi^- )} & \nonumber \\
   \nonumber = \frac{1}{\Lambda^4} \bigg [ & \left ( \frac{s_{13} \sq{54} \ang{46}}{p_{123}^2} + ( 2 \leftrightarrow 4 ) \right ) - \left ( \frac{s_{24} \sq{51} \ang{16}}{p_{156}^2} + ( 1 \leftrightarrow 3 ) \right ) \\
    & - \left ( \left ( \frac{\sq{54} \la 4 | p_{126} | 2 ] \ang{26}}{p_{126}^2} + (1 \leftrightarrow 3 ) \right ) + ( 2 \leftrightarrow 4 ) \right ) + \la 6 | p_{135} | 5 ] \bigg ]\,,
    \label{eq:NLSM_A6_2}\\
    \mathclap{\mathcal A_6 ( 1_Z 2_{\bar Z} 3_\psi^+ 4_\psi^- 5_\psi^+ 6_\psi^- )} & \nonumber \\
    \nonumber = \frac{1}{\Lambda^4} \bigg [ & - \bigg ( \frac{\sq{31} \la 1 | p_{123} | 5 ] \ang{46}}{p_{123}^2} - ( 3 \leftrightarrow 5 ) \bigg ) + \bigg ( \frac{\sq{35} \la 4 | p_{126} | 2 ] \ang{26}}{p_{126}^2} - ( 4 \leftrightarrow 6 ) \bigg ) \\
    & - \bigg ( \bigg ( \frac{\sq{51} \ang{16} \sq{32} \ang{24}}{p_{156}^2} - ( 3 \leftrightarrow 5 ) \bigg ) - ( 4 \leftrightarrow 6 ) \bigg ) \bigg ]\,,
    \label{eq:NLSM_A6_3} \\
    \mathclap{\mathcal A_6 ( 1_\psi^+ 2_\psi^- 3_\psi^+ 4_\psi^- 5_\psi^+ 6_\psi^- )} & \nonumber \\
    = \frac{1}{\Lambda^4} \bigg [ & \bigg ( \frac{\sq{13} \la 2 | p_{123} | 5 ] \ang{46}}{p_{123}^2} - ( 1 \leftrightarrow 5 ) - ( 3 \leftrightarrow 5 ) \bigg ) - ( 2 \leftrightarrow 4 ) - ( 2 \leftrightarrow 6 ) \bigg ]
    \,.
    \label{eq:NLSM_A6_4}
\end{align}
Note that only the pure scalar amplitudes and the 2-fermion amplitudes have local terms. 
The 6-point amplitudes satisfy the NMHV supersymmetry Ward identities in \reef{WI6-1}-\reef{WI6-3}.

\subsection{Supersymmetric Dirac-Born-Infeld Theory}
\label{a:susydbi}
The amplitudes of $\mathcal{N}=1$ supersymmetric Dirac-Born-Infeld theory are all recursively constructible. The 4-point amplitudes are
\begin{align}
\mathcal A_4 ( 1_Z 2_{\bar Z} 3_Z 4_{\bar Z}) & = \frac{1}{\Lambda^4} s_{13}^2\,,
\label{eq:DBI_A4_1} \\
\mathcal A_4 ( 1_Z 2_{\bar Z} 3_\psi^+ 4_\psi^- ) & = \frac{1}{\Lambda^4} s_{13} \sq{32} \ang{24} = \frac{1}{2 \Lambda^4} s_{13} \la 4 | p_1 - p_2 | 3 ]\,,
\label{eq:DBI_A4_2} \\
\mathcal A_4 ( 1_\psi^+ 2_\psi^- 3_\psi^+ 4_\psi^- ) & = - \frac{1}{\Lambda^4} s_{13} \sq{13} \ang{24}\,.
\label{eq:DBI_A4_3}
\end{align}
and the results of soft subtracted recursion for the 6-point amplitudes are
\begin{align}
\mathclap{\mathcal A_6 ( 1_Z 2_{\bar Z} 3_Z 4_{\bar Z} 5_Z 6_{\bar Z} )} \nonumber \\
= \frac{1}{\Lambda^8} & \bigg [ \left ( \frac{s_{13}^2 s_{46}^2}{p_{123}^2} + ( 1 \leftrightarrow 5 ) + ( 3 \leftrightarrow 5 ) \right ) + ( 2 \leftrightarrow 4 ) + ( 2 \leftrightarrow 6 ) - p_{135}^6 \bigg ]\,,
\label{eq:DBI_A6_1} \\
\mathclap{\mathcal A_6 ( 1_Z 2_{\bar Z} 3_Z 4_{\bar Z} 5_\psi^+ 6_\psi^- )} & \nonumber \\
= \frac{1}{\Lambda^8} & \bigg [ \bigg ( \bigg ( \frac{s_{26} s_{35} \sq{54} \la 4 | p_{126} | 1 ] \ang{16}}{p_{126}^2} + (1 \leftrightarrow 3 ) \bigg ) + ( 2 \leftrightarrow 4 ) \bigg ) + \bigg ( \frac{s_{13}^2 s_{46} \sq{54} \ang{46}}{p_{123}^2} + ( 2 \leftrightarrow 4 ) \bigg ) \nonumber \\
& - \bigg ( \frac{s_{15} s_{24}^2 \sq{51} \ang{16}}{p_{156}^2} + ( 1 \leftrightarrow 3 ) \bigg ) + \left ( s_{13} s_{24} - \left ( s_{13} + s_{24} \right ) p_{135}^2 \right ) \la 6 | p_{24} | 5 ] \bigg ]\,,
\label{eq:DBI_A6_2} \\
\mathclap{\mathcal A_6 ( 1_Z 2_{\bar Z} 3_\psi^+ 4_\psi^- 5_\psi^+ 6_\psi^- )} & \nonumber \\
= \frac{1}{\Lambda^8} & \bigg [ \left ( s_{24} + s_{26} \right ) p_{135}^2 \sq{35} \ang{46} - \bigg ( \bigg ( \frac{s_{15} s_{24} \sq{51} \ang{16} \sq{32} \ang{24}}{p_{156}^2} - ( 3 \leftrightarrow 5 ) \bigg ) - ( 4 \leftrightarrow 6 ) \bigg ) \nonumber \\
& - \bigg ( \frac{s_{13} s_{46} \sq{32} \la 2 | p_{123} | 5 ] \ang{46}}{p_{123}^2} - ( 3 \leftrightarrow 5 ) \bigg ) + \bigg ( \frac{s_{26} s_{35} \sq{35} \la 4 | p_{126} | 2 ] \ang{26}}{p_{126}^2} - ( 4 \leftrightarrow 6 ) \bigg ) \bigg ]\,,
\label{eq:DBI_A6_3} \\
\mathclap{\mathcal A_6 ( 1_\psi^+ 2_\psi^- 3_\psi^+ 4_\psi^- 5_\psi^+ 6_\psi^- )} & \nonumber \\
= \frac{1}{\Lambda^8} & \bigg [ \bigg ( \frac{s_{13} s_{46} \sq{13} \la 2 | p_{123} | 5 ] \ang{46}}{p_{123}^2} - ( 1 \leftrightarrow 5 ) - ( 3 \leftrightarrow 5 ) \bigg ) - ( 2 \leftrightarrow 4 ) - ( 2 \leftrightarrow 6 ) \bigg ]\,.
\label{eq:DBI_A6_4}
\end{align}
The 6-point amplitudes satisfy the NMHV supersymmetry Ward identities in \reef{WI6-1}-\reef{WI6-3}. As in the case of the NLSM, only the pure scalar amplitudes and the 2-fermion amplitudes have local terms. 

\subsection{Supersymmetric Born-Infeld Theory}
\label{a:susybi}
In this subsection, we list the amplitudes of Born-Infeld theory.
This theory is the leading order contribution to the effective field theory of a Goldstone $\mathcal N = 1$ vector multiplet. The 4-point amplitudes are 
\begin{align}
\mathcal A_4 ( 1_\psi^+ 2_\psi^- 3_\psi^+ 4_\psi^- ) & = - \frac{1}{\Lambda^4} \sq{13} \ang{24} s_{13}\,,
\label{eq:BI_A4_1} \\
\mathcal A_4 ( 1_\psi^+ 2_\psi^- 3_\gamma^+ 4_\gamma^- ) & = \frac{1}{\Lambda^4} \sq{13} \sq{23} \ang{24}^2 = - \frac{1}{2 \Lambda^4} [ 13 ] \la 4 | p_1 - p_2 | 3 ] \ang{24}\,,
\label{eq:BI_A4_2} \\
\mathcal A_4 ( 1_\gamma^+ 2_\gamma^- 3_\gamma^+ 4_\gamma^- ) & = \frac{1}{\Lambda^4} \sq{13}^2 \ang{24}^2\,.
\label{eq:BI_A4_3}
\end{align}
Except for the all-vector amplitudes, all amplitudes are constructible with soft subtracted recursion. The all-vector amplitudes are the amplitudes of Born-Infeld theory, and they are fixed in terms of the other amplitudes using the supersymmetry Ward identities. In particular, at 6-points, we use \reef{WI6-3} and the remaining five identities in \reef{WI6-1}-\reef{WI6-3} are used as checks. The results are
\begin{align}
\mathclap{\mathcal A_6 ( 1_\psi^+ 2_\psi^- 3_\psi^+ 4_\psi^- 5_\psi^+ 6_\psi^- )} & \nonumber \\
= \frac{1}{\Lambda^8} & \bigg [ \bigg ( \frac{s_{13} s_{46} \sq{13} \la 2 | p_{123} | 5 ] \ang{46}}{p_{123}^2} - ( 1 \leftrightarrow 5 ) - ( 3 \leftrightarrow 5 ) \bigg ) - ( 2 \leftrightarrow 4 ) - ( 2 \leftrightarrow 6 ) \bigg ]\,,
\label{eq:BI_A6_1} \\
\mathclap{\mathcal A_6 ( 1_\gamma^+ 2_\gamma^- 3_\psi^+ 4_\psi^- 5_\psi^+ 6_\psi^- )} & \nonumber \\
= \frac{1}{\Lambda^8} & \bigg [ \bigg (\frac{ s_{46} \sq{13}^2 \la 2 | p_{123} | 5 ] \ang{23} \ang{46}}{p_{123}^2} - ( 3 \leftrightarrow 5 ) \bigg ) + \bigg ( \frac{s_{35} \sq{14} \sq{35} \la 6 | p_{124} | 1 ] \ang{24}^2}{p_{124}^2} - ( 4 \leftrightarrow 6 ) \bigg ) \nonumber \\
& - \bigg ( \bigg ( \frac{\sq{13} \sq{14} \la 4 | p_{134} | 5 ]^2 \ang{52} \ang{26}}{p_{134}^2} - \sq{13} \la 2 | p_{35} | 1 ] \la 6 | p_{46} | 5 ] \ang{24} - ( 3 \leftrightarrow 5 ) \bigg ) - ( 4 \leftrightarrow 6 ) \bigg ) \bigg ]\,,
\label{eq:BI_A6_2} \\
\mathclap{\mathcal A_6 ( 1_\gamma^+ 2_\gamma^- 3_\gamma^+ 4_\gamma^- 5_\psi^+ 6_\psi^- )} & \nonumber \\
= \frac{1}{\Lambda^8} & \bigg [ \bigg ( \frac{\sq{13}^2 \la 2 | p_{123} | 5 ]^2 \ang{54} \ang{46}}{p_{123}^2} + ( 2 \leftrightarrow 4 ) \bigg ) + \bigg ( \frac{\sq{35} \sq{36} \la 6 | p_{124} | 1 ]^2 \ang{24}^2}{p_{124}^2} + ( 1 \leftrightarrow 3 ) \bigg ) \nonumber \\
& + \bigg ( \bigg ( \frac{\sq{15}^2 \sq{36} \la 2 | p_{125} | 3 ] \ang{25} \ang{46}^2}{p_{125}^2} + ( 1 \leftrightarrow 3 ) \bigg ) + ( 2 \leftrightarrow 4 ) \bigg ) + \sq{13}^2 \la 6 | p_{24} | 5 ] \ang{24}^2 \bigg ]\,,
\label{eq:BI_A6_3} \\
\mathclap{\mathcal A_6 ( 1_\gamma^+ 2_\gamma^- 3_\gamma^+ 4_\gamma^- 5_\gamma^+ 6_\gamma^- )} & \nonumber \\
= \frac{1}{\Lambda^8} & \bigg [ \bigg ( \frac{[ 13 ]^2 \la 2 | p_{123} | 5 ]^2 \la 46 \ra^2}{p_{123}^2} + ( 1 \leftrightarrow 5 ) + ( 3 \leftrightarrow 5 ) \bigg ) + ( 2 \leftrightarrow 4 ) + ( 2 \leftrightarrow 6 ) \bigg ]\,.
\label{eq:BI_A6_4}
\end{align}
In this case, only ${\mathcal A_6 ( 1_\gamma^+ 2_\gamma^- 3_\gamma^+ 4_\gamma^- 5_\psi^+ 6_\psi^- )}$ has local terms.

\subsection{Supersymmetric Quartic Galileon Theory}
\label{susygal4amplitudes}
Below, we list the amplitudes of an $\mathcal N = 1$ supersymmetric quartic Galileon. This model 
was discussed in detail in \cite{Elvang:2017mdq} and reviewed in Section \ref{sec:Galileon}. 
The 4-point amplitudes are
\begin{align}
    \mathcal A_4 ( 1_Z 2_{\bar Z} 3_Z 4_{\bar Z}) & = \frac{1}{\Lambda^6} s_{12} s_{13} s_{23}\,,
    \label{eq:Gal_A4_1} \\
    \mathcal A_4 ( 1_Z 2_{\bar Z} 3_\psi^+ 4_\psi^- ) & = \frac{1}{\Lambda^6} s_{12} s_{23} \sq{32} \ang{24} = \frac{1}{2 \Lambda^6} s_{12} s_{23} \la 4 | p_1 - p_2 | 3 ]\,,
    \label{eq:Gal_A4_2} \\
    \mathcal A_4 ( 1_\psi^+ 2_\psi^- 3_\psi^+ 4_\psi^- ) & = - \frac{1}{\Lambda^6} \sq{13} \ang{24} s_{12} s_{23}\,.
    \label{eq:Gal_A4_3}
\end{align}
At 6-point, only the amplitudes with at most two fermions are constructible with soft subtracted recursion relations. The remaining ones are fixed by the supersymmetry Ward identities \reef{WI6-1}-\reef{WI6-3}, and we find 
\begin{align}
    \mathclap{\mathcal A_6 ( 1_Z 2_{\bar Z} 3_Z 4_{\bar Z} 5_Z 6_{\bar Z} )} & \nonumber \\
    = \frac{1}{\Lambda^{12}} \bigg [ & \bigg ( \frac{s_{12} s_{13} s_{23} s_{45} s_{46} s_{56}}{p_{123}^2} + ( 1 \leftrightarrow 5 ) + ( 3 \leftrightarrow 5 ) \bigg ) + ( 2 \leftrightarrow 4 ) + ( 2 \leftrightarrow 6 ) \bigg ]\,,
    \label{eq:Gal_A6_1} \\
   \mathclap{\mathcal A_6 ( 1_Z 2_{\bar Z} 3_Z 4_{\bar Z} 5_\psi^+ 6_\psi^- )} & \nonumber \\
   = \frac{1}{\Lambda^{12}} \bigg [ & \bigg ( \frac{s_{12} s_{13} s_{23} s_{45} s_{56} \sq{54} \ang{46}}{p_{123}^2} + ( 2 \leftrightarrow 4 ) \bigg ) - \bigg ( \frac{s_{16} s_{23} s_{24} s_{34} s_{56} \sq{51} \ang{16}}{p_{156}^2} + ( 1 \leftrightarrow 3 ) \bigg ) \nonumber \\
    & + \bigg ( \bigg ( \frac{s_{12} s_{16} s_{34} s_{45} \sq{53} \la 3 | p_{126} | 2 ] \ang{26}}{p_{126}^2} + (1 \leftrightarrow 3 ) \bigg ) + ( 2 \leftrightarrow 4 ) \bigg ) \bigg ]\,,
    \label{eq:Gal_A6_2} \\
    \mathclap{\mathcal A_6 ( 1_Z 2_{\bar Z} 3_\psi^+ 4_\psi^- 5_\psi^+ 6_\psi^- )} & \nonumber \\
    \nonumber = \frac{1}{\Lambda^{12}} \bigg [ & \bigg ( \frac{\sq{31} \langle 1 | p_{46} | 5 ] \ang{46}}{p_{123}^2} - ( 3 \leftrightarrow 5 ) \bigg ) + \bigg ( \frac{\sq{35} \langle 4 | p_{16} | 2 ] \ang{26}}{p_{126}^2} - ( 4 \leftrightarrow 6 ) \bigg ) \\
    & \qquad - \bigg ( \bigg ( \frac{\sq{32} \ang{24} \sq{51} \ang{16}}{p_{156}^2} - ( 3 \leftrightarrow 5 ) \bigg ) - ( 4 \leftrightarrow 6 ) \bigg ) \bigg ]\,,
    \label{eq:Gal_A6_3} \\
    \mathclap{\mathcal A_6 ( 1_\psi^+ 2_\psi^- 3_\psi^+ 4_\psi^- 5_\psi^+ 6_\psi^- )} & \nonumber \\
    = \frac{1}{\Lambda^{12}} \bigg [ & \bigg ( \frac{\sq{13} \langle 2 | p_{13} | 5 ] \ang{46}}{p_{123}^2} - ( 1 \leftrightarrow 5 ) - ( 3 \leftrightarrow 5 ) \bigg ) - ( 2 \leftrightarrow 4 ) - ( 2 \leftrightarrow 6 ) \bigg ]
    \,.
    \label{eq:Gal_A6_4}
\end{align}
None of the amplitudes have local terms.

\subsection{Chiral Perturbation Theory}

Below, we list the color-ordered amplitudes of the $\frac{U ( N ) \times U ( N )}{U ( N )}$ sigma model, with higher derivative corrections, referred to as chiral perturbation theory in the main text.
Different color orderings are related to the ones listed by momentum relabelling.
At 4-point we have 
\begin{equation}
    \mathcal A_4 [ 1, 2, 3, 4 ] = \frac{g_2}{\Lambda^2} t + \frac{g_6}{\Lambda^6} t \left ( s^2 + t^2 + u^2 \right ) + \frac{g_8}{\Lambda^8} s t^2 u + \mathcal O ( \Lambda^{-10} )
    \label{eq:chiPT_A4}
\end{equation}
and at 6-point
\begin{align}
    & \mathcal A_6 [ 1, 2, 3, 4, 5, 6 ] \nonumber \\
    & \ = \frac{g_2^2}{\Lambda^4} \bigg [ \frac{s_{13} s_{46}}{p_{123}^2} + \frac{s_{24} s_{15}}{p_{234}^2} + \frac{s_{35} s_{26}}{p_{345}^2} - s_{24} - s_{26} - s_{46} \bigg ] \nonumber \\
    & \ + \frac{g_2 g_6}{\Lambda^8} \bigg [ \frac{s_{13} s_{46}}{p_{123}^2} \left ( s_{12}^2 + s_{13}^2 + s_{23}^2 + s_{45}^2 + s_{46}^2 + s_{56}^2 \right ) \nonumber \\
    & \qquad + \frac{s_{24} s_{15}}{p_{234}^2} \left ( s_{23}^2 + s_{24}^2 + s_{34}^2 + s_{56}^2 + s_{15}^2 + s_{16}^2 \right ) + \frac{s_{35} s_{26}}{p_{345}^2} \left ( s_{34}^2 + s_{35}^2 + s_{45}^2 + s_{16}^2 + s_{26}^2 + s_{12}^2 \right ) \nonumber \\
    & \qquad - 2 \big (s_{26}^3 + s_{23} s_{26}^2 + s_{25} s_{26}^2 + s_{34} s_{26}^2 + s_{45} s_{26}^2 + s_{23}^2 s_{26} + s_{25}^2 s_{26} + s_{34}^2 s_{26} + s_{35}^2 s_{26} + s_{45}^2 s_{26} \nonumber \\
    & \qquad \qquad + s_{23} s_{34} s_{26} + s_{23} s_{35} s_{26} + s_{25} s_{35} s_{26} + s_{34} s_{36} s_{26} + s_{23} s_{45} s_{26} + s_{34} s_{45} s_{26} + s_{36} s_{45} s_{26} \nonumber \\
    & \qquad \qquad + s_{46}^3 + s_{24} s_{25}^2 + s_{24} s_{35}^2 + s_{24} s_{45}^2 + s_{23} s_{46}^2 + s_{25} s_{46}^2 + s_{34} s_{46}^2 + s_{35} s_{46}^2 + s_{36} s_{46}^2 \nonumber \\
    & \qquad \qquad + s_{45} s_{46}^2 + s_{24} s_{35} s_{36} + s_{25}^2 s_{46} + s_{34}^2 s_{46} + s_{35}^2 s_{46} + s_{36}^2 s_{46} + s_{45}^2 s_{46} + s_{23} s_{25} s_{46} \nonumber \\
    & \qquad \qquad + s_{25} s_{34} s_{46} + s_{23} s_{45} s_{46} + s_{34} s_{45} s_{46} + s_{35} s_{45} s_{46} + s_{36} s_{45} s_{46} \big ) \nonumber \\
    & \qquad - 4 \big ( s_{24}^3 + s_{25} s_{24}^2 + s_{35} s_{24}^2 + s_{45} s_{24}^2 + s_{23}^2 s_{24} + s_{34}^2 s_{24} + s_{36}^2 s_{24} + s_{23} s_{25} s_{24} + s_{25} s_{34} s_{24} \nonumber \\
    & \qquad \qquad + s_{23} s_{35} s_{24} + s_{25} s_{35} s_{24} + s_{34} s_{35} s_{24} + s_{26} s_{36} s_{24} + s_{23} s_{45} s_{24} + s_{25} s_{45} s_{24} + s_{34} s_{45} s_{24} \nonumber \\
    & \qquad \qquad + s_{35} s_{45} s_{24} + s_{36} s_{45} s_{24} + s_{23} s_{25} s_{26} + s_{25} s_{26} s_{34} + s_{25} s_{26} s_{45} + s_{23}^2 s_{46} + s_{25} s_{26} s_{46} \nonumber \\
    & \qquad \qquad + s_{23} s_{34} s_{46} + s_{23} s_{35} s_{46} + s_{34} s_{35} s_{46} + s_{23} s_{36} s_{46} + s_{25} s_{36} s_{46} + s_{26} s_{36} s_{46} + s_{34} s_{36} s_{46} \nonumber \\
    & \qquad \qquad + s_{35} s_{36} s_{46} + s_{25} s_{45} s_{46} + s_{26} s_{45} s_{46} \big ) \nonumber \\
    & \qquad - 6 \big ( s_{23} s_{24}^2 + s_{34} s_{24}^2 + s_{36} s_{24}^2 + s_{26}^2 s_{24} + s_{46}^2
   s_{24} + s_{23} s_{26} s_{24} + s_{25} s_{26} s_{24} + s_{23} s_{34} s_{24} \nonumber \\
   & \qquad \qquad + s_{26} s_{34} s_{24} + s_{23} s_{36} s_{24} + s_{25} s_{36} s_{24} + s_{26} s_{45} s_{24} + s_{25} s_{46} s_{24} + s_{35} s_{46} s_{24} + s_{45} s_{46} s_{24} \nonumber \\
   & \qquad \qquad + s_{26} s_{46}^2 + s_{25} s_{34} s_{36} + s_{25} s_{36} s_{45} + s_{26}^2 s_{46} + s_{23} s_{26} s_{46} + s_{26} s_{34} s_{46} \big ) \nonumber \\
   & \qquad - 8 s_{24} \big ( s_{24} s_{26} + s_{34} s_{36} + s_{23} s_{46} + s_{24} s_{46} + s_{34} s_{46} + s_{36} s_{46} \big ) -12 s_{24} s_{26} s_{46} \bigg ] + \mathcal O ( \Lambda^{-10} ) \,.
   \label{eq:chiPT_A6}
\end{align}
These amplitudes are discussed in further detail in Section \ref{s:doublecopy}.

\section{Recursion Relations and Ward Identities}
\label{RecWardApp}
We show that if the seed amplitudes of a recursive theory satisfy a set of Ward identities, then all recursively constructible $n$-point amplitudes also satisfy them. For Abelian groups, this follows from two features: 
\begin{itemize}
\item[(a)] additive charges have Ward identities that simply state that the sum of charges of the states in an amplitude must vanish.
\item[(b)] CPT conjugate states sitting on either end of a factorization channel have equal and opposite charges.
\end{itemize}
Hence recursion will result in amplitudes that respect the Abelian symmetry so long as the seed amplitudes do. 

Now consider  Ward identities generated by elements of a semi-simple Lie algebra. In the root space decomposition of the algebra, we can choose a triplet of generators: raising operators 
$\mathcal{T}_+$, lowering operators $\mathcal{T}_-$, and ``diagonal'' $\mathcal{T}_0$ generators, for each positive root that satisfy the algebra
\be
  [\mathcal{T}_+, \mathcal{T}_-]=\mathcal{T}_0\,,~~~~
  [\mathcal{T}_+,\mathcal{T}_0]=-2\mathcal{T}_+\,,~~~~
  [\mathcal{T}_-,\mathcal{T}_0]=2\mathcal{T}_-\,.
\ee
In order for representations of this algebra to be physical, CPT must be an algebra automorphism. The CPT charge conjugation generator $\mathcal{C}$ must also flip the sign of the additive 
$\mathcal{T}_0$-charge. So we determine the action of $\mathcal{C}$ to be
\begin{align}
\mathcal{C}\cdot \mathcal{T}_0\cdot X&=-\mathcal{T}_0\cdot \mathcal{C}\cdot X=-\mathcal{T}_0\cdot\tilde{X}\,,\nonumber\\
\mathcal{C}\cdot \mathcal{T}_+\cdot X&=-\mathcal{T}_-\cdot \mathcal{C}\cdot X=-\mathcal{T}_-\cdot\tilde{X}\,,
\label{Caction}\\
\mathcal{C}\cdot \mathcal{T}_-\cdot X&=-\mathcal{T}_+\cdot \mathcal{C}\cdot X=-\mathcal{T}_+\cdot\tilde{X}\,,\nonumber
\end{align}
where $X$ is a physical state and we have defined the conjugate state $\tilde{X}$ to be the charge conjugate of $X$, i.e.~$\tilde{X}=\mathcal{C}\cdot X$.

If the S-matix is recursively constructible (at some order in the derivative expansion) then each $n$-point amplitude is given as a sum over factorization singularities with residues given in terms of a product of amplitudes with fewer external states
\begin{align}
\mathcal{A}_n(1,\cdots,n)=\sum_I \sum_{X} \underset{z=z_I^\pm}{\text{Res}}\ \frac{\hat{\mathcal{A}}^{(I)}_L(z)\hat{\mathcal{A}}^{(I)}_R(z)}{z\hat{P}_I(z)^2F(z)}\,,
\end{align} 
where $I$ labels all possible factorization channels and 
$X$ the exchanged internal states. 
Since $\mathcal{T}_0$ is diagonal, the Ward identity generated by $\mathcal{T}_0$ works just like in the Abelian case -- charges can be assigned to the physical states and recursion preserves this charge in any $n$-point amplitude. More complicated are the non-diagonal generators $\mathcal{T}_\pm$. For simplicity, we present the argument explicitly for $SU(2)_R$ Ward identities as they apply to the $\mathcal{N}=2$ NLSM described in Section \ref{s:N2NLSM}. For $SU(2)_R$, the action of $\mathcal{T}_+$ on the fermion helicity states is given in \reef{Tplusaction}. The scalar and vectors are singlets under $SU(2)_R$.

The statement of the $SU(2)_R$ Ward identity is that  $\mathcal{T}_+ \cdot \mathcal{A}_n(1,...,n)=0$. The inductive assumption is that this holds true for the lower-point amplitudes in the recursive expression for $\mathcal{A}_n(1,...,n)$. We already know from Section \ref{s:N2NLSM} that   $SU(2)_R$ is a symmetry of the 3- and 4-point amplitudes, so that provides the basis of induction. 

The action of $\mathcal{T}_+$ on the recursive expression for an $n$-point amplitude is
\bea
\label{TplusWI1}
  \mathcal{T}_+ \cdot \mathcal{A}_n(1,...,n) 
  &\equiv& \sum_{i=1}^n (-1)^{P_i}\mathcal{A}_n(1,...,\mathcal{T}_+\cdot i,...,n)  \\\nonumber
  &&\hspace{-40mm}= \sum_I \sum_{X} \underset{z=z_I^\pm}{\text{Res}}\left[ \sum_{i\in I}(-1)^{P_i}\frac{\hat{\mathcal{A}}^{(I)}_L(\ldots,\mathcal{T}_+\cdot i,\ldots,X)\hat{\mathcal{A}}^{(I)}_R(\ldots)}{z\hat{P}_I(z)^2F(z)}\right.\nonumber\\
  &&\hspace{-30mm}\left.+\sum_{i\notin I}(-1)^{P_i}\frac{\hat{\mathcal{A}}^{(I)}_L(\ldots)\hat{\mathcal{A}}^{(I)}_R(\tilde X,\ldots,\mathcal{T}_+\cdot i,\ldots)}{z\hat{P}_I(z)^2F(z)}\right],
\eea
where $P_i= 0$ or $1$ corresponds to the additional signs in the prefactors for the action of $\mathcal{T}_+$ as given in Table \ref{Tplusaction}. We now prove that this expression vanishes channel by channel. Without loss of generality, we will show that the contribution from the $(1\ldots k)^\pm$ channel vanishes independently, where $+$ means the contribution from the $z^\pm$ residue. The argument follows for all other factorization channels by replacing $(1\ldots k)^\pm$ by $I^\pm$. For the $(1\ldots k)$-channel, the relevant part of \reef{TplusWI1} that we want to show vanishes is 
\begin{align}
\label{Tcalc}
&\sum_{X}\bigg[\bigg(\sum_{i=1}^{k}(-1)^{P_i}\hat{\mathcal{A}}_{L}(1,\ldots,\mathcal{T}_+\cdot i,\ldots,k,X)\bigg) \hat{\mathcal{A}}_{R}(\tilde X,k+1,\ldots,n)\nonumber\\
&~~~~~~~~+
\hat{\mathcal{A}}_{L}(1,\ldots,k,X)\bigg(\sum_{i=k+1}^{n}
(-1)^{P_i}\hat{\mathcal{A}}_{R}(\tilde X,k+1,\ldots,\mathcal{T}_+\cdot i,\ldots, n)\bigg)\bigg].
\end{align}
By the inductive assumption, the lower-point amplitudes respect the $\mathcal{T}_+$ Ward identities 
\be
\sum_{i=1}^{k}(-1)^{P_i}\hat{\mathcal{A}}_{L}(1,\ldots,\mathcal{T}_+\cdot i,\ldots,k,X)
~=~(-1)^{P_X+1}\hat{\mathcal{A}}_{L}(1,\ldots, k, \mathcal{T}_+\cdot X)\,,
\ee
and similarly for $\hat{\mathcal{A}}_R$. Using this relation and splitting the sum over particles $X$ allows us to rewrite (\ref{Tcalc}) as
\be
\label{XXp}
\begin{split}
 &-\sum_{X}(-1)^{P_X}\Big[\hat{\mathcal{A}}_{L}(1,\ldots, k, \mathcal{T}_+\cdot X) \hat{\mathcal{A}}_{R}(\tilde X, k+1,\ldots,n) \Big]\\
&-\sum_{X'}(-1)^{P_{\tilde{X}'}}\Big[\hat{\mathcal{A}}_{L}(1,\ldots,k,X')\hat{\mathcal{A}}_{R}(\mathcal{T}_+\cdot\tilde X',k+1,\ldots,n)\Big]\,.
\end{split}
\ee
In the second line we have  made a change of dummy summation variable that we now exploit further. 

It is non-trivial, but turns out to be true for $SU(2)_R$ as we have explicitly checked, that if we define 
$X'=\mathcal{T}_+\cdot X$ and sum over $X$ instead of $X'$, the second line of \reef{XXp} gives exactly the same result. We can then write \reef{XXp} as
\be
\begin{split}
-\sum_{X}\Big[(-1)^{P_X}&\hat{\mathcal{A}}_{L}(1,\ldots, k, \mathcal{T}_+\cdot X) \hat{\mathcal{A}}_{R}(\tilde X,k+1,\ldots,n)\\
&+(-1)^{P_{\tilde{X}'}}\hat{\mathcal{A}}_{L}(1,\ldots,k,\mathcal{T}_+\cdot X)\hat{\mathcal{A}}_{R}(\mathcal{T}_+\cdot C\cdot \mathcal{T}_+\cdot X,k+1,\ldots,n)\Big]\,.
\end{split}
\ee
Since $\mathcal{T}_+\cdot C\cdot \mathcal{T}_+\cdot X=\mathcal{T}_+\cdot \mathcal{T}_-\cdot\tilde{X}$, this becomes
\be
\begin{split}
-\sum_{X}\Big[&(-1)^{P_X}\hat{\mathcal{A}}_{L}(1,\ldots, k, \mathcal{T}_+\cdot X) \hat{\mathcal{A}}_{R}(\tilde X,k+1,\ldots,n)\\
&+(-1)^{P_{\mathcal{T}_-\cdot\tilde{X}}+Q_{\tilde{X}}+1}\hat{\mathcal{A}}_{L}(1,\ldots,k,\mathcal{T}_+\cdot X)\hat{\mathcal{A}}_{R}(\mathcal{T}_+\cdot \mathcal{T}_-\cdot\tilde{X},k+1,\ldots,n)\Big]\,.
\end{split}
\ee
where $Q_X$ refers to the prefactors for the action of $\mathcal{T}_-$ as given in Table \ref{Tplusaction}. This vanishes when $\mathcal{T}_+\cdot \mathcal{T}_-\cdot\tilde{X} = \tilde{X}$ and $P_{\mathcal{T}_-\cdot\tilde{X}}+Q_{\tilde{X}}=0$ for any state $X$ such that $\mathcal{T}_+\cdot X \ne 0$. 
For $SU(2)_R$, we can check explicitly that these conditions are satisfied. The only states for which $\mathcal{T}_+\cdot X \ne 0$ are $X = \psi^{2+}$ and $\psi^{-}_1$. Their conjugates are $\tilde{X} = \psi^{-}_2$ and $\psi^{2+}$, respectively, and by \reef{Tplusaction} we have  
\begin{align}
&\mathcal{T}_+\cdot \mathcal{T}_- \cdot \psi^{1+}=\mathcal{T}_+ \cdot\psi^{2+}=\psi^{1+} &&
\mathcal{T}_+\cdot \mathcal{T}_- \cdot \psi^{-}_2=\mathcal{T}_+ \cdot\psi^-_1 =\psi^-_2,\\
&P_{\mathcal{T}_-\cdot\psi^{1+}}+Q_{\psi^{1+}}=0+0=0 &&
P_{\mathcal{T}_-\cdot\psi^-_2}+Q_{\psi^-_2}=1+1=0 \text{ (mod }2)\,.
\end{align}
If follows that from the inductive step that all amplitudes satisfy the $SU(2)_R$ Ward identities when the seed amplitudes do.

\section{Amplitude Relations in $\mathcal{N}=2$ $\mathds{C}\mathds{P}^1$ NLSM }
\label{WIN2}

Below are explicit formulae, derived from $\mathcal{N}=2$ supersymmetry Ward identities, for all amplitudes in this model with total spin $\leq 2$ expressed as linear combinations of amplitudes with strictly greater total spin. Collectively these formulae allow us to construct every tree-level amplitude in the $\mathcal{N}=2$ $\mathds{C}\mathds{P}^1$ sigma model using unsubtracted recursion. The needed relations are:

\begin{align}
  &\mathcal{A}_{2n}\left(1_Z,2_{\bar{Z}},3_Z,4_{\bar{Z}}...,(2n)_{\bar{Z}}\right) = \sum_{k=1}^{n-1}\frac{\langle 1,2k+1\rangle}{\langle 12\rangle}\mathcal{A}_{2n}\left(1_Z,2_{\psi_1}^-,3_Z,4_{\bar{Z}},...,(2k+1)_{\psi^1}^+,...,(2n)_{\bar{Z}}\right)\nonumber\\
  &\mathcal{A}_{2n}\left(1_{\psi^1}^+,2_{\psi_1}^-,3_Z,4_{\bar{Z}},...,(2n)_{\bar{Z}}\right) = \sum_{k=1}^{n-1}\frac{[2,2k+2]}{[21]}\mathcal{A}_{2n}\left(1_\gamma^+,2_{\psi_1}^-,3_Z,4_{\bar{Z}},...,(2k+2)_{\psi_2}^-,...,(2n)_{\bar{Z}}\right) \nonumber\\
  &\mathcal{A}_{2n+1}\left(1_\gamma^+,2_\gamma^+,3_Z,4_{\bar{Z}},...,(2n+1)_Z\right)  \nonumber\\
  &\hspace{40mm}=\sum_{k=1}^{n-2}\frac{\langle 3,2k+3 \rangle}{\langle 34 \rangle} \mathcal{A}_{2n+1}\left(1_\gamma^+,2_\gamma^+,3_Z,4_{\psi_2}^-,5_Z,...,(2k+3)_{\psi^2}^+,...,(2n+1)_Z\right) \nonumber\\
  &\mathcal{A}_{2n}\left(1_{\psi^1}^+,2_\gamma^-,3_{\psi^2}^+,4_{\bar{Z}},5_Z,...,(2n)_{\bar{Z}}\right) = \sum_{k=1}^{n-1}\frac{[3,2k+2]}{[31]}\mathcal{A}_{2n}\left(1_\gamma^+,2_\gamma^-,3_{\psi^2}^+,4_{\bar{Z}},...,(2k+2)_{\psi_2}^-,...,(2n)_{\bar{Z}}\right) \nonumber\\
  &\mathcal{A}_{2n}\left(1_\gamma^+,2_\gamma^-,3_Z,4_{\bar{Z}},5_Z,...,(2n)_{\bar{Z}}\right) = \sum_{k=1}^{n-1}\frac{[1,2k+2]}{[13]}\mathcal{A}_{2n}\left(1_\gamma^+,2_\gamma^-,3_{\psi^2}^+,4_{\bar{Z}},...,(2k+2)_{\psi_2}^-,...,(2n)_{\bar{Z}}\right) \nonumber\\
  &\mathcal{A}_{2n+1}\left(1_\gamma^+,2_{\psi^1}^+,3_{\psi^2}^+,4_Z,5_{\bar{Z}},...,(2n+1)_{\bar{Z}}\right) = -\frac{\langle 42 \rangle}{\langle 45\rangle} \mathcal{A}_{2n+1}\left(1_\gamma^+,2_\gamma^+,3_{\psi^2}^+,4_Z,5_{\psi_2}^-,6_Z,...,(2n+1)_{\bar{Z}}\right) \nonumber\\
  &\hspace{40mm}+ \sum_{k=1}^{n-2}\frac{\langle 4,2k+4\rangle}{\langle 45 \rangle }\mathcal{A}_{2n+1}\left(1_\gamma^+,2_{\psi^1}^+,3_{\psi^3}^+,4_Z,5_{\psi_2}^-,6_Z,...,(2k+4)_{\psi^2}^+,...,(2n+1)_{\bar{Z}}\right) \nonumber\\
  &\mathcal{A}_{2n}\left(1_{\psi^1}^+,2_{\psi_1}^-,3_{\psi^2}^+,4_{\psi_2}^-,5_Z,6_{\bar{Z}},...,(2n)_{\bar{Z}}\right) = \frac{[32]}{[31]} \mathcal{A}_{2n}\left(1_\gamma^+,2_\gamma^-,3_{\psi^2}^+,4_{\psi_2}^-,5_Z,6_{\bar{Z}},...,(2n)_{\bar{Z}}\right) \nonumber\\
  &\hspace{40mm}+ \sum_{k=1}^{n-2}\frac{[3,2k+4]}{[31]}\mathcal{A}_{2n}\left(1_\gamma^+,2_{\psi_1}^-,3_{\psi^2}^+,4_{\psi_2}^-,5_Z,...,(2k+4)_{\psi_2}^-,...,(2n)_{\bar{Z}}\right)\nonumber\\
  &\mathcal{A}_{2n}\left(1_{\psi^1}^+,2_{\psi_1}^-,3_{\psi^1}^+,4_{\psi_1}^-,5_Z,6_{\bar{Z}},...,(2n)_{\bar{Z}}\right) = \frac{[42]}{[41]}\mathcal{A}_{2n}\left(1_\gamma^+,2_\gamma^-,3_{\psi^1}^+,4_{\psi_1}^-,5_Z,6_{\bar{Z}},...,(2n)_{\bar{Z}}\right) \nonumber\\
  &\hspace{40mm}+ \sum_{k=1}^{n-2}\frac{[4,2k+4]}{[41]}\mathcal{A}_{2n}\left(1_\gamma^+,2_{\psi_1}^-,3_{\psi^1}^+,4_{\psi_1}^-,5_Z,...,(2k+4)_{\psi_2}^-,...,(2n)_{\bar{Z}}\right)\nonumber\\
  &\mathcal{A}_{2n+1}\left(1_{\psi^1}^+,2_{\psi^1}^+,3_{\psi^2}^+,4_{\psi^2}^+,5_{\bar{Z}},6_Z,...,(2n+1)_{\bar{Z}}\right)\nonumber\\
  &\hspace{40mm} = -\frac{\langle 21 \rangle}{\langle 25 \rangle}\mathcal{A}_{2n+1}\left(1_\gamma^+,2_{\psi^1}^+,3_{\psi^2}^+,4_{\psi^2}^+,5_{\psi_2}^-,6_Z,7_{\bar{Z}},...,(2n+1)_{\bar{Z}}\right) \nonumber\\
  &\hspace{40mm}+\sum_{k=1}^{n-2}\frac{\langle 2,2k+4\rangle}{\langle 25 \rangle}\mathcal{A}_{2n+1}\left(1_{\psi^1}^+,2_{\psi^1}^+,3_{\psi^2}^+,4_{\psi^2}^+,5_{\psi_2}^-,6_Z,...,(2k+4)_{\psi^2}^+,...,(2n+1)_{\bar{Z}}\right).
\end{align}


\bibliographystyle{utphys}
\bibliography{SoftRecBib}

\end{document}